\documentclass[aps,prd,preprint,11pt,superscriptaddress,nofootinbib,floatfix]{revtex4}  
\pdfoutput=1
\usepackage{fancyhdr}
\usepackage{latexsym}
\usepackage{amsfonts}

\usepackage[latin1]{inputenc}
\usepackage{graphicx}
\usepackage{mathrsfs}
\usepackage{amssymb}
\usepackage{amsmath}
\usepackage{verbatim}
\usepackage{multirow}
\usepackage{soul}
\usepackage{subfigure}
\usepackage[usenames,dvipsnames]{color}

\newcommand{\nn}{\nonumber}
\newcommand{\TeV}{{\ensuremath\rm TeV}}
\newcommand{\GeV}{{\ensuremath\rm GeV}}
\newcommand{\MeV}{{\ensuremath\rm MeV}}

\newcommand{\pb}{{\ensuremath\rm pb}}
\newcommand{\eqn}{equation}
\newcommand{\al}{\alpha}
\newcommand{\be}{\beta}
\newcommand{\lb}{\left(}
\newcommand{\rb}{\right)}
\newcommand{\mO}{\mathcal{O}}
\newcommand{\lam}{\lambda}

\def\D0{\slash\!\!\!\!\!\!\!\!\!\:D0}
\newcommand{\HS}{\texttt{HiggsSignals}}
\newcommand{\HB}{\texttt{HiggsBounds}}
\newcommand{\HSv}[1]{\texttt{HiggsSignals-#1}}
\newcommand{\HBv}[1]{\texttt{HiggsBounds-#1}}

\newcommand{\cp}{\mathcal{CP}}
\newcommand{\MHexp}{125\,\GeV}

\newcommand{\csqa}{\cos^2 \alpha}
\newcommand{\sa}{\sin \alpha}
\newcommand{\ssqa}{\sin^2 \alpha}

\newcommand{\tb}{\tan\beta}

\newcommand{\oblique}{Altarelli:1990zd,Peskin:1990zt,Peskin:1991sw,Maksymyk:1993zm}

\begin{document}
\date{\today}
\title{Status of the Higgs Singlet Extension of the {Standard Model} after LHC Run 1\vspace{0.5cm}}
\vspace*{1.0truecm}
\author{Tania Robens}
\email{Tania.Robens@tu-dresden.de}
\affiliation{TU Dresden, Institut f\"ur Kern- und Teilchenphysik,
Zellescher Weg 19, D-01069 Dresden, Germany\vspace{0.2cm}}
\author{Tim Stefaniak\vspace{0.2cm}}
\email{tistefan@ucsc.edu}
\affiliation{Department of Physics and Santa Cruz Institute for Particle Physics, University of California, Santa Cruz, CA 95064, USA\vspace{0.5cm}}
\renewcommand{\abstractname}{\vspace{0.5cm} Abstract}

\begin{abstract}
\vspace{0.5cm}

We discuss the current status of theoretical and experimental constraints on the real Higgs singlet extension of the Standard Model. For the second neutral (non-standard) Higgs boson we consider the full mass range from $1~\GeV$ to $1~\TeV$ accessible at past and current collider experiments. We separately discuss three scenarios, namely, the case where the second Higgs boson is lighter than, approximately equal to, or heavier than the discovered Higgs state {at around $125~\GeV$}. We investigate the impact of constraints from perturbative unitarity, electroweak precision data with a special focus on higher order contributions to the $W$ boson mass, perturbativity of the couplings as well as vacuum stability. The latter two are tested up to a scale of $\sim\,4 \times 10^{10}\,\GeV$ using renormalization group equations. Direct collider constraints from Higgs signal rate measurements at the LHC and $95\%~\mathrm{C.L.}$ exclusion limits from Higgs searches at LEP, Tevatron and LHC are included via the public codes \HS\ and \HB, respectively. We identify the strongest constraints in the different regions of parameter space. We comment on the collider phenomenology of the remaining viable parameter space and the prospects for a future discovery or exclusion at the LHC. 

\end{abstract}

\preprint{SCIPP 15/02}
\maketitle
\tableofcontents

\newpage


\section{Introduction}
\label{Sec:Intro}
\noindent
The LHC discovery~\cite{atlres,cmsres} of a Higgs boson in July 2012 has been a major breakthrough in modern particle physics. {The first runs of the LHC at $7$ and $8~\TeV$ are now completed and the main results from various experimental analyses of the Higgs boson properties have been presented at the 2014 summer conferences.} So far, the discovered state is well compatible~\cite{Aad:2013xqa, Aad:2014eha, Aad:2014eva,Aad:2014xzb,Khachatryan:2014ira, Chatrchyan:2014vua, Chatrchyan:2013mxa,Chatrchyan:2013iaa}  with the interpretation in terms of the scalar boson of the Standard Model (SM) Higgs mechanism~\cite{Higgs:1964ia,Higgs:1964pj,Englert:1964et, Guralnik:1964eu, Kibble:1967sv}. {A simple combination of the Higgs mass measurements performed by ATLAS~\cite{Aad:2014aba} and CMS~\cite{CMS:2014ega} yields a central value of 
\begin{align}
m_H = (125.14 \pm 0.24)~\GeV.
\label{Eq:mhexp}
\end{align}
}
If the discovered particle is indeed the Higgs boson predicted by the SM, its mass constitutes the last unknown ingredient to this model, as all other properties of the electroweak sector then follow directly from theory. The current and future challenge for the theoretical and experimental community is to thoroughly investigate the Higgs boson's properties in order to identify whether the SM Higgs sector is {indeed} complete, or {instead,} the structure of a more involved Higgs sector is realized. On the experimental side, this requires detailed and accurate measurements of its {coupling strengths and $\cp$ structure} at the LHC and {ultimately} at future experimental facilities for Higgs boson precision {studies}, such as the International Linear Collider (ILC)~\cite{Asner:2013psa}. {A complementary and equally important strategy} is to perform collider searches for additional Higgs bosons. {Such a finding would provide clear evidence for a non-minimal Higgs sector. This road needs to be continued within the full mass range that is accessible to current and future experiments.}

In this work, we consider the simplest extension of the SM Higgs sector, where an additional real singlet field is added, which is neutral under all quantum numbers of the SM gauge group~\cite{Schabinger:2005ei,Patt:2006fw} and acquires a vacuum expectation value (VEV). This model has been widely studied in the literature~\cite{Barger:2007im, Bhattacharyya:2007pb, Dawson:2009yx, Bock:2010nz,Fox:2011qc, Englert:2011yb,
  Englert:2011us,Batell:2011pz, Englert:2011aa, Gupta:2011gd, Dolan:2012ac, Bertolini:2012gu,Batell:2012mj,Lopez-Val:2013yba,Heinemeyer:2013tqa,Chivukula:2013xka,Englert:2013tya,Cooper:2013kia,Caillol:2013gqa,Coimbra:2013qq,Pruna:2013bma,Dawson:2013bba, Lopez-Val:2014jva,Englert:2014aca,Englert:2014ffa,Chen:2014ask,Karabacak:2014nca,Profumo:2014opa}. Here, we present a complete {exploration of the model parameter space in the light of} the latest experimental and theoretical {constraints}. We consider masses of the second (non-standard) Higgs boson in the whole mass range up to $1\,\TeV$, thus extending and updating the findings of previous work~\cite{Pruna:2013bma}. This minimal setup can be interpreted as a limiting case for more generic BSM scenarios, e.g.~models with additional gauge sectors~\cite{Basso:2010jm} or additional matter content~\cite{Strassler:2006im,Strassler:2006ri}. 
  
 In our analysis, we study the implications of various constraints: We take into account bounds from perturbative unitarity and electroweak (EW) precision measurements, {in particular} focus{sing} on higher order corrections to the $W$ boson mass~\cite{Lopez-Val:2014jva}. Furthermore, we study the impact of requiring perturbativity, vacuum stability and correct minimization of the model up to a high energy scale using renormalization group evolved couplings\footnote{The value of this high energy scale is chosen to be larger than the energy scale where the running SM Higgs quartic coupling turns negative. This will be made more precise in Section~\ref{Sect:Constraints}.}. We include the exclusion limits from Higgs searches at the LEP, Tevatron and LHC experiments via the public tool \HB~\cite{Bechtle:2008jh,Bechtle:2011sb,Bechtle:2013gu,Bechtle:2013wla}, and use the program \HS~\cite{Bechtle:2013xfa} (cf.~also Ref.~\cite{Bechtle:2014ewa}) to test the compatibility of the model with the signal strength measurements of the discovered Higgs state. 
  
We separate the discussion of the parameter space into three different mass regions: \emph{(i)} the high mass region, $m_H \in [130, 1000]\,\GeV$, where the lighter Higgs boson $h$ is interpreted as the discovered Higgs state; \emph{(ii)} the intermediate mass region, where both Higgs bosons $h$ and $H$ are located in the mass region $[120,130]\,\GeV$ and potentially contribute to the measured signal rates and \emph{(iii)} the low mass region, $m_h \in [1,120]\,\GeV$, where the heavier Higgs boson $H$ is interpreted as the discovered Higgs state.
  
 We find that the most severe constraints in the whole parameter space for the second {Higgs} mass {$m_H \lesssim 300~\GeV$} are {mostly} given by limits from {collider searches for a SM Higgs boson} as well as {by} the {LHC} Higgs {boson} signal strength {measurements}. For $m_H\gtrsim 300\,\GeV$ limits from higher order contributions to the $W$ boson mass prevail, followed by the requirement of perturbativity of the couplings which is tested via renormalization group equation (RGE) evolution. {For the remaining viable parameter space} we present predictions for {signal} cross sections of the {yet undiscovered second} Higgs boson {for the LHC at center-of-mass (CM) energies of} $8$ and $14\,\TeV${, discussing both the SM Higgs decay signatures and the novel Higgs-to-Higgs decay mode $H\to hh$.} {We furthermore} present our results in terms of a global suppression factor $\kappa$ for SM-like channels as well as the total width $\Gamma$ of the {second} Higgs {boson}, and {show} regions which are allowed in the $(\kappa,\Gamma)$ plane.

The paper is organized as follows: In Section~\ref{Sec:Model} we briefly review the model {and the chosen parametrization}. {In} Section~\ref{Sect:Constraints} {we elaborate upon the various theoretical and experimental constraints and discuss their impact on the model parameter space.} In Section~\ref{sec:results} a scan {of the full} model parameter space {is presented}, {in which all relevant constraints are combined.} {This is followed by a discussion of the collider phenomenology of the viable parameter space.} We summarize {and conclude} in Section~\ref{Sec:Conclusions}.



\section{The model}
\label{Sec:Model}
\subsection{Potential and couplings}
The real Higgs singlet extension of the SM is described in detail in Refs.~\cite{Schabinger:2005ei,Patt:2006fw, Bowen:2007ia,Pruna:2013bma}. Here, we only briefly review the theoretical setup as well as the main features relevant to the work presented here.

We consider the extension of the SM electroweak sector containing a complex $SU(2)_L$ doublet, in the following denoted by $\Phi$, by an additional real scalar $S$ which is a singlet under all SM gauge groups. The most generic renormalizable Lagrangian is then given by 
\begin{equation}\label{lag:s}
\mathscr{L}_s = \left( D^{\mu} \Phi \right) ^{\dagger} D_{\mu} \Phi + 
\partial^{\mu} S \partial_{\mu} S - V(\Phi,S ) \, ,
\end{equation}
with the scalar potential
\begin{eqnarray}\label{potential}\nonumber
V(\Phi,S ) &=& -m^2 \Phi^{\dagger} \Phi - \mu ^2  S ^2 +
\left(
\begin{array}{cc}
\Phi^{\dagger} \Phi &  S ^2
\end{array}
\right)
\left(
\begin{array}{cc}
\lambda_1 & \frac{\lambda_3}{2} \\
\frac{\lambda_3}{2} & \lambda _2 \\
\end{array}
\right)
\left(
\begin{array}{c}
\Phi^{\dagger} \Phi \\  S^2 \\
\end{array}
\right) \\
\nonumber \\ 
&=& -m^2 \Phi^{\dagger} \Phi -\mu ^2 S ^2 + \lambda_1
(\Phi^{\dagger} \Phi)^2 + \lambda_2  S^4 + \lambda_3 \Phi^{\dagger}
\Phi S ^2.
\end{eqnarray}
Here, we implicitly impose a $Z_2$ symmetry which forbids all linear or cubic terms of the singlet field $S$ in the potential. The scalar potential $V(\Phi,S )$ {is} bounded from below if the following conditions are fulfilled:
\begin{eqnarray}
4 \lambda_1 \lambda_2 - \lambda_3^2 &>& 0 , \label{bound_pot}\\
\lambda_1, \lambda_2 &>& 0 \label{pos_pot},
\end{eqnarray}
cf.~Appendix \ref{app:hpot}. If the first condition, Eq.~\eqref{bound_pot}, is fulfilled, the extremum is a local minimum. The second condition, Eq.~\eqref{pos_pot}, guarantees that the potential is bounded from below for large field values. We assume that both Higgs fields {$\Phi$} and $S$ have a non-zero vacuum expectation value (VEV), {denoted by $v$ and $x$, respectively.}
In the unitary gauge, the Higgs fields are given by
\begin{equation}\label{unit_higgs}
\Phi \equiv
\left(
\begin{gathered}
0 \\
\frac{\tilde{h}+v}{\sqrt{2}}
\end{gathered} \right), 
\hspace{2cm}
S \equiv \frac{h'+x}{\sqrt{2}}.
\end{equation} 
Expansion around the minimum leads to the {squared} mass matrix
\begin{align}
\mathcal{M}^2 =  \left( \begin{array}{cc}  
      2\lambda_1v^2 & \lambda_3 vx \\
      \lambda_3vx & 2\lambda_2x^2  
     \end{array} \right)
\label{eq:mass-matrix}
\end{align}
with the mass eigenvalues 
\begin{align}\label{mh1}
m^2_{h} &= \lambda_1 v^2 + \lambda_2 x^2 - \sqrt{(\lambda_1 v^2 -
  \lambda_2 x^2)^2 + (\lambda_3 x v)^2}, \\
\label{mh2}
m^2_{H} &= \lambda_1 v^2 + \lambda_2 x^2 + \sqrt{(\lambda_1 v^2 -
  \lambda_2 x^2)^2 + (\lambda_3 x v)^2},
\end{align}
where $h$ and $H$ are the scalar fields with masses
$m_{h}$ and $m_{H}$ respectively, {and $m^2_{h} \le m^2_{H}$ by convention}. Note that $m_h^2\,\geq\,0$ follows from Eq.~\eqref{bound_pot} and we assume {Eqs.~\eqref{bound_pot} and~\eqref{pos_pot} to be fulfilled} in all following definitions. The gauge and mass eigenstates are related via the mixing matrix
\begin{equation}\label{eigenstates}
\left(
\begin{array}{c}
h \\
H
\end{array}
\right) = \left(
\begin{array}{cc}
\cos{\alpha} & -\sin{\alpha} \\
\sin{\alpha} & \cos{\alpha}
\end{array}
\right) \left(
\begin{array}{c}
\tilde{h} \\
h'
\end{array}
\right),
\end{equation}
where {the mixing angle} $-\frac{\pi}{2} \leq \alpha \leq \frac{\pi}{2}$ {is given by}
\begin{align}\label{sin2a}
\sin{2\alpha} &= \frac{\lambda_3 x v}{\sqrt{(\lambda_1 v^2 -
    \lambda_2 x^2)^2 + (\lambda_3 x v)^2}}, \\
\label{cos2a}
\cos{2\alpha} &= \frac{\lambda_2 x^2 - \lambda_1
  v^2}{\sqrt{(\lambda_1 v^2 - \lambda_2 x^2)^2 + (\lambda_3 x v)^2}}.
\end{align}
It follows from Eq.~\eqref{eigenstates} that the light (heavy) Higgs boson couplings to SM particles are suppressed by $\cos\al\,(\sin\al)$.

Using Eqs.~\eqref{mh1},~\eqref{mh2} and \eqref{sin2a}, we can express the couplings in terms of the mixing angle $\alpha$, the Higgs VEVs $x$ {and $v$} and the two Higgs boson masses, $m_h$ and $m_H$:
\begin{eqnarray}\label{isomorphism}
\lambda_1&=&\frac{m_{h}^2}{2 v^2} + \frac{ \left(
m_{H}^2 - m_{h}^2 \right)}{2 v^2}\sin^2{\alpha} =
\frac{ m_{h}^2}{2v^2}\cos^2{\alpha} +  \frac{m_{H}^2}{2
v^2}\sin^2{\alpha},\nonumber \\
\lambda_2&=&\frac{m_{h}^2}{2 x^2} + \frac{ \left(
m_{H}^2 - m_{h}^2 \right)}{2 x^2}\cos^2{\alpha} =
\frac{ m_{h}^2}{2x^2}\sin^2{\alpha} +  \frac{m_{H}^2}{2
x^2}\cos^2{\alpha}, \nonumber \\
\lambda_3&=&\frac{ \left(
m_{H}^2 - m_{h}^2 \right)}{ 2vx}\sin{(2\alpha)}.
\end{eqnarray}

If kinematically allowed, the additional decay channel $H\to hh$ is present. Its {partial} decay width is given by~\cite{Schabinger:2005ei, Bowen:2007ia}
\begin{align}\label{eq:gtot}
\Gamma_{H\rightarrow hh}\,=\,\frac{|\mu'|^2}{8\pi m_{H}}\,\sqrt{1-\frac{4 m^2_{h}}{m_{H}^2}} \, ,
\end{align}
where the coupling strength $\mu'$ of the $H \rightarrow hh$ decay reads
 \begin{align}
\mu'=-\frac{\sin\lb 2\al \rb}{2vx}\,\lb \sin\al v+ \cos\al\,x\rb\,\lb m_h^2+\frac{m_H^2}{2} \rb.
\label{eq:muprime}
\end{align}
We therefore obtain as branching ratios for the {\sl heavy} Higgs mass eigenstate $m_H$ 
\begin{eqnarray}\label{eq:brdefs}
\text{BR}_{H\rightarrow hh}&=&\frac{\Gamma_{H\rightarrow hh}}{\Gamma_\text{tot}},\nn \\
\text{BR}_{H\rightarrow \text{SM}}&=&\sin^2\al \times \frac{\Gamma_{\text{SM}, H\rightarrow\text{SM}}}{\Gamma_\text{tot}},
\end{eqnarray}
where $\Gamma_{\text{SM},\,H\rightarrow\text{SM}}$ denotes the partial decay width of the SM Higgs boson {and $H\to\text{SM}$ represents any SM Higgs decay mode}. The total width is given by
\begin{\eqn*}
\Gamma_\text{tot}\,=\,\sin^2\al\,\times\,\Gamma_\text{SM, tot}+\Gamma_{H\to hh},
\end{\eqn*}
where $\Gamma_\text{SM, tot}$ denotes the total width of the SM Higgs boson with mass $m_H$. The suppression by $\sin^2\,\al$ directly follows from the suppression of all SM--like couplings, cf.~Eq.~\eqref{eigenstates}. For $\mu'\,=\,0$, we recover the SM Higgs boson branching ratios.

For collider phenomenology, two features are important:

\begin{itemize}
\item the suppression of the {\sl production {cross section}} of the
  two {Higgs states induced by the mixing}, which {is} given by $\sin^2\al\,(\cos^2\al)$ for the heavy (light) Higgs, respectively;
\item {the} suppression of the {\sl {Higgs} decay modes to SM particles}, which is {realized} if the competing decay mode $H\to hh$ is kinematically accessible. 

\end{itemize}
For the high mass (low mass) scenario, {i.e.~the case where the light (heavy) Higgs boson is identified with the discovered Higgs state at $\sim 125~\GeV$}, $|\sin\al|\,=\,0\,(1)$ corresponds to the complete decoupling of the second Higgs {boson} and therefore the SM-like scenario.

\subsection{Model parameters}
\noindent
At the Lagrangian level, the model has five free parameters,
\begin{\eqn}
\lambda_1,\,\lambda_2,\,\lambda_3,\,v,\,x,
\end{\eqn}
while the values of the additional parameters $\mu^2,\,m^2$ are fixed by the minimization conditions 
to the values 
\begin{align}
m^2 &= \lam_1\,v^2+\frac{\lam_3}{2}x^2,\\ 
\mu^2&=\lam_2\,x^2+\frac{\lam_3}{2}v^2,
\end{align}
cf.~Appendix~\ref{app:hpot}. In this work, we choose {to parametrize the model in terms of the independent physical quantities} 
\begin{\eqn}\label{eq:pars}
m_h,\,m_H,\,\alpha,\,v,\,\tan\beta\,\equiv\,\frac{v}{x}.
\end{\eqn}
The couplings $\lambda_1, \lambda_2$ and $\lambda_3$ can then be expressed via Eq.~\eqref{isomorphism}. The vacuum expectation value of the Higgs doublet {$\Phi$} is given by the SM value $v~\sim~246~\GeV$. Unless otherwise stated, we fix one of the Higgs masses to be {$m_{h/H}\,=\,125.14\,\GeV$}, hence interpreting the Higgs boson {$h/H$} as the discovered Higgs state at the LHC. In this case, we are left with only three independent parameters, ${m\equiv m_{H/h}},\,\sin\alpha,\,\tan\be$, where the latter enters the collider phenomenology only via the additional decay channel\footnote{In fact, all Higgs self-couplings depend on $\tan\be$. However, in the factorized leading-order description of production and decay followed here, and as long as no experimental data exists which constrains the Higgs boson self-couplings, only the $Hhh$ coupling needs to be considered.} $H\to hh$.


\section{Theoretical and experimental constraints} 
\label{Sect:Constraints}
\noindent
We now discuss the various theoretical and experimental constraints on the singlet extension model. In our analysis, we impose the following constraints:
\begin{enumerate}
\item[(1.)]{}limits from perturbative unitarity,
\item[(2.)]{} limits from from EW precision data in form of the $S,\,T,\,U$ parameters~\cite{\oblique} as well as the singlet--induced NLO corrections to the $W$ boson mass as presented in Ref.~\cite{Lopez-Val:2014jva},
\item[(3.)]{} perturbativity of the couplings as well as the requirement on the potential to be bounded from below, Eqs.~\eqref{bound_pot} and \eqref{pos_pot},
\item[(4.)]{} limits from perturbativity of the couplings as well as vacuum stability up to a certain scale $\mu_\text{run}$, where we chose $\mu_\text{run}\,\sim\,4\,\times\,10^{10}\,\GeV$ as benchmark point (these constraints will only be applied in the high mass region, see Section~\ref{sec:rge} for further discussion),
\item[(5.)]{} upper cross section limits at $95\%~\mathrm{C.L.}$ from null results in Higgs searches at the LEP, Tevatron and LHC experiments,
\item[(6.)]{} consistency with the Higgs boson signal rates measured at the LHC experiments. 
\end{enumerate}

The constraints {(1.) -- (4.)} have already been discussed extensively in a previous publication~\cite{Pruna:2013bma}, where the scan was however restricted to the case that $m_H \geq 600\,\GeV$. In the following, we will therefore briefly recall the definition of the theoretically motivated bounds and comment {on} their importance in the whole mass range $m_{h/H}\,\in\,[1,\,{1000}]\,\GeV$.

\subsection{Perturbative unitarity}
\label{Sect:unitarity}

Tree-level perturbative unitarity~\cite{Lee:1977eg,Luscher:1988gc} puts a constraint on the Higgs masses via a relation 
on the partial wave amplitudes $a_J(s)$ of all possible $2\,\rightarrow\,2$ scattering processes:
\begin{eqnarray}\label{condition}
|\textrm{Re}(a_J(s))|\leq \frac{1}{2},
\end{eqnarray}
where the partial wave amplitude $a_0$ poses the strongest constraint.
Following Ref.~\cite{Pruna:2013bma}, we consider all $2\,\rightarrow\,2$ processes
 $X_1\,X_2\,\rightarrow\,Y_1\,Y_2$, with $(X_1,X_2),\,(Y_1,Y_2)\,\in\,(W^+\,W^-,ZZ,hh,hH,HH)$, and impose the condition of Eq.~\eqref{condition} to the eigenvalues of the diagonalized scattering matrix.
Note that the unitarity constraint based on the consideration of  $W_L\,W_L\,\rightarrow\,W_L\,W_L$ scattering alone, leading to $m_H\lesssim 700\,\GeV$ (as e.g.~in Ref.~\cite{Englert:2011yb}), is much loosened when all scattering channels are taken into account~{\cite{Bowen:2007ia,Basso:2010jt}}.

In general, perturbative unitarity poses an upper limit on $\tan\be$. In the decoupling case, which corresponds to $\sin\alpha \to 0~(1)$ for the light (heavy) Higgs being SM-like, it is given by~\cite{Pruna:2013bma}
\begin{\eqn}\label{eq:tbpu}
\tan^2\be \leq \frac{16\pi v^2}{3m^2}\,+\,\mO\lb \al \rb\quad \text{for }\quad a_0(hh\rightarrow hh)\leq 0.5,
\end{\eqn}
where $h$ and $m$ refer to the purely singlet Higgs state and its respective mass.

\begin{figure}
\centering
\includegraphics[width=0.55\textwidth]{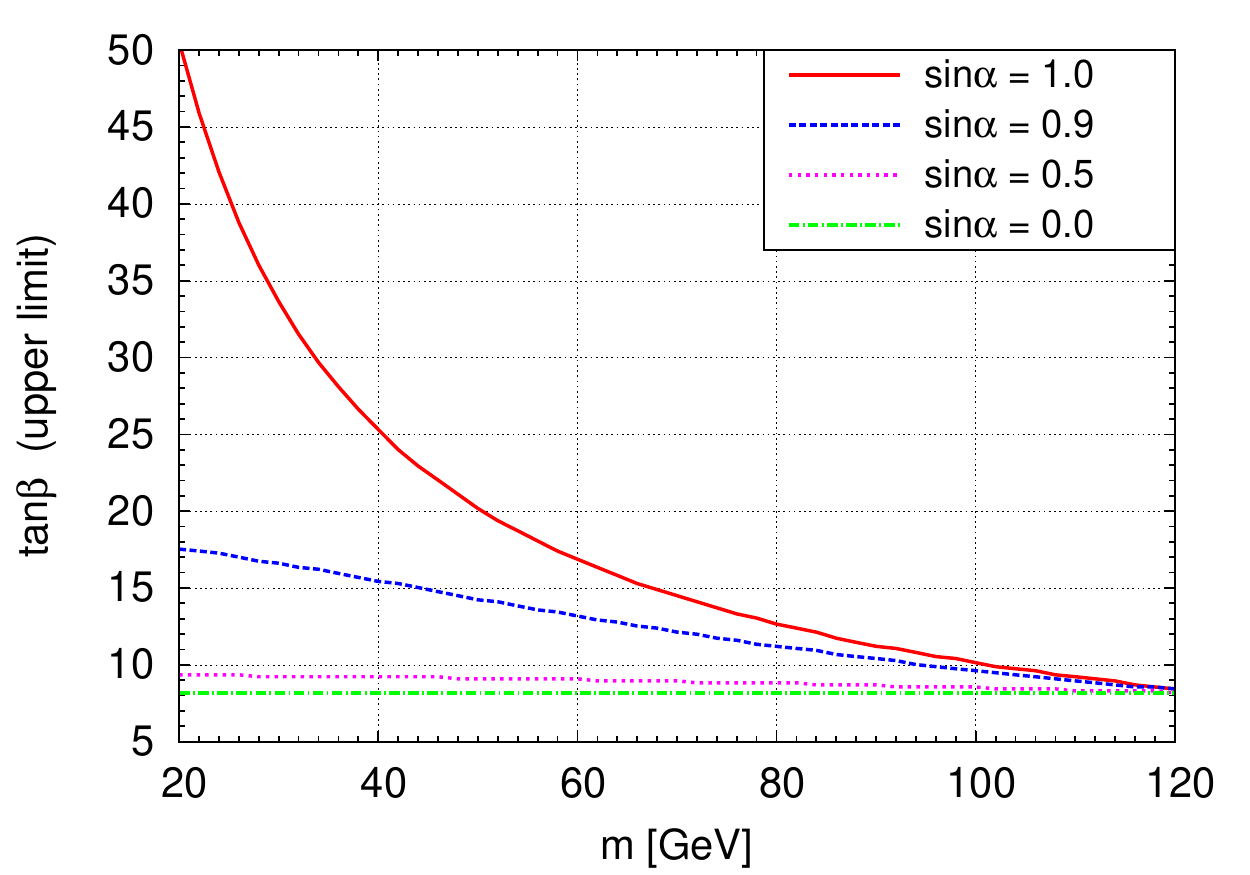}

\caption{\label{fig:tbmax} {Maximally allowed values for $\tan\be$ in the low mass range, $m \in [20, 120]~\GeV$, for various values of $\sin\alpha = 1.0, 0.9, 0.5, 0.0$, from considering only perturbative unitarity.}
}
\end{figure}

While in the high mass scenario this bound is always {superseded} by bounds from perturbativity of the couplings, {cf.~Section~\ref{sec:rge}}, in the low mass scenario this poses the strongest theoretical bound on $\tan\be$. We exemplarily show the upper limits {on $\tan\be$ derived from perturbative unitarity} in Fig.~\ref{fig:tbmax} {for the low mass range $m \in [20, 120]$ for the four values of $\sin\al = 1.0, 0.9, 0.5, 0.0$}.
{The bounds on $\tan\be$ are strongest for small values of $\sin\al$.} However, values too far from the decoupling case $\sin\alpha \approx 1$ are highly constrained by Higgs searches at LEP as well as by the LHC signal strength measurements of the heavier Higgs at $\sim\MHexp$, cf.~Sections~\ref{Sect:HB} and~\ref{Sect:HS} for more details.

\subsection{Perturbativity of the couplings}
For perturbativity of the couplings, we require that
\begin{\eqn}\label{eq:lampert}
|\lam_i| \leq 4\pi,\;i \in (1,2,3).
\end{\eqn}
At the electroweak scale, these bounds do not pose additional constraints on the parameter space after limits from perturbative unitarity have been taken into account. 

\subsection[RGE evolution of the couplings]{Renormalization group equation evolution of the couplings}\label{sec:rge}

While perturbativity as well as vacuum stability and the existence of a local minimum at the electroweak scale are necessary ingredients for the validity of a parameter point, it is instructive to investigate up to which energy scale these requirements remain valid. In particular, we study whether the potential is bounded from below and features a local minimum at energy scales above the electroweak scale. In order to achieve this, we promote the requirements of Eqs.~\eqref{bound_pot}, \eqref{pos_pot}, and \eqref{eq:lampert} to be valid at an arbitrary scale $\mu_\text{run}$, where $\lam_i\lb \mu_\text{run} \rb$ are evolved according to the one-loop {renormalization group equations} (RGEs) (see~e.g.~Ref.~\cite{Lerner:2009xg})
\begin{align}
\frac{d}{dt}\lambda_1&=\frac{1}{16\,\pi^2}\left\{\frac{1}{4}\lam_3^2+12\,\lam_1^2+6\,\lam_1\,y_t^2-3\,y_t^4-\frac{3}{2}\lam_1\,\lb 3\,g^2+g_1^2 \rb\,+\,\frac{3}{16}\left[ 2\,g^4+\lb g^2+g_1^2\rb^2  \right]  \right\},\\
\frac{d}{dt}\lam_2&=\frac{1}{16\,\pi^2}\left[ \lam_3^2+9\,\lam_2^2 \right],\\
\frac{d}{dt}\lam_3&=\frac{1}{16\,\pi^2}\,\lam_3\,\left[ 6\,\lam_1+3\,\lam_2+2\,\lam_3+3\,y_t^2-\frac{3}{4}\lb 3\,g^2+g_1^2 \rb \right].
\end{align}
Here we introduced $t\,=\,2\,\log\,\lb \mu_\text{run}/{v}\rb$ as a dimensionless running parameter. The initial conditions at the electroweak scale require that $\lam_i\lb \mu_\text{res}\,\equiv\,v \rb$ are given by Eqn.~\eqref{isomorphism}. The top Yukawa coupling $y_t$ as well as the SM gauge couplings $g,\,g_1$ evolve according to the {one-loop} SM RGEs, cf.~Appendix~\ref{sec:rge_sm}. For the decoupling case as well as to cross-check the implementation of the running of the gauge couplings we reproduced the results of~Ref.~\cite{Degrassi:2012ry}.

As in Ref.~\cite{Pruna:2013bma}, we require all RGE-dependent constraints to be valid at a scale which is slightly higher than the breakdown scale of the SM, {such that the singlet extension of the SM improves the stability of the electroweak vacuum. The SM breakdown scale is defined} as the scale where the potential becomes unbounded from below in the {decoupled, SM-like} scenario. With the input values of $m_h={125.14}~\GeV$ and $v=246.22~\GeV$, a top mass of $173.0\,\GeV$ as well as a top-Yukawa coupling $y_t(m_t)\,=\,0.93587$ and {strong coupling constant} $\al_s(m_Z)\,=\,0.1184$, we obtain as a SM breakdown scale\footnote{As has been discussed in e.g.~Ref.~\cite{Degrassi:2012ry}, the scale where $\lam_1\,=\,0$ in the decoupling case strongly depends on the initial input parameters. However, as we are only interested in the {\sl difference} of the running in the case of a non-decoupled singlet component {with respect to the Standard Model}, we do not need to determine this scale to the utmost precision. For a more thorough discussion of the behavior of the RGE-resulting constraints in case of varying input parameters, see e.g.~Ref.~\cite{Pruna:2013bma}.}
\begin{\eqn*}
\mu_\text{run, SM bkdw}\,\sim\,{2.5}\,\times\,10^{10}\,\GeV .
\end{\eqn*}
We therefore chose as a {slightly higher} test scale the value $\mu_\text{run, stab}\,\sim\,{4.0}\,\times\,10^{{10}}\,\GeV$. Naturally, we only apply this test to points in the parameter space which have passed constraints from perturbative unitarity as well as perturbativity of the couplings at the electroweak scale. 
Changing the scale to higher (lower) values leads to more (less) constrained regions in the models parameter space~\cite{Pruna:2013bma}. 

\begin{figure}[!tb]
\begin{minipage}{0.49\textwidth}
\includegraphics[width=1.1\textwidth]{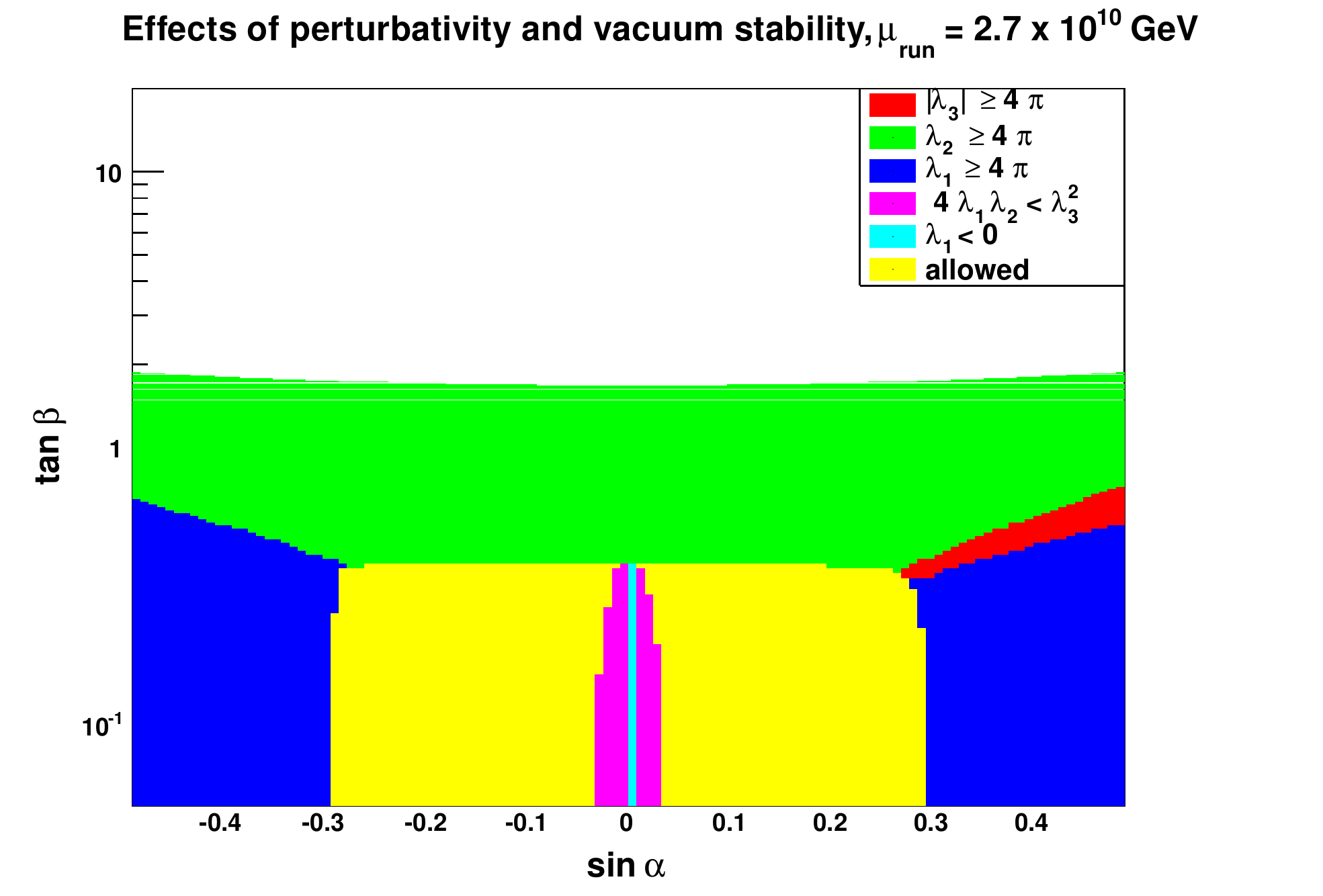}
\end{minipage}
\caption{\label{fig:mh600_scales} Limits {in the ($\sin\al, \tan\be$) plane} for $m_H=600~\GeV$ {from requiring perturbativity and vacuum stability at a scale} $\mu_\text{run}\,=\,2.7\,\times\,10^{10}\,\GeV$ {using RGE evolution}. Taken from Ref.~\cite{Pruna:2013bma}. }
\label{Fig:RGE_sinatb_old}
\end{figure}

In the high mass scenario {we see the behavior studied in Ref.~\cite{Pruna:2013bma} for Higgs masses $\ge 600~\GeV$ continuing to the lower mass ranges.} The {strongest} constraints {that impact} different parts of the $(\sin\al,\tan\be)$ parameter space {are displayed in Fig.~\ref{Fig:RGE_sinatb_old} for a heavy Higgs mass of $m_H=600~\GeV$ (taken from Ref.~\cite{Pruna:2013bma}). Two main features can be observed:}

First, the upper value of $\tan\be$ for fixed Higgs masses is determined by {requiring} perturbativity of $\lam_2$ as well as perturbative unitarity, cf.~Section~\ref{Sect:unitarity}.
Second, the allowed range of the mixing angle $\sin\alpha$ is determined by perturbativity of the couplings as well as the requirement of vacuum stability, especially when these are required at renormalization scales $\mu_\text{run}$, which are significantly larger than the electroweak scale. {\sl Small} mixings are excluded by the requirements of vacuum stability\footnote{For the requirement of vacuum stability, we found that in some cases the coupling strengths vary very mildly over large variations of the RGE running scale. In these regions the inclusion of higher order corrections in the spirit of Ref.~\cite{Degrassi:2012ry} seems indispensable. Therefore, all lower limits on the mixing angle originating from RGE constraints need to be viewed in this perspective. In fact, such higher order contributions to the scalar-extended RGEs have recently been presented in Ref.~\cite{Costa:2014qga}. However, the authors did not specifically investigate the higher order effects on parameter points which exhibit small variations over large energy scales at NLO.} as well as minimization of the scalar potential. This corresponds to the fact that we enter an unstable vacuum for $\mu_\text{run}\,\gtrsim\,\mu_\text{SM,bkdw}$ for $\sin\al\,\sim\,0$.

In summary, the constraints from RGE evolution of the couplings pose the strongest bounds on the minimally allowed value $|\sin\al|$ and the maximal value of $\tan\be$ {in the high mass scenario}. {Note that,} for lower $m_H$, the $(\sin\al,\,\tan\be)$ parameter space is less constrained, as will be discussed in Section~\ref{sec:highmassres}.

In the low mass scenario, i.e.~where the heavier Higgs state is considered to be the discovered Higgs boson, none of the points in our scan fulfilled vacuum stability above the electroweak scale. 
This is due to the fact that for a relatively {low} $m_h$, the value of $\lam_1$ at the electroweak scale is quite small, cf.~Eq.~\eqref{isomorphism}. In the non-decoupled case, $|\sin\al|\ne1$, $\lam_1$ then receives negative contributions in the RG evolution towards higher scales, leading to $\lam_1(\mu_\text{run})\,\leq\,0$ already at relatively low scales $\mu_\text{run}$, corresponding to the breakdown of the electroweak vacuum. 
Hence, in the low mass scenario, the theory breaks down even earlier than in the SM case. In the analysis presented here, we will therefore refrain from taking limits from RGE running into account in the low mass scenario. 
Then, the theoretically maximally allowed value of $\tan\be$ is determined from perturbative unitarity and rises to quite large values, where we obtain $\tan\be_\text{max}\,\lesssim 50$, depending on the value of the light Higgs mass $m_h$. 

Further constraints on $\tan\be$ in the low mass scenario stem from the Higgs signal rate observables {through the potential decay $H\rightarrow hh$}, as {will be} discussed in Section~\ref{Sect:HS}.

\subsection[The $W$ boson mass and $S$, $T$, $U$]{The $W$ boson mass {and electroweak oblique parameters $S$, $T$, $U$}}
\label{sec:Wmass}

Recently, the one--loop corrections to the $W$ boson mass, $m_W$, for this model have been {calculated} in Ref.~\cite{Lopez-Val:2014jva}. In that analysis, $m_W$ is required {to agree within $2\sigma$ with} the experimental value $m_W^\text{exp}\,=\,80.385\,\pm\,0.015\,\GeV$~\cite{Alcaraz:2006mx, Aaltonen:2012bp, D0:2013jba}, leading to {an allowed} range for the purely singlet-induced corrections of $\Delta\,m_W^\text{sing}\,\in\,\left[-5\,\MeV;\,55\,\MeV\right]$. {T}heoretical uncertainties due to contributions {at even higher orders} have been estimated to be $\mO\lb 1\,\MeV\rb$. The one-loop corrections are independent\footnote{In the electroweak gauge sector, $\tan\be$ only enters at the 2--loop level when the Higgs mass sector is renormalized in the on--shell scheme.} of $\tan\be$ and give rise to additional constraints on $|\sin\al|$, which in the high mass scenario turn out to be much more stringent~\cite{Lopez-Val:2014jva} than the constraints obtained from the oblique parameters $S$, $T$ and $U$~\cite{\oblique}.

\begin{figure}
\includegraphics[width=0.55\textwidth]{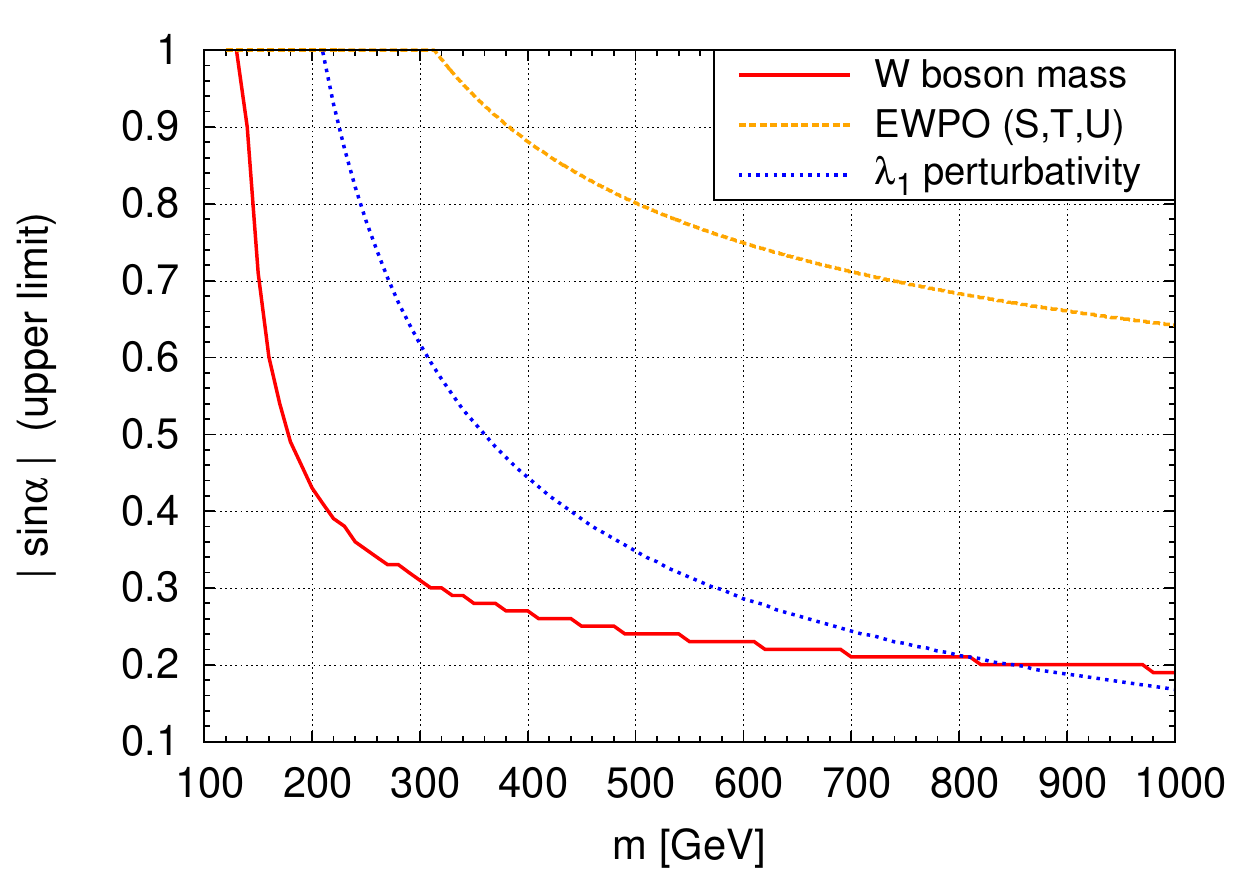}
\caption{\label{fig:sinamw} Maximal allowed values for $| \sin\al |$ in the high mass region, $m_H\in [130, 1000]\,\GeV$, from  {NLO} calculations of the $W$ boson mass (\emph{red, solid})~\cite{Lopez-Val:2014jva}, electroweak precision observables (EWPOs) tested via the oblique parameters $S$, $T$ and $U$ (\emph{orange, dashed}), as well as from the perturbativity requirement of the RG-evolved coupling $\lam_1$ (\emph{blue, dotted}), evaluated at $\tan\be\,=\,0.1$. For Higgs masses $m_H\lesssim 800\,\GeV$ the NLO corrections to the $W$ boson mass {yield the strongest constraint}.}
\end{figure}

Figure \ref{fig:sinamw} shows the maximally allowed mixing angle {obtained} from the $m_W$ {constraint as a function of the heavy Higgs mass $m_H$} in the high mass scenario. For comparison, we also included the limit stemming from {the electroweak oblique parameters $S$, $T$, and $U$ (see below), as well as from requiring} perturbativity of $\lam_1\lb \mu_\text{run} \rb$, evaluated at $\tan\be = 0.1$. We see that for $m_H\lesssim 800\,\GeV$ constraints from $m_W$ {yield the strongest constraint}. {The oblique parameters $S$, $T$ and $U$ do not pose additional limits on the allowed parameter space.}

In the low mass region, as discussed in Ref.~\cite{Lopez-Val:2014jva}, the NLO contributions within the Higgs singlet extension model even tend to decrease the current $\sim\,20\,\MeV$ discrepancy between the theoretical value $m_W$ {in the SM}~\cite{Awramik:2003rn} and the experimental measurement~\cite{Alcaraz:2006mx,Aaltonen:2012bp,D0:2013jba}. {However, substantial reduction of the discrepancy only occurs if the light Higgs has a sizable doublet component. Hence, this possibility is strongly constrained by exclusion limits from LEP and/or LHC Higgs searches (depending on the light Higgs mass) as well as by the LHC Higgs signal rate measurements.}

In the low mass region the electroweak {oblique parameters} pose non-negligible constraints, as will be shown in Section~\ref{Sect:lowmass}. However, these {constraints} are again superseded once the Higgs signal strength as well as direct search limits {from LEP} are taken into account, cf. Section~\ref{Sect:HB} and~\ref{Sect:HS} respectively.

{In our analysis} we test the constraints from the electroweak oblique parameters $S$, $T$ and $U$ by evaluating
\begin{align}
\chi^2_\mathrm{STU} = \mathbf{x}^T \mathbf{C}^{-1} \mathbf{x},
\end{align}
with $\mathbf{x}^T = (S - \hat{S}, T - \hat{T}, U - \hat{U})$, where the observed parameters are given by~\cite{Baak:2014ora}
\begin{align}
\hat{S} = 0.05,\quad \hat{T} = 0.09,\quad \hat{U} = 0.01,
\end{align}
and the \emph{unhatted} quantities denote the model predictions~\cite{Lopez-Val:2014jva}.\footnote{The exact one--loop quantities from Ref.~\cite{Lopez-Val:2014jva} render qualitatively the same constraints as the $S,T,U$ values used in Ref.~\cite{Pruna:2013bma}, which were obtained from rescaled SM expressions~\cite{Hagiwara:1994pw}.} The covariance matrix reads~\cite{Baak:2014ora}
\begin{align}
(\mathbf{C})_{ij} = \left(\begin{array}{ccc}
0.0121 & 0.0129 & -0.0071 \\
0.0129 & 0.0169 & -0.0119 \\
-0.0071 & -0.0119 & 0.0121 \\
\end{array}\right).
\end{align}
We then require $\chi^2_\mathrm{STU} \le 8.025$, corresponding to a {maximal} $2\sigma$ deviation given the three degrees of freedom.

\subsection[Exclusion limits from Higgs searches\dots]{Exclusion limits from Higgs searches at LEP and LHC}
\label{Sect:HB}

Null results from Higgs searches at collider experiments limit the signal strength of the second, non SM-like Higgs boson. Recall that its signal strength is given by the SM Higgs signal rate scaled by $(\cos \alpha)^2$ in the low mass region and, in the absence of Higgs-to-Higgs decays, $(\sin\alpha)^2$ in the high mass region. Thus, the exclusion limits can easily be translated into lower or upper limits on the mixing angle $|\sin \alpha|$, respectively.\footnote{{Here we neglect the possible influence of interference effects in the production of the light and heavy Higgs boson and its successive decay. Recent studies~\cite{Hagiwara:2005wg, Uhlemann:2008pm, Kalinowski:2008fk, Goria:2011wa, Wiesler:2012rkl,Kauer:2012hd,Englert:2014aca, Logan:2014ppa, Maina:2015ela} have shown that interference and finite width effect can lead to sizable deviations in the invariant mass spectra of prominent LHC search channels such as $gg\to H\to ZZ^{*} \to 4\ell$ in the high mass region and thus should be taken into account in accurate experimental studies of the singlet extended SM at the LHC. However, the inclusion of these effects is beyond the scope of the work presented here.\label{Footnote:interference}}}


We employ \HBv{4.2.0}~\cite{Bechtle:2008jh,Bechtle:2011sb,Bechtle:2013gu,Bechtle:2013wla} to derive the exclusion limits from collider searches. The exclusion limits from the LHC experiments\footnote{{\HB\ also contains limits from the Tevatron experiments. In the singlet extended SM, however, these limits are entirely superseded by LHC results.}} are usually given at the $95\%~\mathrm{C.L.}$. For most of the LEP results we employ the $\chi^2$ extension~\cite{Bechtle:2013wla} of the \HB\ package.\footnote{The LEP $\chi^2$ information is available for Higgs masses $\ge 4~\GeV$. For lower masses, we take the conventional $95\%~\mathrm{C.L.}$ output from \HB.} The obtained $\chi^2$ value will later be added to the $\chi^2$ contribution from the Higgs signal rates, cf.~Section~\ref{Sect:HS}, to construct a global likelihood.

\begin{figure}[!tb]
\subfigure[\label{fig:sinaexp_a} Mass region $m~{\in}~[1,100{]}~\GeV$. 
]{\includegraphics[width=0.495\textwidth]{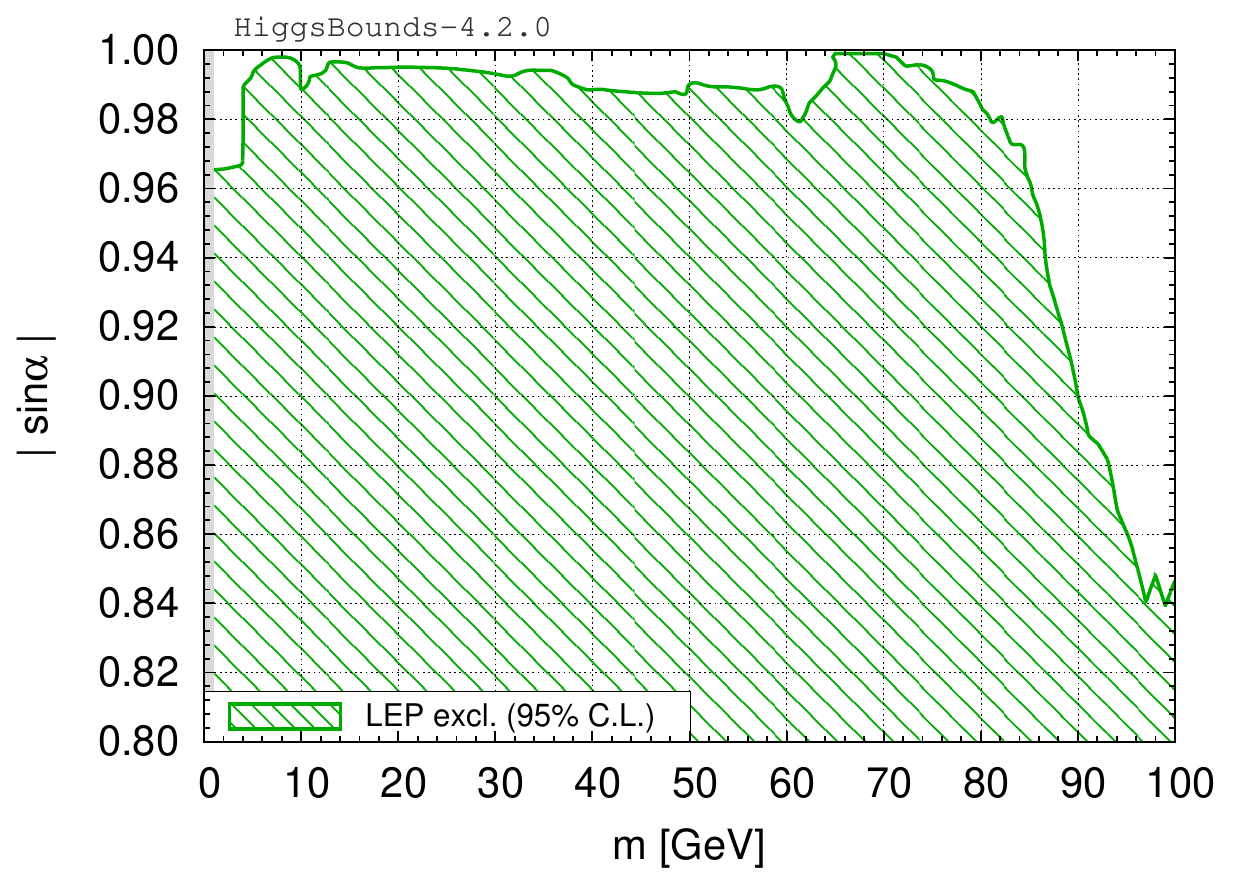}}
\hfill
\subfigure[\label{fig:sinaexp_b} Mass region $m~{\in}~[100,1000{]}~\GeV$, assuming a vanishing Higgs-to-Higgs decay mode, $\Gamma_{H\to hh}=0$.]
{\includegraphics[width=0.495\textwidth]{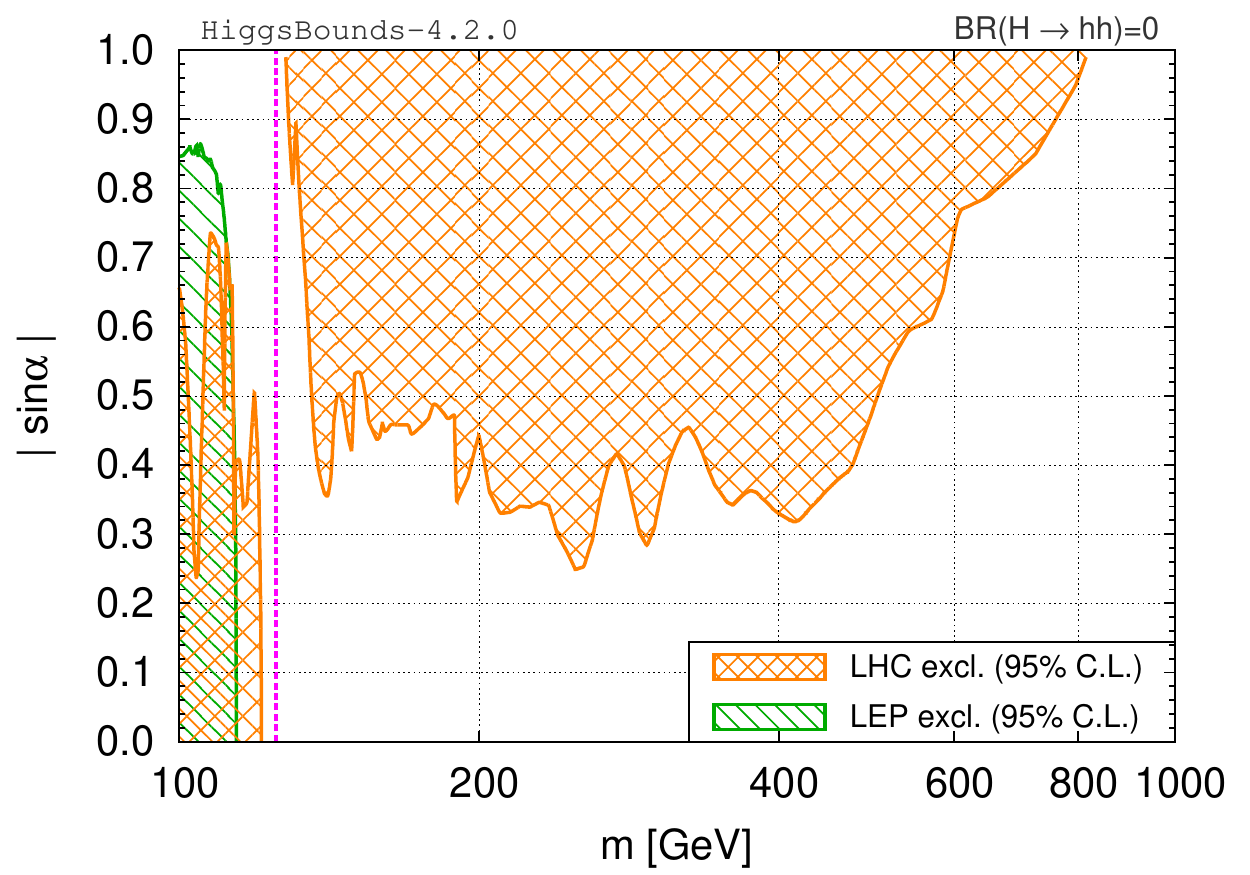}}
\caption{\label{fig:sinaexp} {$95\%$~C.L.~excluded values of $| \sin\al |$ from LEP and LHC Higgs searches, evaluated with \HBv{4.2.0} in the mass regions $m~{\in}~[1,100{]}~\GeV$ (\emph{a}) and $m~{\in}~[100,1000{]}~\GeV$ (\emph{b}). We assume a vanishing decay width for the Higgs-to-Higgs decay mode, $\Gamma_{H \rightarrow hh}\,=\,0$, hence the displayed results in the high mass region correspond to the most stringent upper limit on $| \sin\al |$ that can be obtained from current LHC Higgs searches. The other Higgs boson mass is set to $125.14~\GeV$ and is indicated by the dashed, magenta line in Fig.~\ref{fig:sinaexp_b}.}}
\end{figure}

The $95\%$~C.L.~excluded regions of $|\sin\alpha|$ derived with \HB\ are shown in Fig.~\ref{fig:sinaexp} as a function of the second Higgs mass, assuming a vanishing decay width of the Higgs-to-Higgs decay mode $H\rightarrow hh$. Since {all Higgs boson production modes} are reduced with respect to their SM prediction by a universal factor, limits from LHC Higgs search analyses for a SM Higgs boson can be applied straight-forwardly~\cite{Bechtle:2013wla}. In particular, the exclusion limits obtained from combinations of SM Higgs boson searches with various final states are highly sensitive. {However,} so far, such combinations have only been presented by ATLAS and CMS for the full $7~\TeV$ dataset~\cite{Chatrchyan:2012tx,Aad:2012an} {and for a subset of the $8~\TeV$ data~\cite{CMS:aya}.}
The strongest exclusions are therefore obtained mostly from the single search analyses of the full $7/8~\TeV$ dataset, in particular from the channel $H\to ZZ\to 4\ell$~\cite{ATLAS:2013nma,Chatrchyan:2012dg,CMS:xwa} in the mass region $m \in [130,150]~\GeV$ and for $m \gtrsim 190~\GeV$, as well as from the $H\to WW\to \ell\nu\ell\nu$ channel~\cite{ATLAS:2013wla,Chatrchyan:2013iaa,CMS:bxa} in the mass region $m \in [160,170]~\GeV$ due to the irreducible $ZZ$ background in the $H\to ZZ\to 4\ell$ analyses. For Higgs masses $m \in [65, 110]~\GeV$ the only LHC exclusion limits currently available are from the ATLAS search for scalar diphoton resonances~\cite{Aad:2014ioa}. However, these constraints are weaker than the LEP limits from the channel $e^+e^- \to HZ \to (b\bar{b})Z$~\cite{Schael:2006cr}, as can be seen in Fig.~\ref{fig:sinaexp_b}. In the {remaining} mass regions with $m \ge 110~\GeV$ the CMS limit~\cite{CMS:aya} from the combination of SM Higgs analyses yields the strongest constraint. For very low Higgs masses, $m \lesssim 10~\GeV$, the LEP constraints come from Higgs pair production processes, $e^+e^- \to hh \to \tau^+\tau^-\tau^+\tau^-~\mbox{and}~\tau^+\tau^-b\bar{b}$~\cite{Schael:2006cr}, as well as from the decay-mode independent analysis of $e^+e^- \to Zh$ by OPAL~\cite{Abbiendi:2002qp}. The latter analysis provides limits for Higgs masses as low as $1~\mathrm{keV}$.

In the presence of Higgs-to-Higgs decays, $\mathrm{BR}(H \rightarrow hh)\ne 0$, additional constraints arise.  In case of very low masses, $m \,\lesssim\,3.5\,\GeV$, these stem from the CMS search in the $H\rightarrow hh \rightarrow \mu^+\mu^-\mu^+\mu^-$ channel~\cite{CMS:2013lea}, and for large masses, $m \in [260, 360]~\GeV$, from the CMS search for $H\rightarrow hh$ with multileptons and photons in the final state~\cite{CMS:2013eua}. These limits will be discussed separately in Section~\ref{sec:results}. {Note that the limit from SM Higgs boson searches in the mass range $m\gtrsim 250~\GeV$, as presented in Fig.~\ref{fig:sinaexp_b}, will diminish in case of non-vanishing $\mathrm{BR}(H \rightarrow hh)$ due to a suppression of the SM Higgs decay modes.} 
{We find {in the full scan (see Section~\ref{sec:results})} that, in general, $\mathrm{BR}(H \rightarrow hh)$ can be as large as $\sim 70\%$ in this model. Neglecting the correlation between $\sin\alpha$ and $\mathrm{BR}(H \rightarrow hh)$ for a moment, such large branching fractions could lead to a reduction of the upper limit on $|\sin\alpha|$ obtained from  SM Higgs searches by a factor of $\sim\,1/\sqrt{1-\text{BR}(H\rightarrow hh)}\,\lesssim\, 1.8$.}
However, once all other constraints (in particular from the NLO calculation of $m_W$) are taken into account, only {$\text{BR}({H\rightarrow hh})$} values of up to $40\%$ are found,
see Section~\ref{sec:highmassres}, Fig.~\ref{fig:brHtohh_mH}.
Moreover, in the mass region $m_H \sim 270-360~\GeV$ where the largest values of $\text{BR}(H\to hh)$ appear, the $m_W$ constraint on $|\sin\al|$ is typically stronger than the constraints from SM Higgs searches, even if $\text{BR}(H\to hh) = 0$ is assumed in the latter.
Therefore, given the present Higgs search exclusion limits, the signal rate reduction currently does not have a visible impact on the viable parameter space\footnote{Note, however, that this may change in future with significantly improved exclusion limits from SM Higgs searches.}.

\subsection[Higgs boson signal rates\dots]{Higgs boson signal rates measured at the LHC}
\label{Sect:HS}

The compatibility of the predicted signal rates for the Higgs state at $\sim\MHexp$ with the latest measurements from ATLAS~\cite{Aad:2014eha,Aad:2014eva,ATLAS-WW-Note,ATLAS-tautau-Note,Aad:2014xzb} and CMS~\cite{Khachatryan:2014ira,Chatrchyan:2013mxa,Chatrchyan:2013iaa,Chatrchyan:2014vua} is evaluated with \HSv{1.3.0} by means of a statistical $\chi^2$ measure. {The implemented observables are listed in Tab.~\ref{Tab:HSobservables}.} In the following we denote this $\chi^2$ value by $\chi^2_\mathrm{HS}${, which also includes the $\chi^2$ contribution from the Higgs mass observables evaluated within \HS. The latter, however, only yields non-trivial constraints on the parameter space if the fit allows a varying Higgs mass in the vicinity of $125~\GeV$.}
{In the low mass scenario, where one of the Higgs bosons is within the kinematical range of the LEP experiment, the $\chi^2$ value obtained from the \HB\ LEP $\chi^2$ extension, denoted as~$\chi^2_{\mathrm{LEP}}$, is added to the \HS\ $\chi^2$ to construct the global likelihood
\begin{align}
\chi^2_\mathrm{tot} = \chi^2_\mathrm{HS} + \chi^2_\mathrm{LEP}.
\end{align}
The $68\%$ and $95\%$ confidence level ($\mathrm{C.L.}$) parameter regions of the model are approximated by the $\chi^2$ difference to the minimal $\chi^2$ value found at the best-fit point, $\Delta\chi^2 = \chi^2 - \chi_\text{min}^2$, taking on values of $1~(2.30)$ and $4~({6.18})$ in the case of a $1~(2)$-dimensional projected parameter space, respectively.

\begin{table}
 \begin{tabular}{| r l c c |}
 \toprule
experiment & channel & obs.~signal rate  & obs.~mass [$\GeV$]\\
 \colrule
ATLAS &$ h\to WW\to \ell\nu\ell\nu$~\cite{ATLAS-WW-Note} & $  1.08\substack{+  0.22\\ -  0.20}$ & -- \\
ATLAS & $ h\to ZZ\to 4\ell$~\cite{Aad:2014eva} & $  1.44\substack{+  0.40\\ -  0.33}$ & $124.51 \pm 0.52$ \\
ATLAS & $ h\to \gamma\gamma$~\cite{Aad:2014eha} & $  1.17\substack{+  0.27\\ -  0.27}$ & $125.98 \pm 0.50$ \\
ATLAS & $ h\to \tau\tau$~\cite{ATLAS-tautau-Note} & $  1.42\substack{+  0.43\\ -  0.37}$ & -- \\
ATLAS & $ Vh\to V(b\bar{b})$~\cite{Aad:2014xzb} & $  0.51\substack{+  0.40\\ -  0.37}$ & --\\
CMS &$ h\to WW\to \ell\nu\ell\nu$~\cite{Chatrchyan:2013iaa} & $  0.72\substack{+  0.20\\ -  0.18}$ & -- \\
CMS &$ h\to ZZ\to 4\ell$~\cite{Chatrchyan:2013mxa} & $  0.93\substack{+  0.29\\ -  0.25}$ & $125.63 \pm 0.45 $ \\
CMS &$ h\to \gamma\gamma$~\cite{Khachatryan:2014ira} & $  1.14\substack{+  0.26\\ -  0.23}$ & $124.70 \pm 0.34$ \\
CMS &$ h\to \tau\tau$~\cite{Chatrchyan:2014vua} & $  0.78\substack{+  0.27\\ -  0.27}$ & -- \\
CMS &$ Vh\to V(b\bar{b})$~\cite{Chatrchyan:2014vua} & $  1.00\substack{+  0.50\\ -  0.50}$ & -- \\
 \botrule
 \end{tabular}
\caption{Higgs boson signal rate and mass observables from the LHC experiments, as implemented in \HSv{1.3.0} and used in this analysis. For the mass measurements we combined the systematic and statistical uncertainty in quadrature.}
\label{Tab:HSobservables}
\end{table}

When both Higgs masses are fixed, the fit depends on two free
parameters, namely $\sin\al$ and $\tan\be$. The latter can only
  influence the signal rates of the Higgs boson $H$ if the additional
  decay mode $H\to hh$ is accessible. The branching
  fraction $\mathrm{BR}(H\to hh)$ then leads to a decrease of all other decay modes and hence to a reduction of the predictions for the measured signal rates, cf.~Eq.~\eqref{eq:brdefs}. The sensitivity to $\tan\be$ via the signal
  rate measurements is thus only given if the heavier Higgs state is
  interpreted as the discovered particle, $m_H \sim \MHexp$ (low mass region), and the second Higgs state is sufficiently light, $m_h \lesssim 62\,\GeV$. If the $H\to hh$ decay is not kinematically accessible, or in the case where the light Higgs is considered as the discovered Higgs state at $\sim \MHexp$, there are no relevant experimental constraints on $\tan\be$.
    
\begin{figure}[t]
\centering
\subfigure[\label{sf:lowmass_fixedmassBRHtohh} ($\sin\al$, $\tan\be$) plane.]{\includegraphics[width=0.48\textwidth]{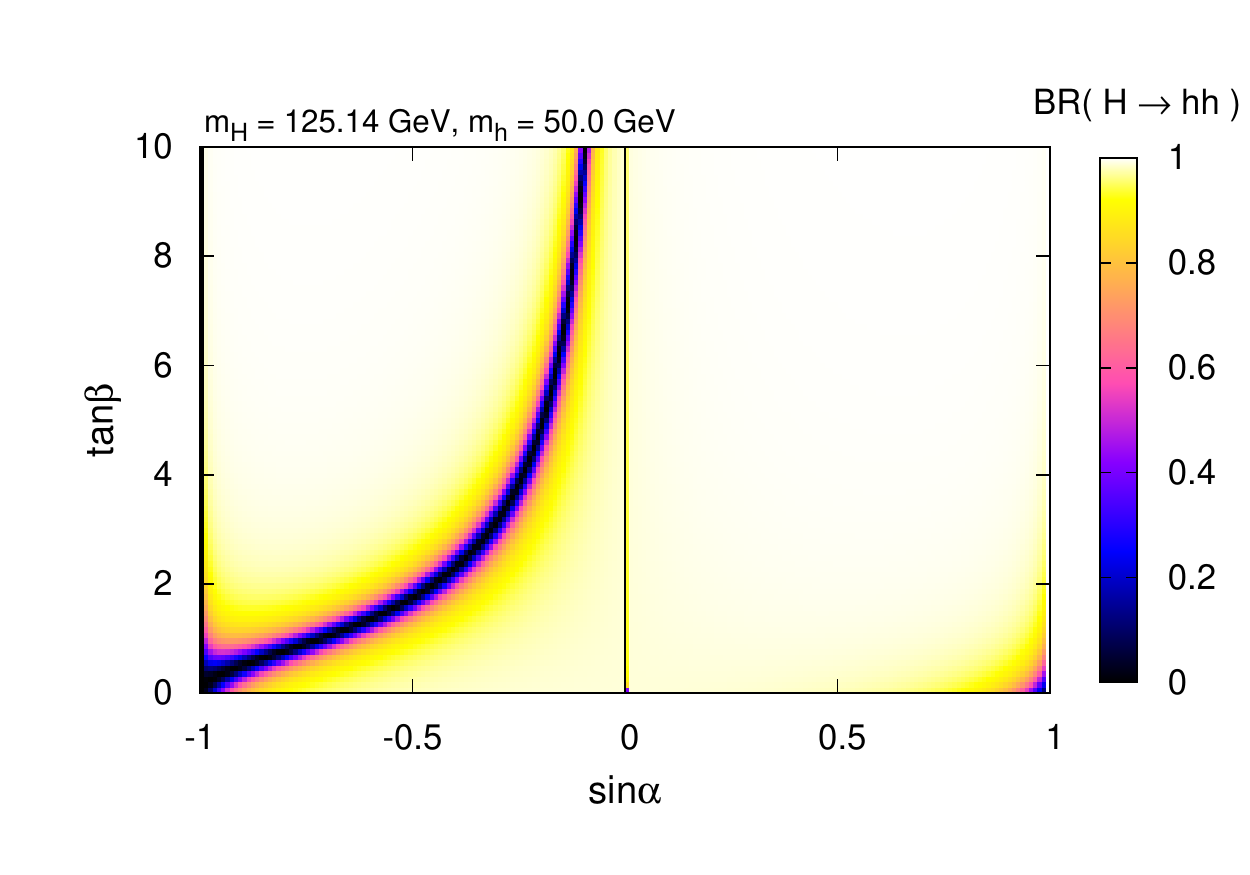}
}
\hfill
\subfigure[\label{sf:lowmass_fixedmassBRHtohh_zoom} Zoomed region of the ($\sin\al$, $\tan\be$) plane. The gray contours indicate the $1$, $2$ and $3\sigma$ regions preferred by the signal rates; the green, dashed line displays the $95\%~\mathrm{C.L.}$ limit from LEP, cf.~Section~\ref{Sect:HB}.]{ \includegraphics[width=0.48\textwidth]{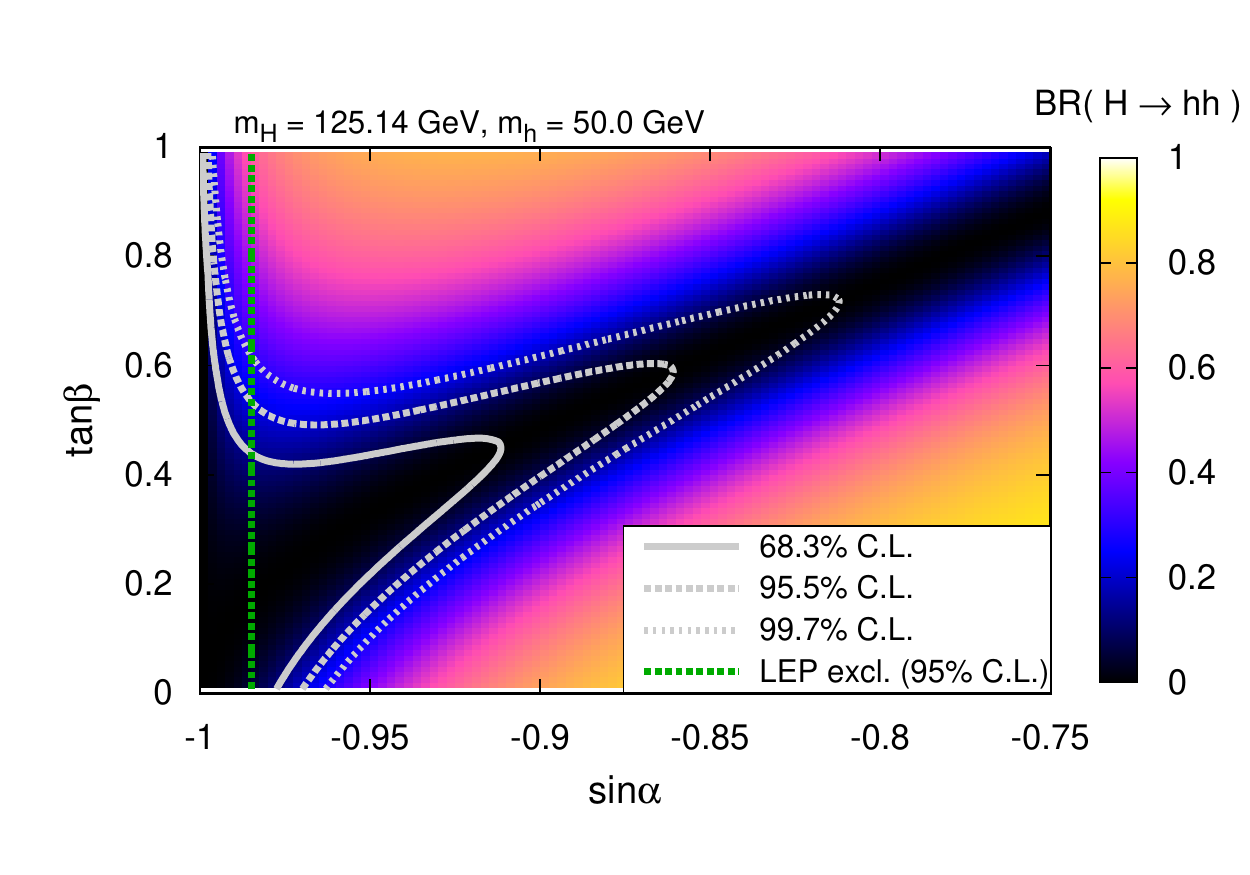}
}
\caption{\label{fig:lowmass_fixedmassBRHtohh} Branching ratio $\mathrm{BR}(H\to hh)$ in the ($\sin\al$, $\tan\be$) plane for fixed Higgs masses $m_h = 50\,\GeV$ and $m_H = 125.14\,\GeV$. {It} becomes minimal for either $\sin\al\,=\,0$, $\cos\al\,=\,0$ or {$\tan\be=-\cos\al/\sin\al$}.}
\end{figure}

{In the low mass region, t}he Higgs signal rate measurements constrain the modulus of the mixing angle $\sin\al$ to be close to $1$, such that the heavy Higgs boson has nearly the same coupling strengths as the SM Higgs boson. Moreover, in order to obtain sizable predictions for the measured signal rates, the branching ratio $\mathrm{BR}(H\to hh)$ must not be too large.
We illustrate its dependence on $\tan\be$ in Fig.~\ref{fig:lowmass_fixedmassBRHtohh}, where we exemplarily show the branching ratio $\mathrm{BR}(H\to hh)$ in the ($\sin\al$,~$\tan\be$) plane for fixed Higgs boson masses of $m_h = 50~\GeV$ and $m_H = 125.14~\GeV$. As can be seen in Fig.~\ref{sf:lowmass_fixedmassBRHtohh}, the decay $H\to hh$ is dominant over large regions of the parameter space, {with the exception of three distinct cases: The branching ratio $\mathrm{BR}(H\to hh)$ exactly vanishes in the case that 
\begin{align}
(\textit{i})\;\sin\alpha = 0,\quad(\textit{ii})\;\cos\alpha = 0\quad\mbox{or}\quad (\textit{iii})\;\tan\beta  = - \cos\al/ \sin\al.
\end{align}
 In the first case (\textit{i}) all couplings of the heavy Higgs boson to SM particles vanish completely, thus this case is highly excluded by observations. The second case  (\textit{ii}) corresponds to the complete decoupling of the lighter Higgs boson, such that the heavier Higgs is identical to the SM Higgs boson. In the third and more interesting case (\textit{iii}) the branching fraction can be expanded in powers of $(\tan\beta  + \cos\al/ \sin\al)$:}
\begin{\eqn*}
\text{BR} ({H\rightarrow hh})=\frac{\sqrt{1-\frac{4\,m_h^2}{m_H^2}}\lb  m_h^2+\frac{m_H^2}{2} \rb^2\,}{8\pi v^2\, m_H\, \Gamma_{\text{SM,~tot}}(m_H)} \cos^2\al \sin^4\al\lb \tan\be+\frac{\cos\al}{\sin\al} \rb^2+\mathcal{O}\left[\lb \tan\be+\frac{\cos\al}{\sin\al} \rb^3\right],
\end{\eqn*}
{where $\Gamma_{\text{SM,~tot}}(m_H)$ is the total width in the SM for a Higgs boson at mass $m_H$.}

In Fig.~\ref{sf:lowmass_fixedmassBRHtohh_zoom} we show a zoom of the ($\sin\al$,~$\tan\be$) plane, {focussing on} the low-$\mathrm{BR}(H\to hh)$ valley and $\sin\al$ values close to $-1$. We furthermore indicate the parameter regions which are allowed at the $1$, $2$ and $3\sigma$ level by the Higgs signal rate measurements by the gray contour lines. The maximally values of $\mathrm{BR}(H\to hh) \approx 26\%$ allowed by the Higgs signal rate measurements at $95\%~\mathrm{C.L.}$ are found for $\sin\al$ very close to $-1$ and large $\tan\be$ values, i.e.~in the vicinity of case (\emph{ii}) discussed above. In the given example with $m_h =50~\GeV$, the $95\%~\mathrm{C.L.}$ exclusion from LEP searches, as discussed in Section~\ref{Sect:HB} (cf.~Fig.~\ref{fig:sinaexp_a}), imposes $\sin\al \lesssim -0.985$ and is indicated in Fig.~\ref{sf:lowmass_fixedmassBRHtohh_zoom} by the green, dashed line.

\begin{figure}[t]
\centering
\subfigure[\label{sf:lowmass_fixedmass_kappaH_a} ($\sin\al$, $\tan\be$) plane. The black solid and dashed line indicate constant values of $\kappa_H^2 = 1$ and $2$, respectively. The other contour lines are the same as in Fig.~\ref{sf:lowmass_fixedmassBRHtohh_zoom}.]{\includegraphics[width=0.48\textwidth]{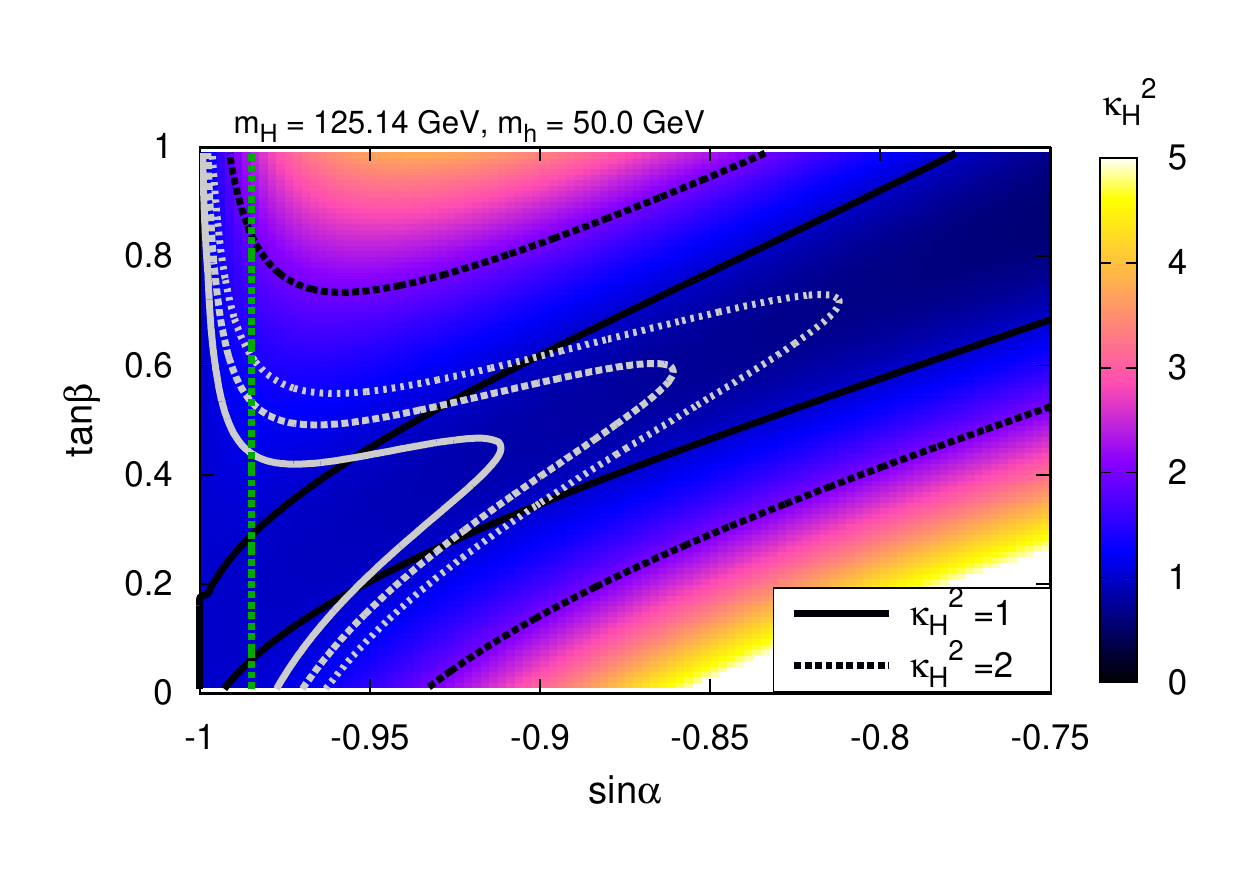}}
\hfill
\subfigure[\label{sf:lowmass_fixedmass_kappaH_b} Total width scale factor $\kappa_H^2$ for the $1\sigma$, $2\sigma$ and $3\sigma$ regions favored by the Higgs boson signal rates.] {\includegraphics[width=0.48\textwidth]{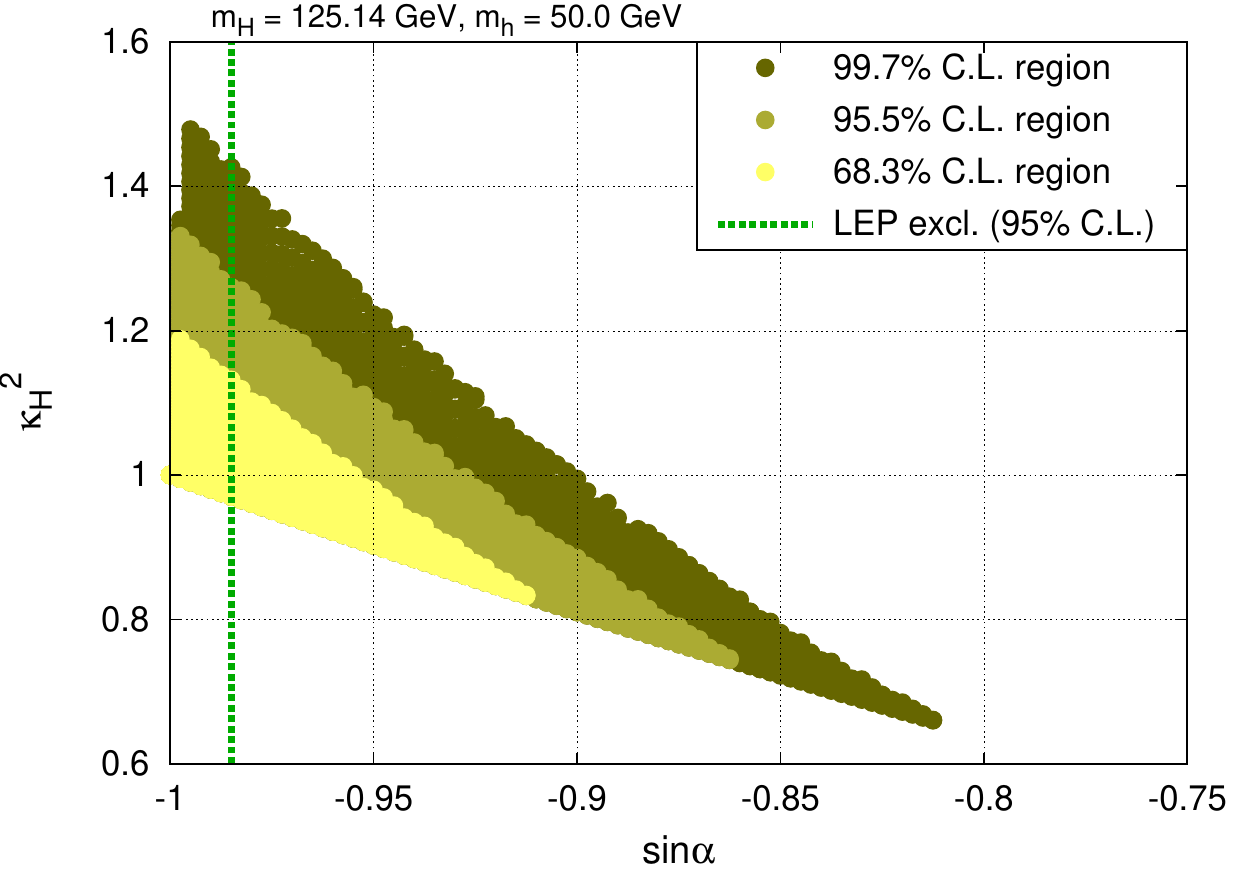}}
\caption{\label{fig:lowmass_kappaH} 
Total width scale factor $\kappa_H^2 = \Gamma_\mathrm{tot} / \Gamma_\mathrm{tot, SM}$ for fixed Higgs masses $m_h = 50~\GeV$ and $m_H = 125.14~\GeV$: (\emph{a}) in the ($\sin\al$, $\tan\be$) plane and (\emph{b}) for the regions favored by the Higgs boson signal rate measurements only.}

\end{figure}    
Finally, we plot the total width scaling factor, defined by $\kappa^2_H =\Gamma_\text{tot}/\Gamma_\text{tot, SM}$, {in the ($\sin\al$, $\tan\be$) plane in Fig.~\ref{sf:lowmass_fixedmass_kappaH_a}. We furthermore plot $\kappa_H^2$ for the parameter regions favored by the Higgs signal rates in Fig.~\ref{sf:lowmass_fixedmass_kappaH_b}. The largest values of $\kappa_H^2$ allowed by both the signal rates and the LEP constraints at $95\%~\mathrm{C.L.}$ are obtained for $\sin\alpha$ close to $-1$. In the example of $m_h = 50~\GeV$ discussed here, the total width is increased by up to around $34\%$ with respect to the SM.} This maximal value of total width enhancement is independent of the light Higgs mass (assuming that the channel $H\to hh$ is kinematically accessible).

We now want to draw the attention to the intermediate mass range, where both mass eigenstates can contribute to the signal strength measurements at the LHC. If the masses of the two Higgs bosons are well separated, the signal yields measured in the LHC Higgs analyses can be assumed to be solely due to the one Higgs boson lying in the vicinity of the signal, $m\sim \MHexp$. However, in analyses with a poor mass resolution, as is typically the case in search analyses for the decay modes $H\to W^+W^-$, $H\to \tau^+\tau^-$ and $VH\to b\bar{b}$, the signal contamination from the second Higgs boson needs to be taken into account if its mass is not too far away from $125~\GeV$. While a proper treatment of this case can only be done by the experimental analyses, \HS\ employs a \textit{Higgs boson assignment procedure} to approximately account for this situation~\cite{Bechtle:2013xfa}. Based on the experimental mass resolution of the analysis and the difference between the predicted mass and the mass position where the measurement has been performed, \HS\ decides whether the signal rates of multiple Higgs states need to be combined. Hence, superpositions of the two Higgs signal rates considered here are possible if the second Higgs mass lies within $100~\GeV \lesssim m \lesssim 150~\GeV$.

\begin{figure}[t]
\centering

\includegraphics[width=0.8\textwidth]{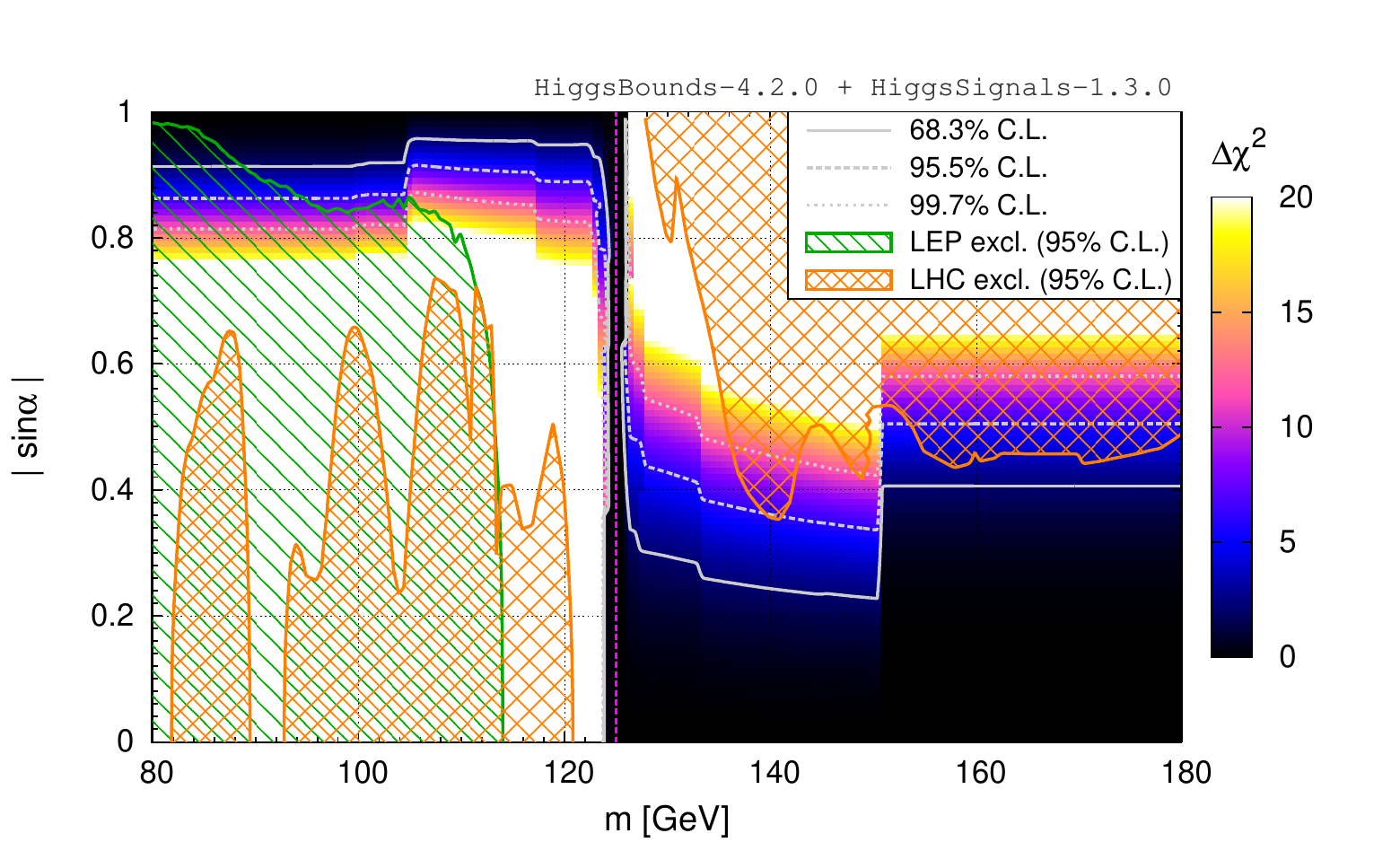}
\caption{\label{fig:massoverlap_chi2} $\Delta \chi^2$ distribution from the Higgs signal rate observables, obtained from \HSv{1.3.0}, as a function of the second Higgs boson mass $m$ and the mixing angle $\sin\al$. The mass of the other Higgs boson is fixed at $m=125.14~\GeV$, indicated by the dashed, magenta line. The gray contour lines indicate the favored parameter space at $68.3\%$, $95.5\%$ and $99.7\%$~C.L., solely based on the Higgs signal rate observables. 
The green striped (orange patterned) region is excluded by LEP (LHC) Higgs searches at $95\%$ C.L., cf.~also Fig.~\ref{fig:sinaexp}. For Higgs masses $m$ below $100~\GeV$ and beyond around $152~\GeV$ the signal rate constraints are independent of $m$.
}
\end{figure}    

In Fig.~\ref{fig:massoverlap_chi2} we show the \HS\ $\Delta\chi^2$ value obtained from the signal rate observables as a function of the second Higgs boson mass $m$ and the mixing angle $\sin\alpha$. The mass of the other Higgs boson is fixed at $m = 125.14~\GeV$. The scan range for $m$ extends over both the low mass and high mass region. 
Since the Higgs boson at $\sim \MHexp$ needs sufficiently large signal rates {to accommodate for the observed SM-like Higgs signal strength}, small (large) values of $\sin\al$ are favored in the high (low) mass region, such that the second Higgs boson is rather decoupled. We furthermore show the parameter space excluded at $95\%~\mathrm{C.L.}$ by LEP and LHC searches, as previously discussed in Section~\ref{Sect:HB} in Fig.~\ref{fig:sinaexp}.

In the case of nearly mass degenerate Higgs bosons, $m_h \approx m_H = 125.14~\GeV$, the sensitivity on the mixing angle $\sin\al$ significantly decreases, as the signal rates of the two Higgs states are always superimposed. There remains a slight dependence {of the total signal rate on the Higgs masses}, though, since the production cross sections {and branching ratios are mass dependent.} Moreover, depending on the actual mass splitting and mixing angle, potential effects {may possibly be seen in the invariant mass distributions of the high-resolution LHC channels $pp \to H\to \gamma\gamma$~\cite{Khachatryan:2014ira} and $pp\to H\to ZZ^*\to 4\ell$, at a future linear collider like the ILC~\cite{Asner:2013psa,Dawson:2013bba} or eventually a muon collider~\cite{Alexahin:2013ojp,Dawson:2013bba}.} However, the sensitivity on $\sin\al$ completely vanishes in the case of exact mass degeneracy, $m_h = m_H$, such that the singlet extended SM becomes indistinguishable from the SM.

The weak $\Delta \chi^2$ dependence on $m$ outside of the mass degenerate region, i.e.~for $m \gtrsim 128~\GeV$ and $m\lesssim 122~\GeV$, is caused by the superposition of the signal rates of both Higgs bosons in some of the $H\to W^+W^-, \tau^+\tau^-$ and $b\bar{b}$ channels, as discussed above. These structures depend on the details of the implementation within \HS, in particular on the assumed experimental resolution for each analysis. {For Higgs masses $m$ below $100~\GeV$ and beyond around $152~\GeV$ the $\sin\alpha$ limit from the signal rates is independent\footnote{This statement is only true if the Higgs state at $\sim125~\GeV$ does not decay to the lighter Higgs. As discussed above, at low light Higgs masses $m_h < m_H/2$, the branching ratio $\mathrm{BR}(H\to hh)$ can reduce the signal rates of the heavy Higgs decaying to SM particles.} of $m$. }

{We see that for Higgs masses $m$ in the range between $\sim 100~\GeV$ and $150~\GeV$, the constraints from the Higgs signal rates are more restrictive than the exclusion limits from Higgs searches at LEP and LHC. For lower Higgs masses, $m < 100~\GeV$, the LEP limits (cf.~Fig.~\ref{fig:sinaexp_a}) generally yield stronger constraints on the parameter space. For higher Higgs masses, $m \in [150,500]~\GeV$, the direct LHC limits (cf.~Fig.~\ref{fig:sinaexp_b}) are slightly stronger than the constraints from the signal rates, however, this picture reverses again for Higgs masses beyond $500~\GeV$, where direct heavy Higgs searches become less sensitive.}

\section{Results of the Full Parameter Scan}
\label{sec:results}

In this section we investigate the {interplay} of {all theoretical and experimental} constraints discussed in the previous section on the real singlet extended SM parameter space, {specified by} 
\begin{\eqn*}
m \equiv m_{h/H},\,\sin\al,\,\tan\be.
\end{\eqn*}
{We separate the  discussion into the high mass, the low mass and the intermediate (or degenerate) mass region of the parameter space.}
{In the high and low mass region, we keep one of the Higgs masses fixed at ${125.14}\,\GeV$ and vary the other, while in the intermediate mass region we treat both Higgs masses as scan parameters}. In the following we first present results for fixed mass $m$ in order to facilitate the understanding of the respective parameter space in dependence of $\sin\al,\,\tan\be$. These discussions will then be {extended} by a {more general} scan, where all parameters are allowed to vary simultaneously. For {each of} the{se} scan{s}, we generate around $\mathcal{O}(10^5-10^6)$ points. We close the discussion of each mass region by commenting on the relevant collider phenomenology.}


\subsection{High mass region}\label{sec:highmassres}

In this section, we explore the parameter space of the high mass region, $m \in [130, 1000]\,\GeV$. 
In general, for masses $m \geq 600~\GeV$, our results agree with those presented in Ref.~\cite{Pruna:2013bma}. However, we obtain {stronger} bounds on the maximally allowed value of $|\sin\al|$ due to the constraints from the NLO calculation of $m_W$~\cite{Lopez-Val:2014jva}, which has not been available for the previous analysis~\cite{Pruna:2013bma}. As has been discussed in Section~\ref{sec:Wmass},~Fig.~\ref{fig:sinamw}, the constraints from $m_W$ are much more stringent than those obtained from the oblique parameters $S$, $T$, and $U$ in the high mass region.

\begin{figure}
\centering
\includegraphics[width=1.0\textwidth]{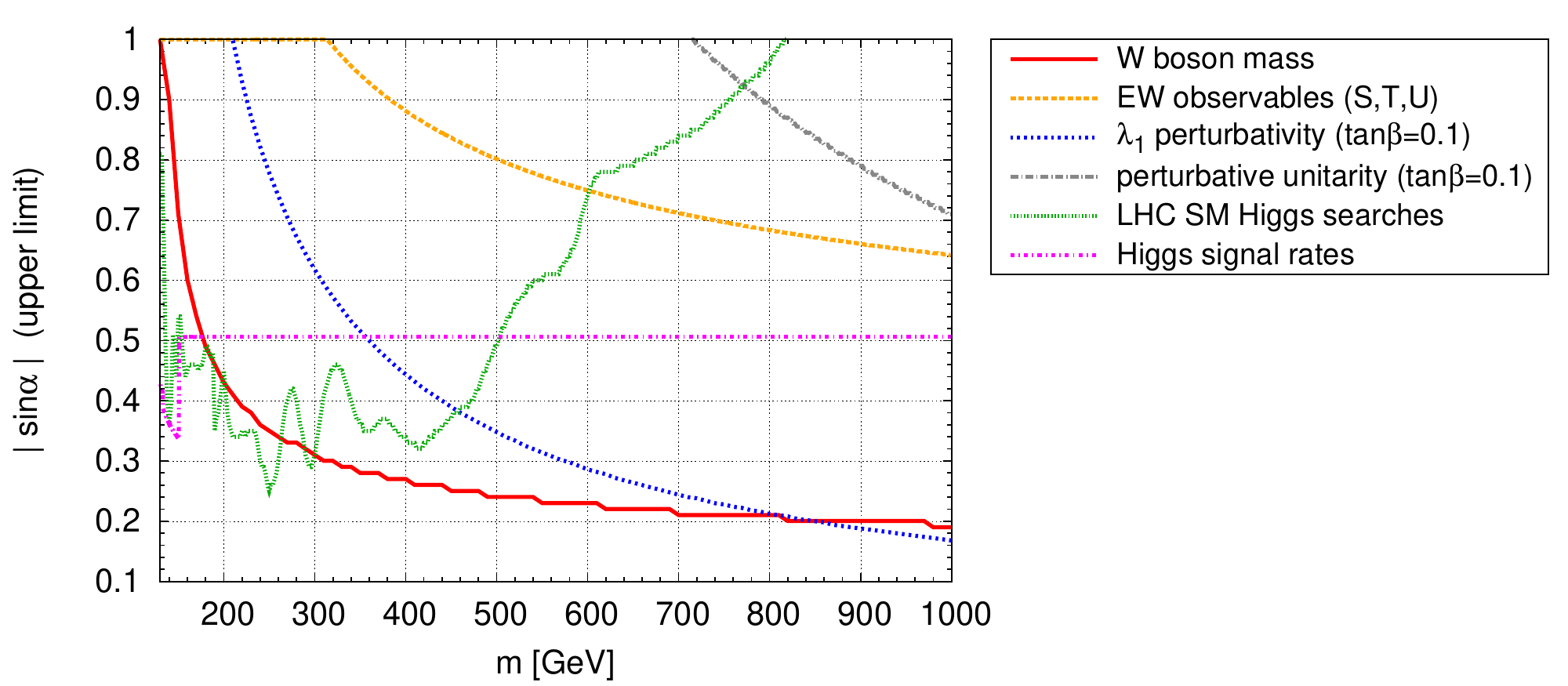}
\caption{\label{fig:comb} Comparison of all constraints on $|\sin\al|$ as a function of the heavy Higgs mass $m$ in the high mass region. {The $\lam_1$ perturbativity and perturbative unitarity constraint have been evaluated for $\tan\be = 0.1$.}}
\end{figure}

\begin{table}[t]
\begin{tabular}{| c || c | c | c |}
\toprule
$m~[\GeV]$ & $|\sin\al|$& source upper limit &$(\tan\be)_\text{max}$\\
\colrule
{$1000$} &{$[0.018, 0.17]$}&{$\lam_1$ perturbativity}&$ {0.23}$\\
{$900$}&{$[0.022, 0.19]$}&{$\lam_1$ perturbativity}&${0.26}$ \\
{$800$}&{$[0.027, 0.21]$}&{$m_W$ at NLO}& ${0.29}$\\
{$700$}&{$[0.031, 0.21]$}&{$m_W$ at NLO}& ${0.33}$ \\
$600$					    &  $[{0.038}, {0.23}]$	&	$m_W$ at NLO		&${0.39}$\\
$500$					    &  $[{0.046}, {0.24}]$	& 	$m_W$ at NLO			&${0.47}$\\
$400$					    &  $[{0.055}, {0.27}]$	&	$m_W$ at NLO				&${0.59}$ \\
$300$					    &  $[0.067, {0.31}]$		& 	$m_W$ at NLO					  &${0.78}$\\
$200$					    &  $[{0.090}, {0.43}]$	&	{$m_W$ at NLO} 					   &${1.17}$\\
$180$					    &  $[0.10, {0.46}]$		&      signal rates				    &$1.30$\\
$160$					    &  $[{0.11}, {0.46}]$		&      signal rates				     &${1.46}$\\
$140$					    &  $[{0.16}, 0.31]$	       & {signal rates}& ${1.67}$\\
\botrule
\end{tabular}
\caption{\label{tab:highm} Allowed ranges for $\sin\alpha$ and $\tan\beta$ in the high mass region for fixed Higgs masses $m$. The allowed interval of $\sin\alpha$ was determined at $\tan\be=0.15$. The $95\%~\mathrm{C.L.}$ limits on $\sin\al$ from the Higgs signal rates are derived from one-dimensional fits and taken at $\Delta\chi^2 = 4$. The lower limit on $\sin\al$ always stems from vacuum stability, and the upper limit on $\tan\be$ always from perturbativity of $\lam_2$, evaluated at $\sin\al\,=\,0.1$. The source of the most stringent upper limit on $\sin\alpha$ is named in the third column. We fixed $m_h={125.14}~\GeV$, and the stability and perturbativity were tested at a scale of $\sim\,4\,\times\,10^{10}\,\GeV$.
}
\end{table}

We compile {\sl all} previously discussed constraints on the maximal mixing angle in Fig.~\ref{fig:comb}. Furthermore, the (one-dimensional) allowed regions in $|\sin\alpha|$ and $\tan\beta$ {are given} in Tab.~\ref{tab:highm} for fixed values of $m$.\footnote{{Note, that the upper limit on $|\sin\alpha|$ from the Higgs signal rates is based on a two-dimensional $\Delta\chi^2$ profile (for floating $m_h$) in Fig.~\ref{fig:comb}, whereas in Tab.~\ref{tab:highm} the one-dimensional $\Delta\chi^2$ profile (for fixed $m_h$) is used. This leads to small differences in the obtained limit.}} {Here,} the allowed range of $|\sin\alpha|$ is evaluated for fixed $\tan\be = 0.15$ and we explicitly specify the relevant constraint that provides in the upper limit on $|\sin\alpha|$. We find the following generic features: For Higgs masses $m \gtrsim 200 - 300~\GeV$, the $W$ boson mass NLO calculation provides the upper limit on $|\sin\alpha|$, at lower masses the LHC constraints at $95\%~\mathrm{C.L.}$ from direct Higgs searches and the signal rate measurements are most relevant. The purely theory-based limits from perturbativity of $\lam_1$ only become important for $m\,\gtrsim\,800\,\GeV$. Furthermore, in the whole mass range, the minimal value of $|\sin\al|$ and the maximal value of $\tan\be$ are determined by vacuum stability and perturbativity of the couplings.

\begin{figure}[t]
\subfigure[~Higgs masses below $200~\GeV$.]{
\includegraphics[width=0.48\textwidth]{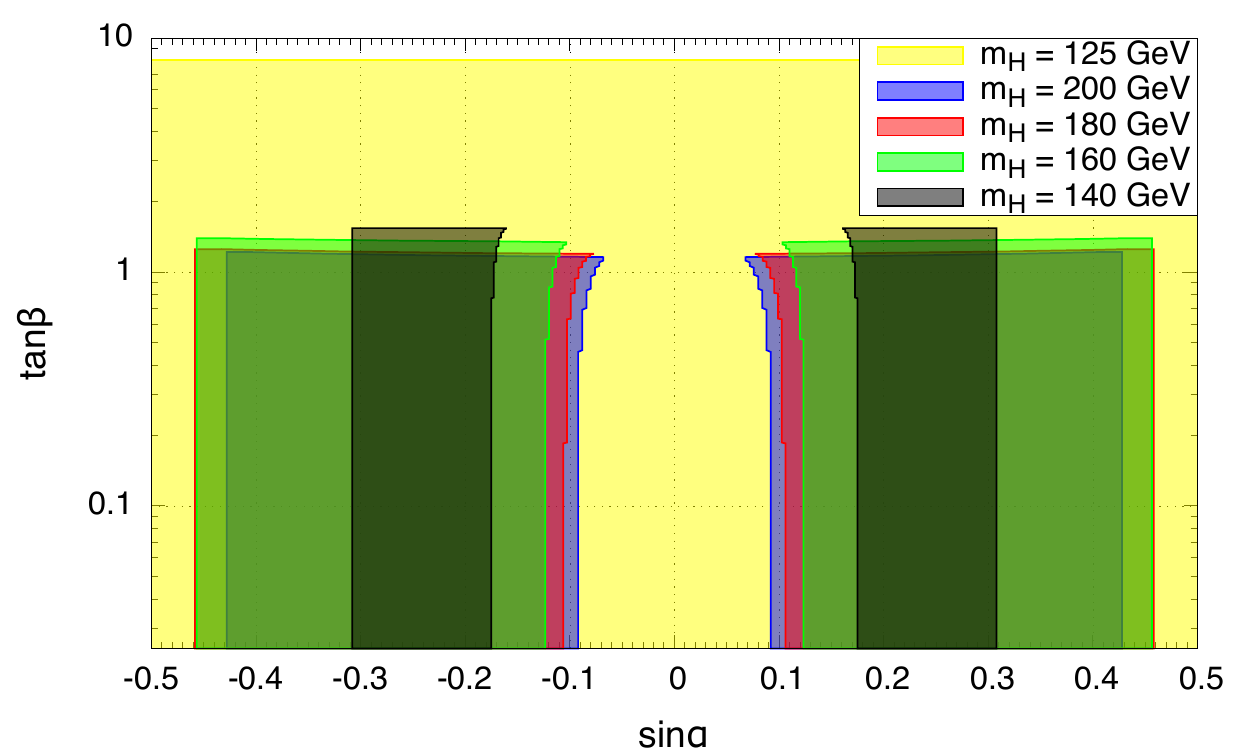}}
\hfill
\subfigure[~Higgs masses above $200~\GeV$.]{
\includegraphics[width=0.48\textwidth]{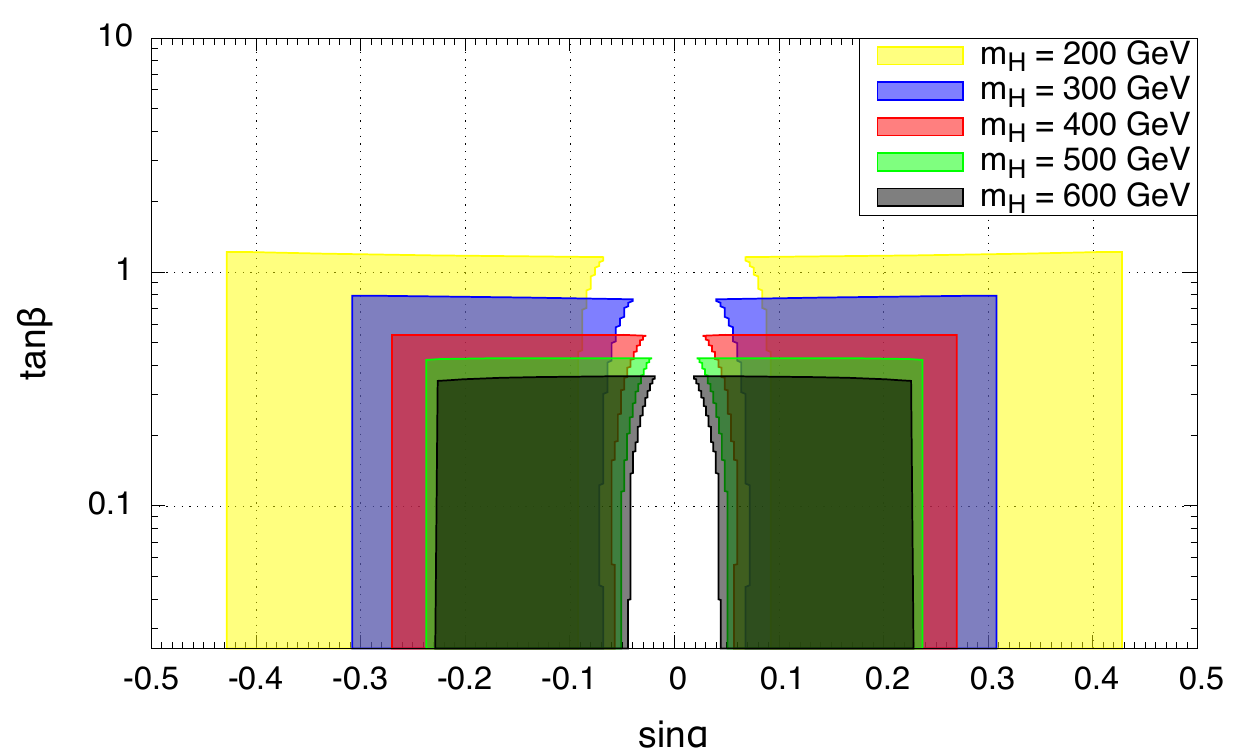}}
\caption{\label{fig:contours}Allowed regions in the ($\sin\al,\,\tan\be$) plane {in the high mass region for fixed Higgs masses $m$}. For $m_H\gtrsim200\,\GeV$, the {upper limit on the} mixing angle stems from $m_W$, while for $m_H\lesssim 200~\GeV$ the upper limit is given by the signal strength measurements as well as experimental searches (cf. also Fig.~\ref{fig:comb}). 
}
\end{figure}

The corresponding {(two-dimensional)} {allowed regions} in the ($\sin\al,\,\tan\be$) plane {for fixed Higgs masses} are shown in Fig.~\ref{fig:contours}. 
{Their} shapes are largely dictated by the perturbativity and vacuum stability requirements of the RGE evolved couplings, thus basically resembling the features observed before in Fig.~\ref{fig:mh600_scales} for $m_H = 600~\GeV$. Here, however, the maximally allowed values for the mixing angle $\sin\alpha$ stem now from {the NLO calculation of} $m_W$ {or, at rather low masses $m_H \lesssim 200~\GeV$,} from the Higgs signal rates {and/or exclusion limits}. In all cases, the upper limit on $\tan\be$ stems from the perturbativity requirement of RGE evolved couplings. For the degenerate case, {$m_h \approx m_H \approx 125~\GeV$}, we \emph{a priori} find no {upper or lower} limit on the mixing angle. In the degenerate case we do not take limits from RGE running into account, hence the only constraint stems from perturbative unitarity which renders an upper limit on $\tan\be$. 

{We now extend the discussion and} treat $m_H$ as a free model parameter. The results are presented {as scatter plots} using the following color scheme:
\begin{itemize}
\setlength{\itemsep}{0pt}
\item {\sl light gray} {points} include all scan points which are not further classified,
\item {\sl dark gray} points fulfill constraints from perturbative unitarity, perturbativity of the couplings, RGE running and the $W$ boson mass, as discussed in Sect.~\ref{Sect:Constraints}A--D,
\item{} {\sl blue} points {additionally pass} the $95\%~\mathrm{C.L.}$ exclusion limits from Higgs searches,
\item{} {\sl red/ yellow} points {fulfill all criteria above and furthermore} lie within a $1/\,2\,\sigma$ regime {favored by the Higgs signal rate observables}.
\end{itemize}
\begin{figure}[t]
\centering
\subfigure[\label{sf:mHressatb} ($\sin\al$, $\tan\be$) plane.]{
\includegraphics[width=0.46\textwidth]{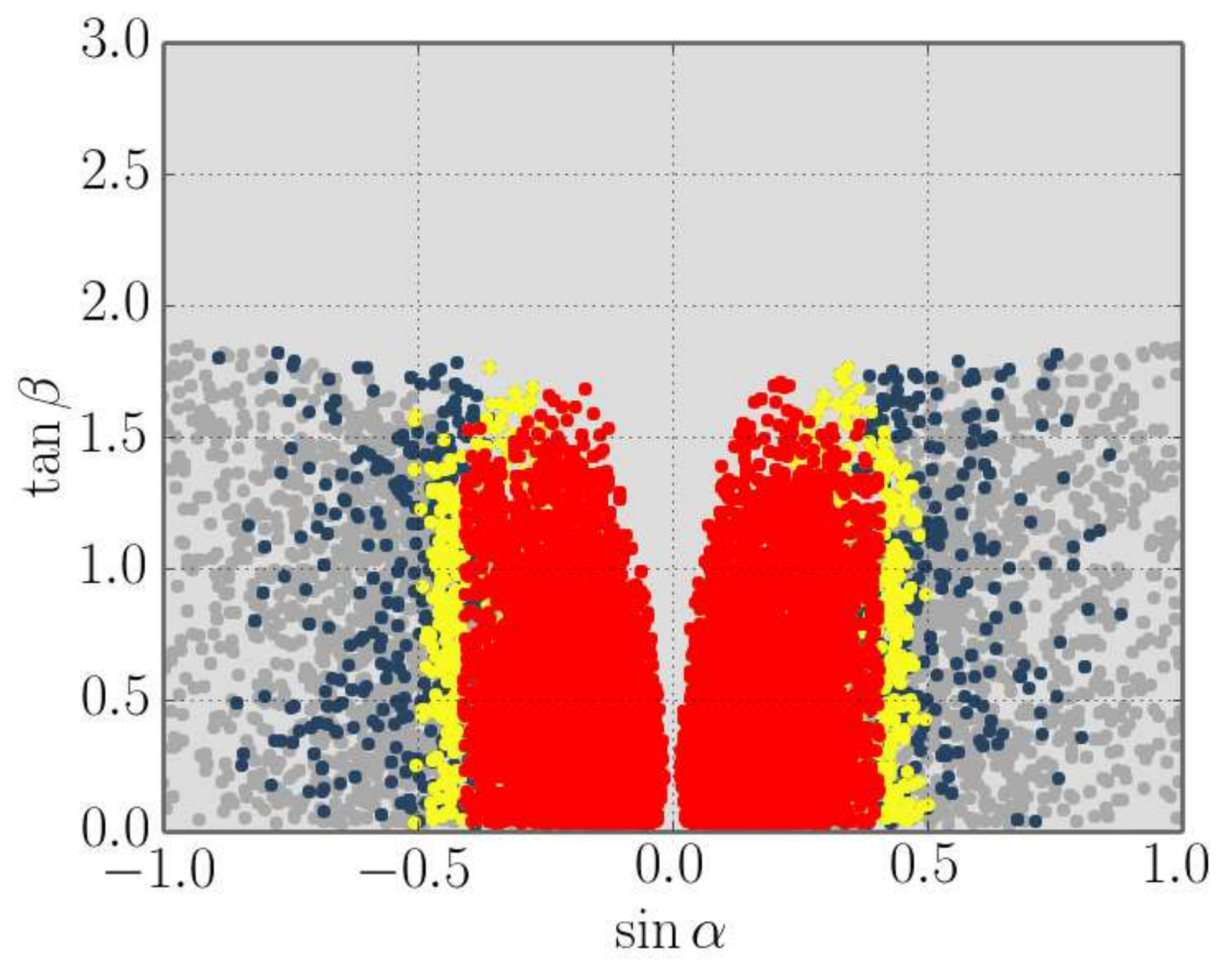}
}
\hfill
\subfigure[\label{sf:mHresmHtb} ($m_H$, $\tan\be$) plane.]{
 \includegraphics[width=0.46\textwidth]{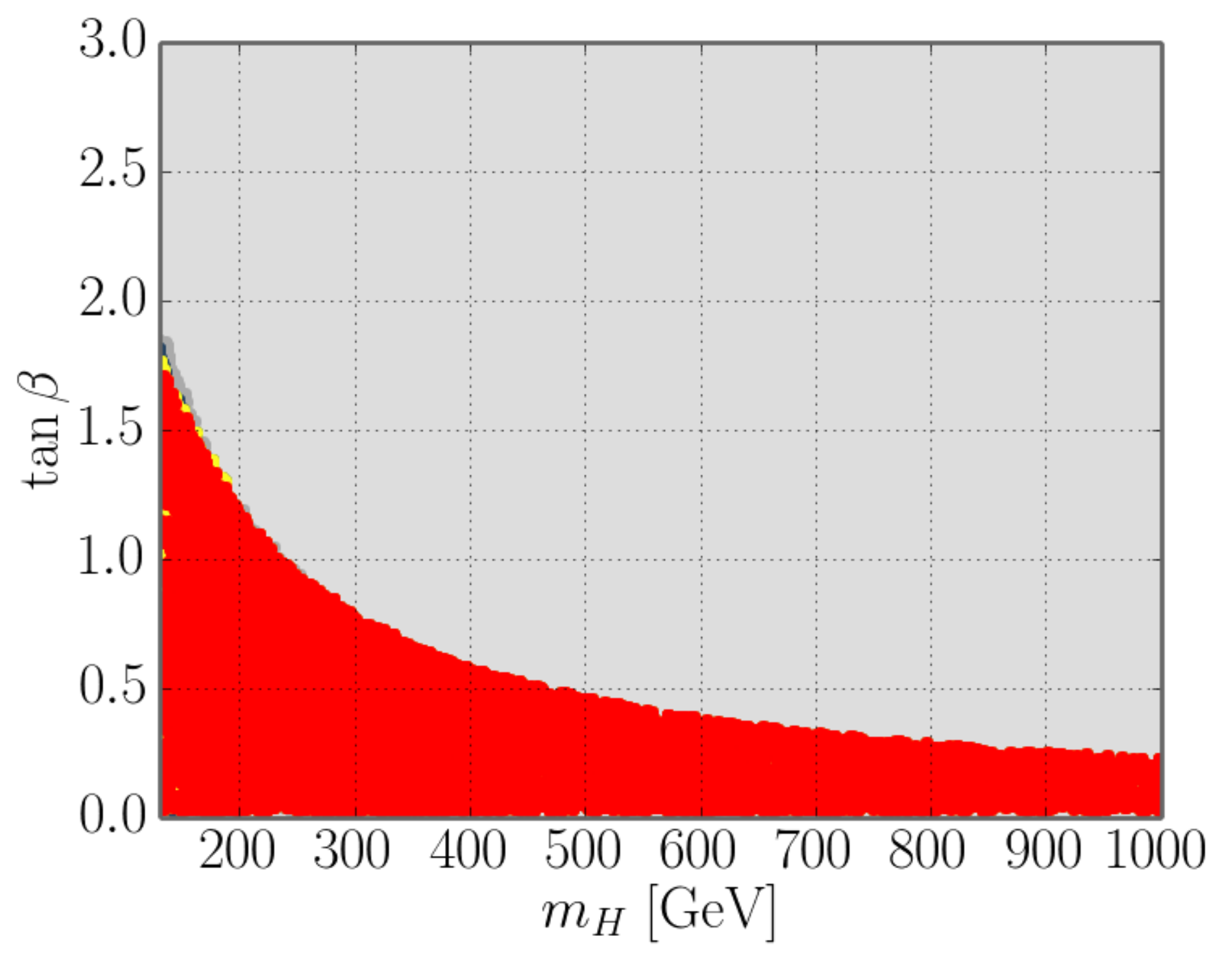}
}
\subfigure[\label{sf:mHresmHsa} ($m_H$, $\sin\al$) plane.]{ 
\includegraphics[width=0.46\textwidth]{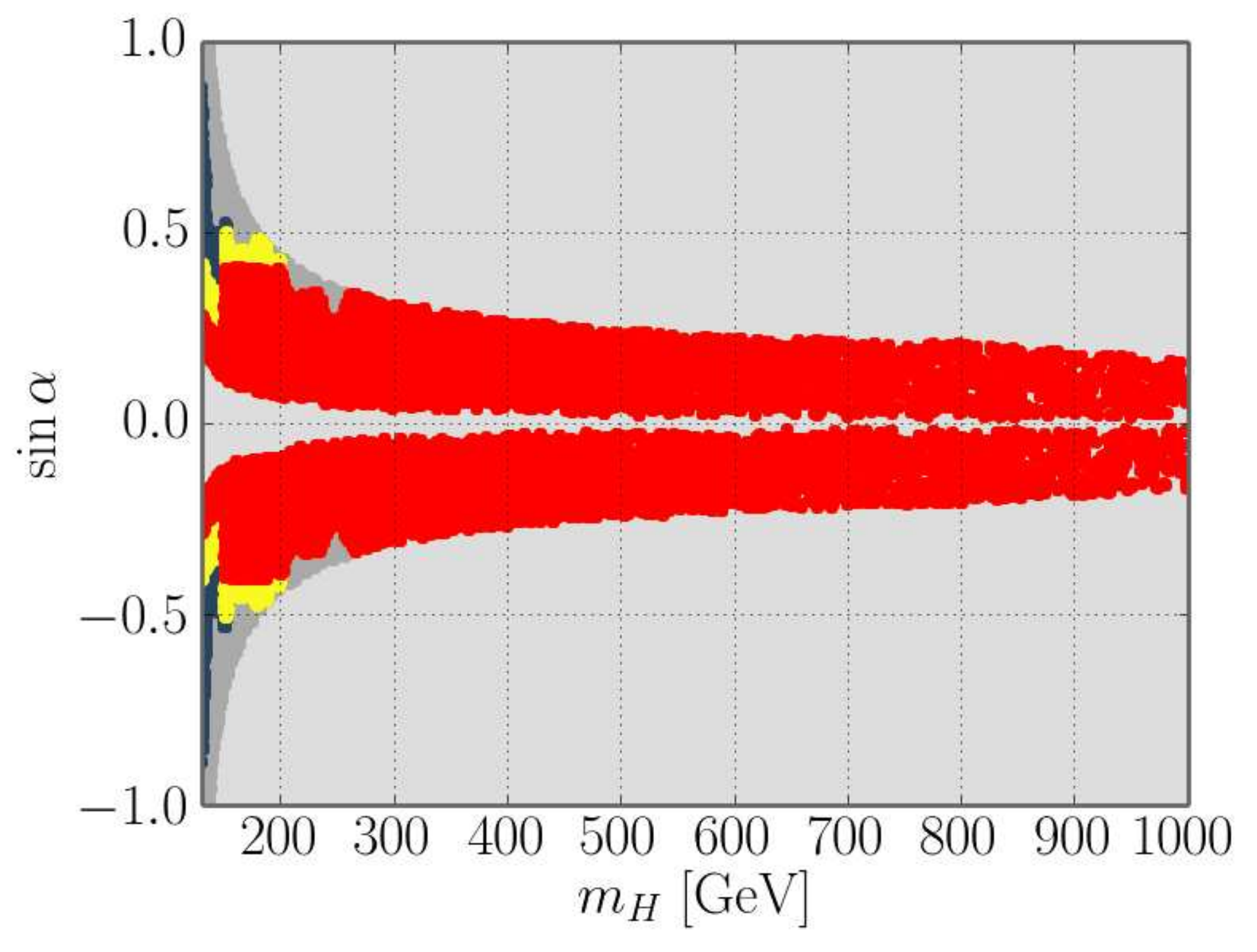}
}
\caption{\label{fig:mHres} Two-dimensional parameter correlations between $m_H$, $\tb$ and $\sa$ in the high mass region. See text for a description of the color coding.}
\end{figure}

The results are presented in Fig.~\ref{fig:mHres} in terms of two--dimensional scatter plots in the three scan parameters. The point distribution in the ($\sa$, $\tan\be$) plane shown in Fig.~\ref{sf:mHressatb} neatly resembles the features of Fig.~\ref{fig:contours} discussed above: Small mixings are forbidden from the requirement of vacuum stability, while {the maximal value for the mixing angle, $|\sin\al| \lesssim 0.50$}, {is limited by} the Higgs signal rate observables. Fig.~\ref{sf:mHresmHtb} illustrates how the upper limit on $\tan\be$, which stems from the perturbativity {requirement} of $\lam_2$, roughly follows the expected $\sim\,m_H^{-1}$ scaling, cf.~Eq.~\eqref{eq:tbpu}.
Finally, we can easily recognize the upper limit on the mixing angle $\sin\al$ from the $m_W$ constraint and the perturbativity requirement of $\lam_1$, cf.~Fig.~\ref{fig:sinamw}, in the point distribution in the ($m_H$, $\sin\al$) plane shown in Fig.~\ref{sf:mHresmHsa}. These constraints provide the most stringent upper limit on $|\sin\alpha|$ for Higgs masses $m_H \gtrsim 260~\GeV$. At lower Higgs masses, the upper limit is set by the Higgs signal rate measurements and exclusion limits from Higgs searches at the LHC, {cf.~Fig.~\ref{fig:comb}}. Here it is interesting to see that the favored $|\sin\alpha|$ region at Higgs masses $m_H$ between $130~\GeV$ and $\sim 152~\GeV$ is more restricted than at higher Higgs masses. Two effects play a role here: Firstly, the lower limit on $|\sin\al|$ from the vacuum stability requirement is stronger than at larger Higgs masses; And secondly, the heavy Higgs boson lies still in the vicinity of the discovered Higgs state, such that their signal rates are combined in the $H\to\tau\tau$, $H\to WW$ and $VH\to V(b\bar{b})$ channels, where the mass resolution is poor. In total, these channels however favor a slightly lower signal strength than obtained for a SM Higgs, thus the fit slightly prefers larger Higgs masses $m_H$, where the signal rates are not added for these observables within \HS, cf.~also Fig.~\ref{fig:massoverlap_chi2}.

We now turn to the discussion of the collider phenomenology of the high mass region. Experimentally, the model can be probed by searches for a SM--like Higgs boson with a reduced signal rate and total decay widths, or by direct searches for the Higgs-to-Higgs decay mode $H\rightarrow hh$, where $h$ is the light Higgs boson at around $125~\GeV$.

\begin{figure}[!tb]

\subfigure[\label{fig:brHtohh_sina} Allowed values of $\text{BR}(H\to hh)$ as function of $\sin\alpha$ for fixed Higgs masses $m_H$.]{\includegraphics[width=0.5\textwidth, trim = -1.2cm -0.6cm 1.2cm 0.6cm]{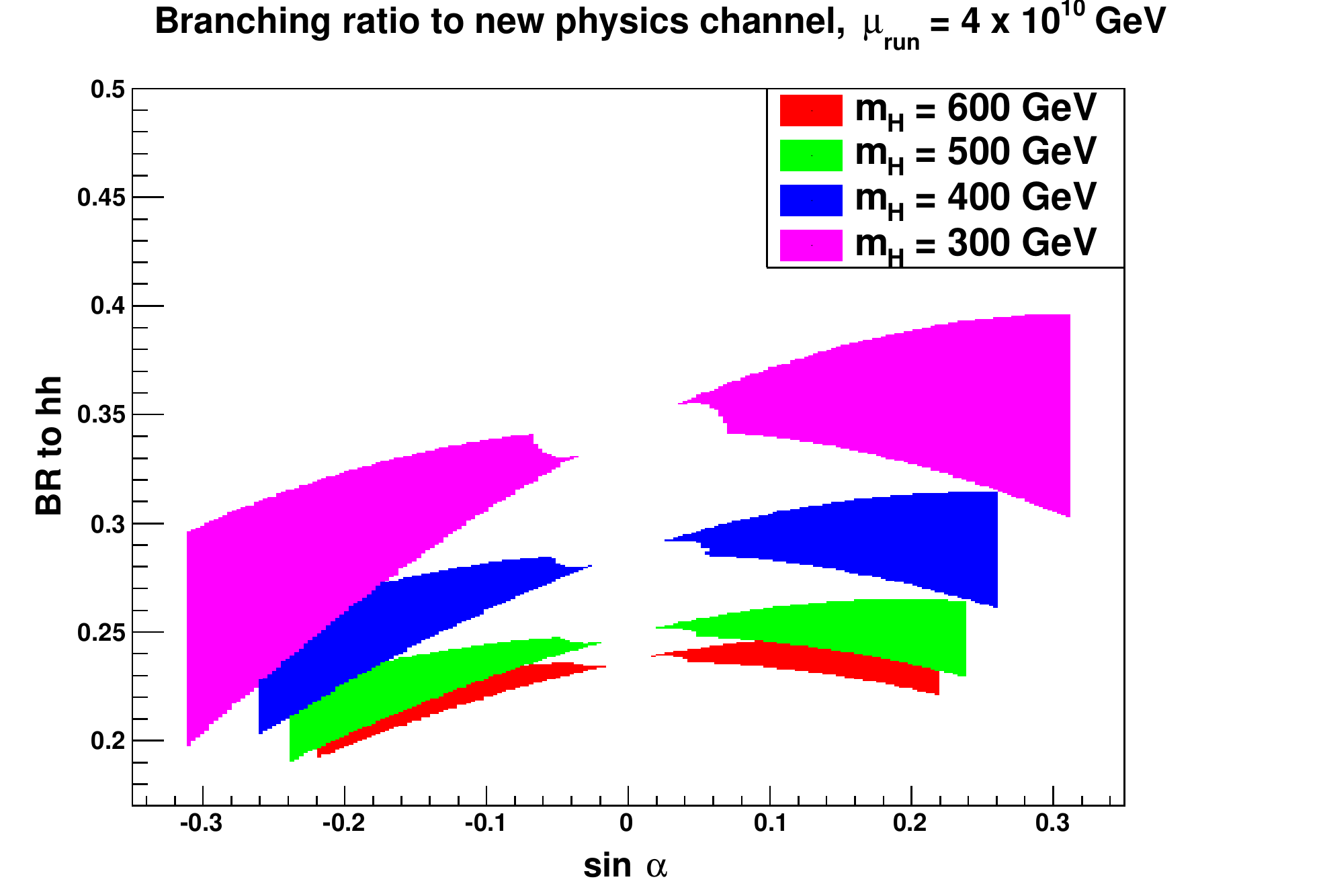}}
\hfill
\subfigure[\label{fig:brHtohh_mH} $\text{BR}(H\to hh)$ as a function of $m_H$, using the same color code as in Fig.~\ref{fig:mHres}.]
{\includegraphics[width=0.46\textwidth]{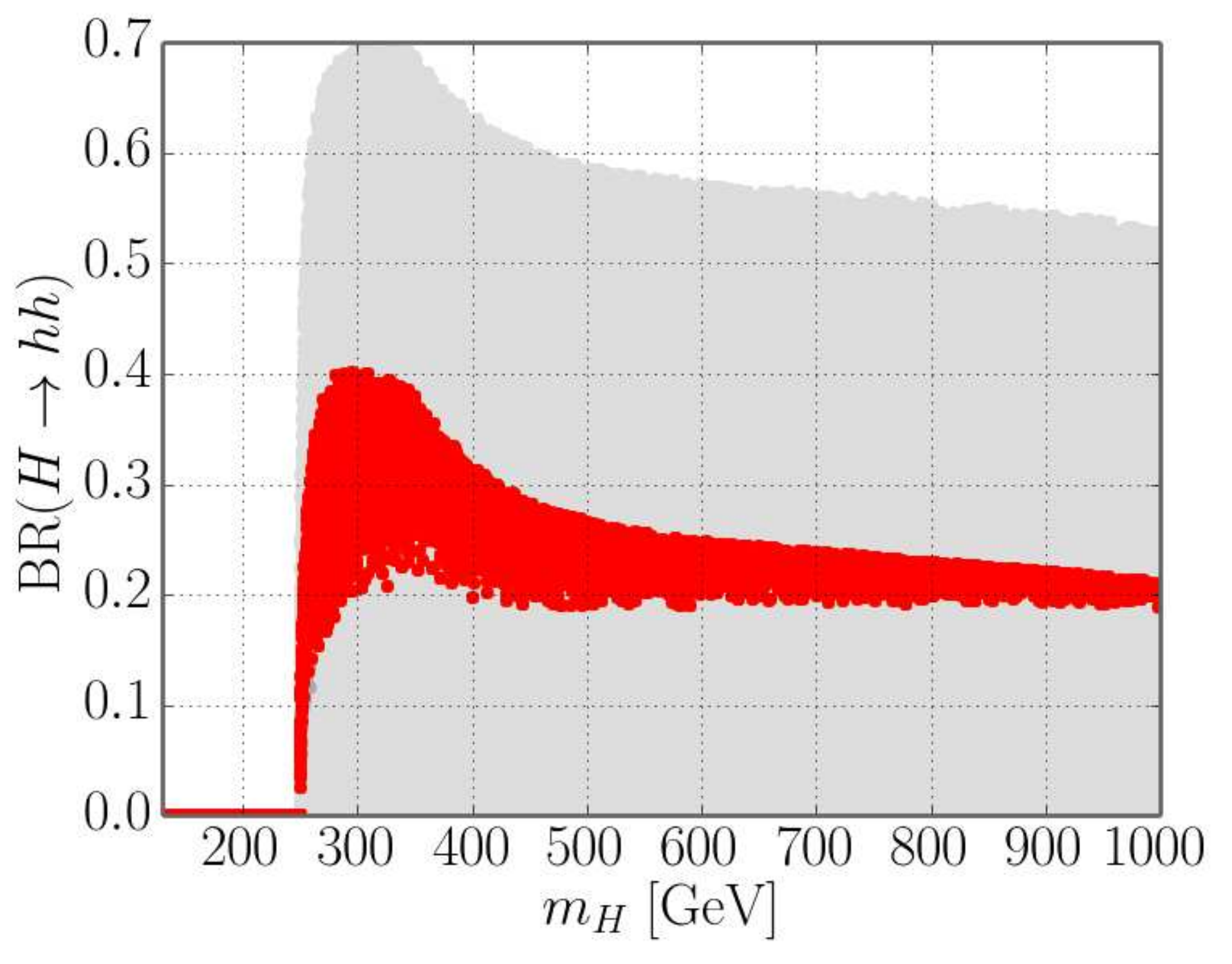}}
\caption{\label{fig:brHtohh_high}Allowed branching ratios of the Higgs-to-Higgs decay channel $H\rightarrow hh$ in the high mass scenario. }
\end{figure}

{We show} the allowed values of the branching ratio $\text{BR} (H\rightarrow hh)$, given by Eq.~\eqref{eq:brdefs}, in Fig.~\ref{fig:brHtohh_high}. In Fig.~\ref{fig:brHtohh_sina} we show the dependence on $\sin\al$ exemplarily for fixed Higgs masses $m_H$, whereas the full $m_H$ dependence is displayed in Fig.~\ref{fig:brHtohh_mH}, using the same color code as above. We observe that the maximal values of $\text{BR} (H\rightarrow hh)$ are $\sim 40\%$, reached for large, positive $\sin\al$ values~\cite{Pruna:2013bma}, and low Higgs masses {$m \sim 300~\GeV$}. {At higher Higgs masses} the branching ratio $\text{BR} (H\rightarrow hh)$ {is} around $20\%$ or slightly higher.

\begin{figure}
\centering
\subfigure[~Heavy Higgs signal rate with SM particles in the final state. The orange solid (dashed) curves indicate the observed (expected) $95\%~\text{C.L.}$ limits from the latest CMS combination of SM Higgs searches~\cite{CMS:aya}.]{
\includegraphics[width=0.46\textwidth]{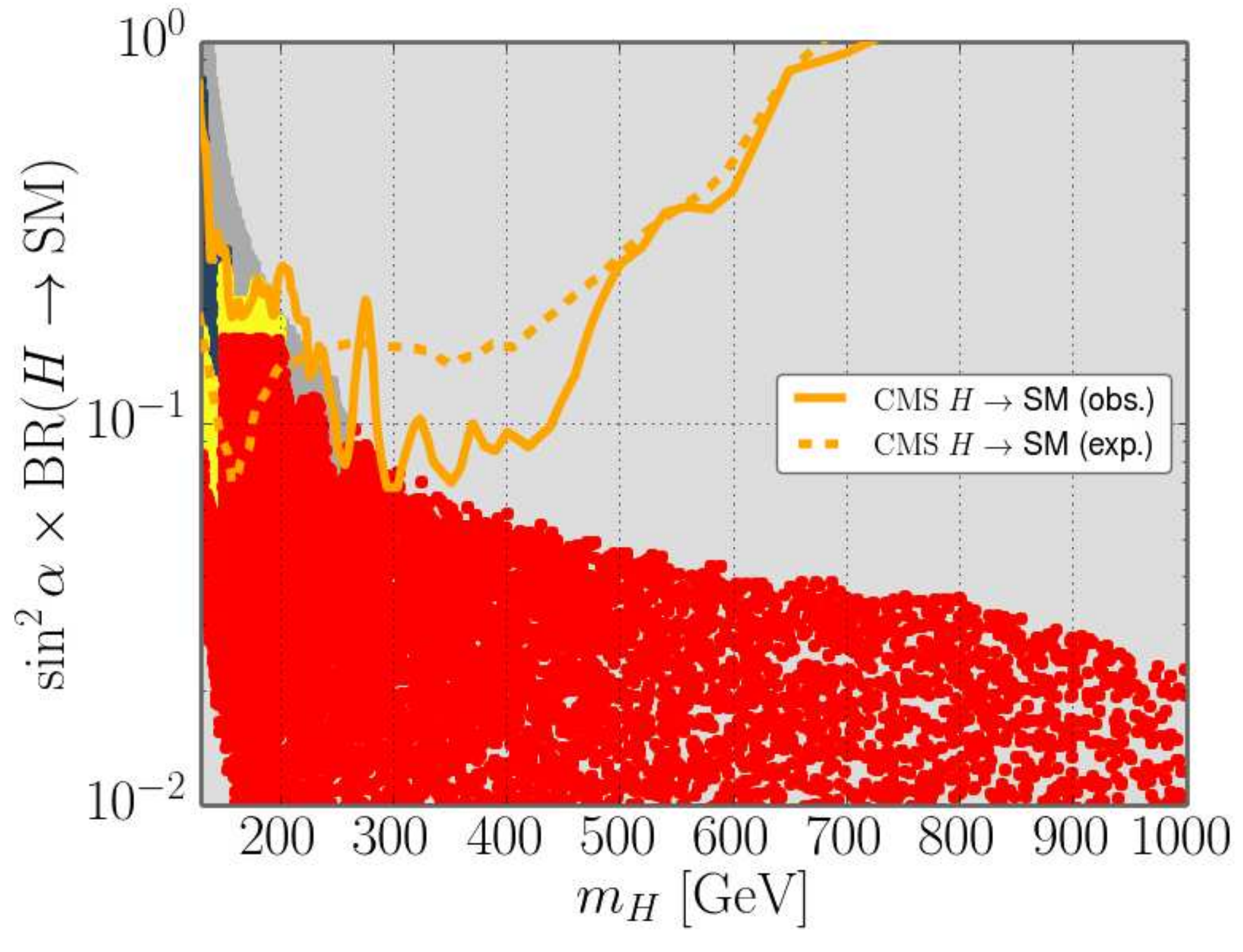}
}
\hfill
\subfigure[~Heavy Higgs signal rate with light Higgs bosons in the final state. We display the current expected and observed $95\%~\text{C.L.}$ limits from CMS $H\to hh$ searches with $\gamma\gamma b\bar{b}$~\cite{CMS:2014ipa} and $b\bar{b}b\bar{b}$~\cite{CMS:2014eda} final states.]{
\includegraphics[width=0.458\textwidth]{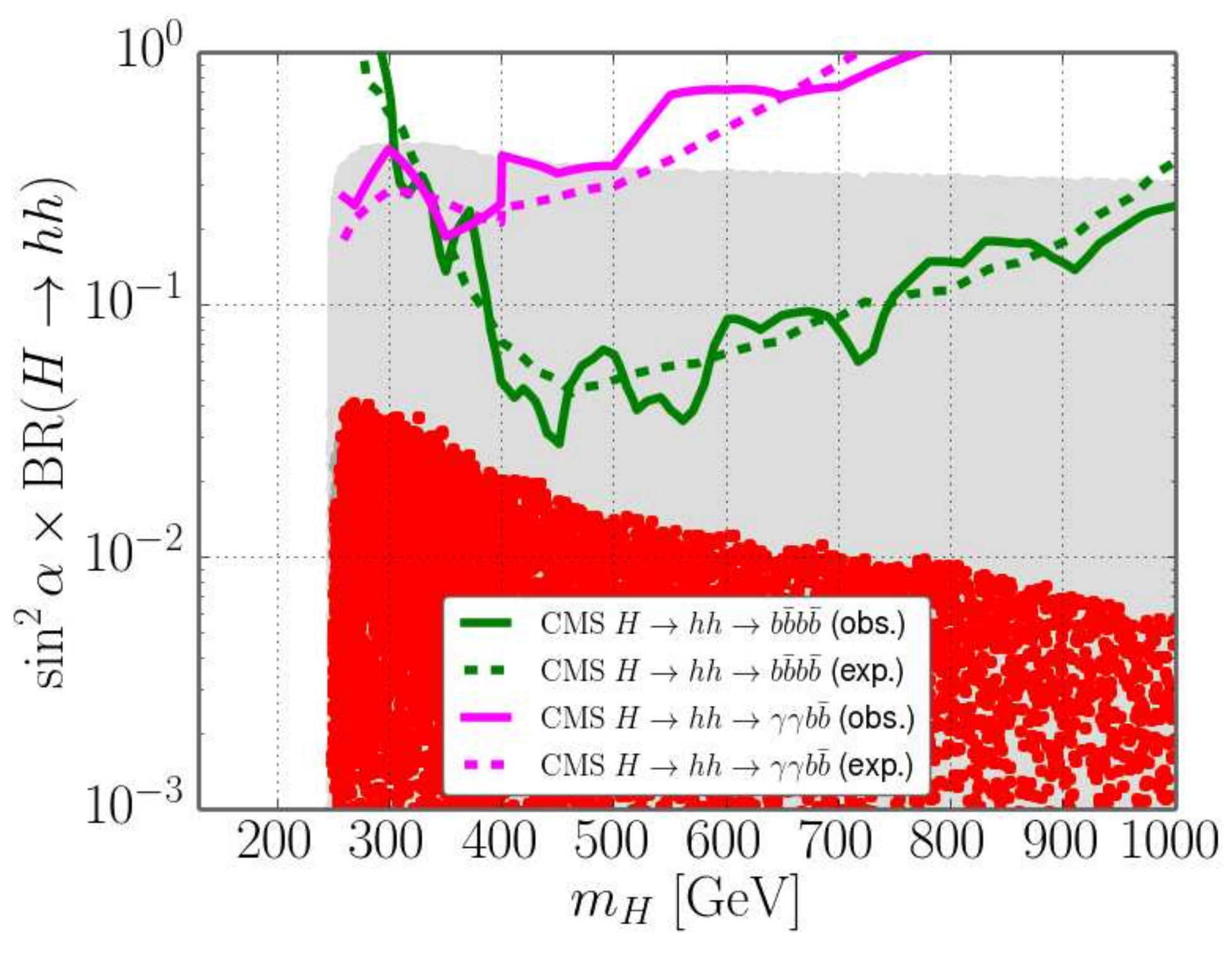}
}
\caption{\label{fig:lhcsig} Collider signal rates of the heavy Higgs boson $H$ decaying into SM particles (\emph{a}) or into two light Higgs bosons, $H\to hh$, (\emph{b}), in dependence of the heavy Higgs mass, $m_H$. The rates are normalized to the inclusive SM Higgs production cross section at the corresponding mass value~\cite{Dittmaier:2011ti,Dittmaier:2012vm,Heinemeyer:2013tqa}.}
\end{figure}

\begin{figure}
\centering
\subfigure[~Heavy Higgs signal rate with SM particles in the final state {for the LHC at $8~\TeV$}. The orange solid (dashed) curves indicate the observed (expected) $95\%~\text{C.L.}$ limits from the latest CMS combination of SM Higgs searches~\cite{CMS:aya}.]{
\includegraphics[width=0.46\textwidth]{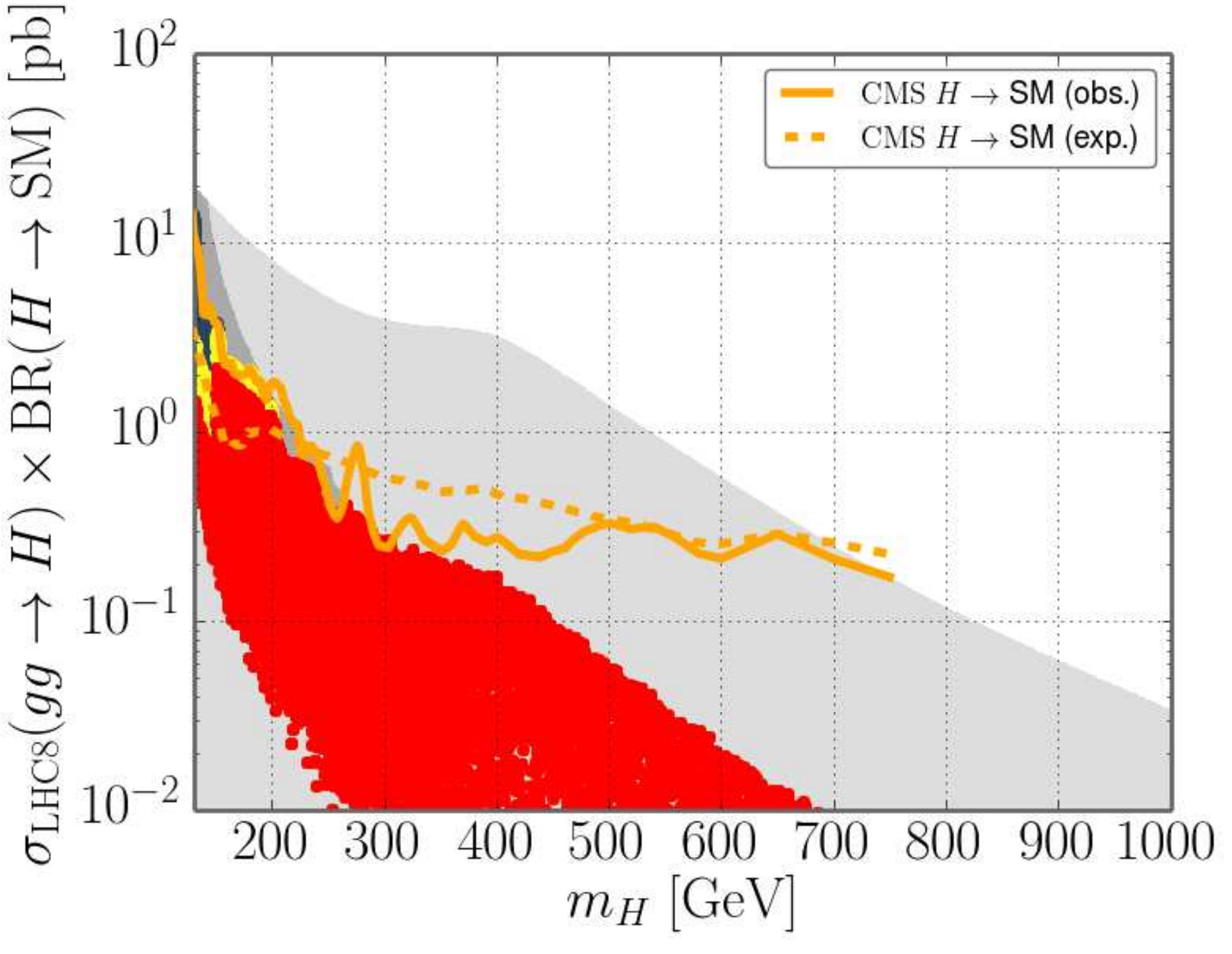}
}
\hfill
\subfigure[~Heavy Higgs signal rate with light Higgs bosons in the final state {for the LHC at $8~\TeV$}. We display the current expected and observed $95\%~\text{C.L.}$ limits from CMS $H\to hh$ searches with $\gamma\gamma b\bar{b}$~\cite{CMS:2014ipa} and $b\bar{b}b\bar{b}$~\cite{CMS:2014eda} final states.]{
\includegraphics[width=0.458\textwidth]{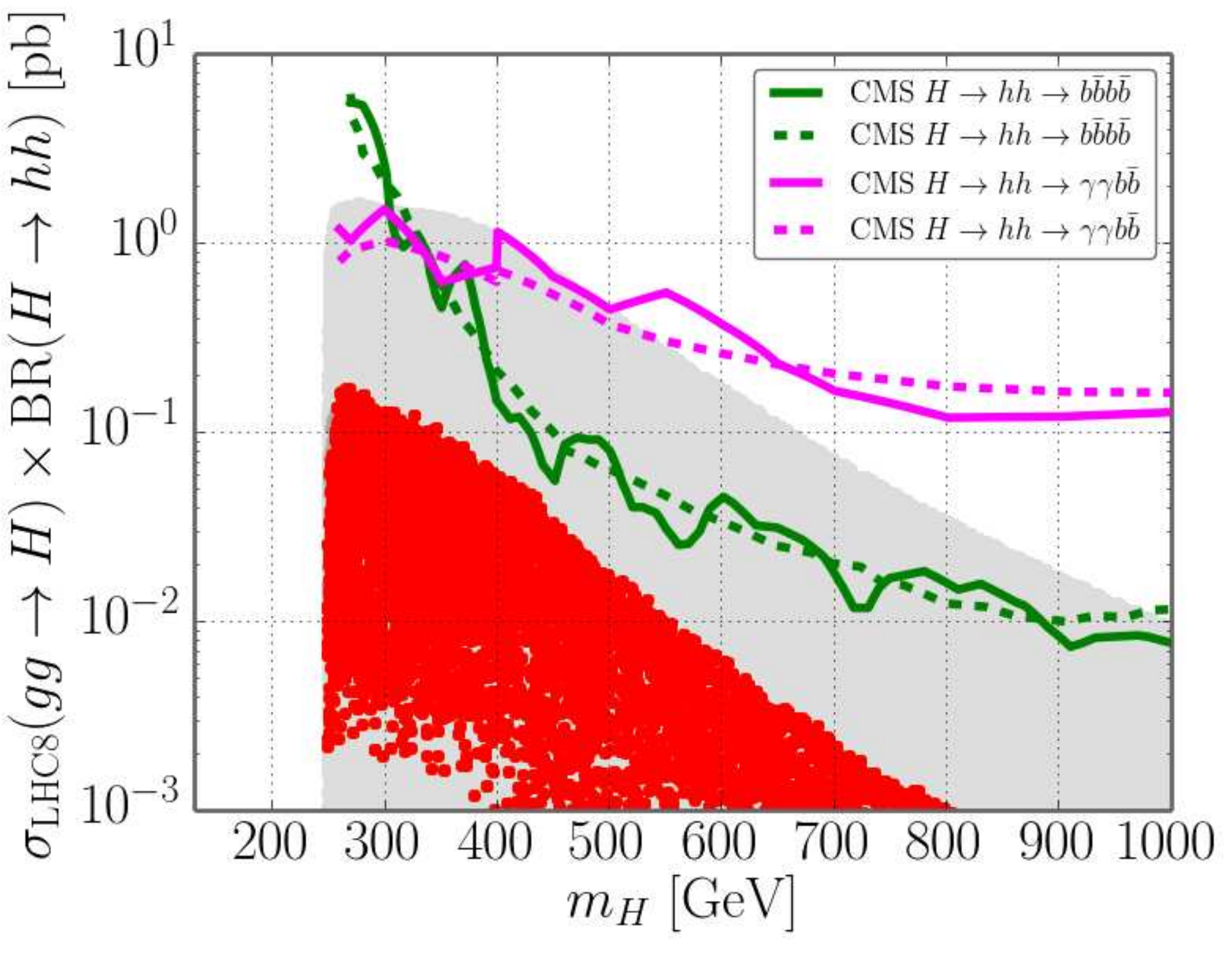}
}
\subfigure[~{Heavy Higgs signal rate with SM particles in the final state for the LHC at $14~\TeV$}.]{
\includegraphics[width=0.46\textwidth]{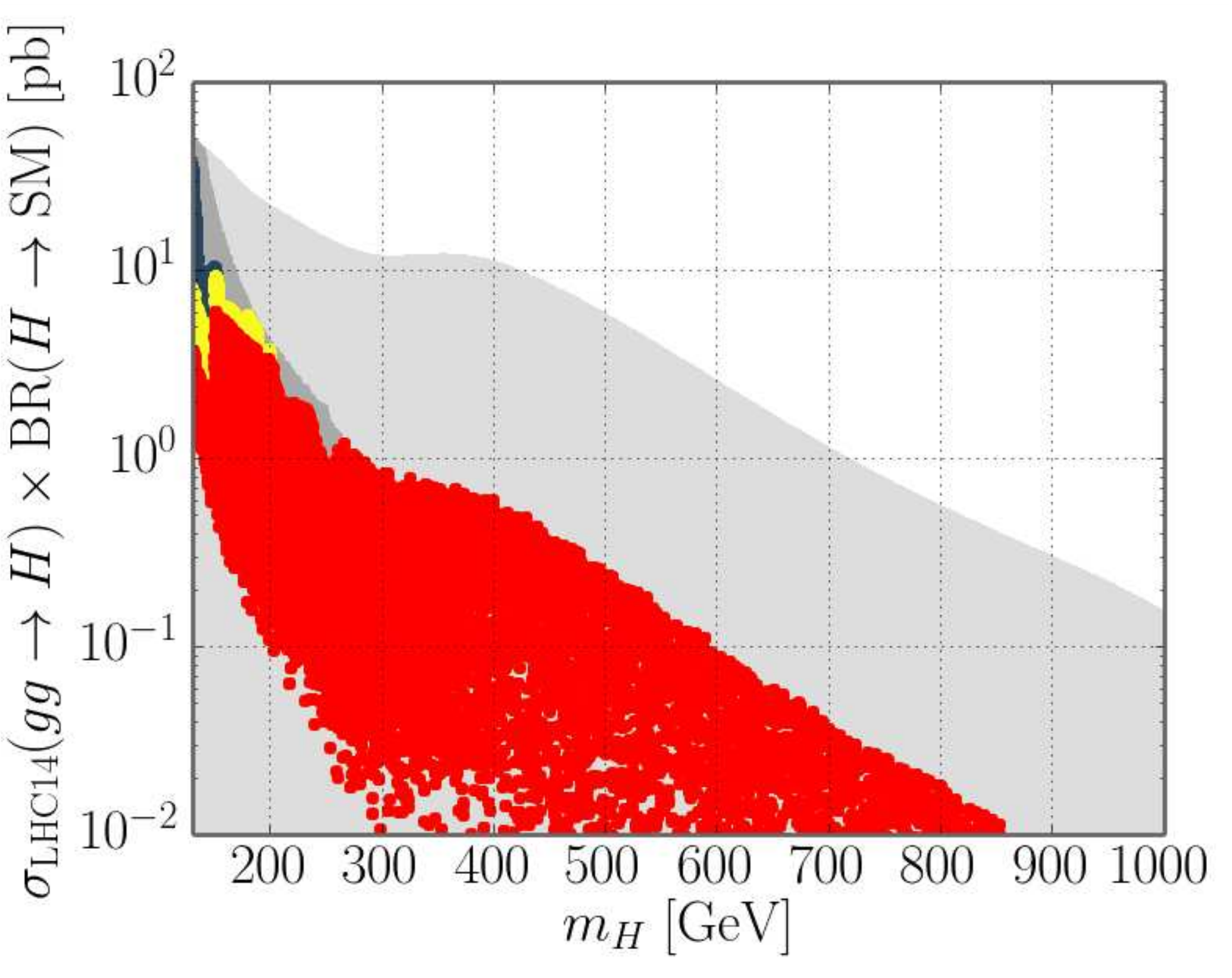}
}
\hfill
\subfigure[~{Heavy Higgs signal rate with light Higgs bosons in the final state for the LHC at $14~\TeV$}.]{
\includegraphics[width=0.46\textwidth]{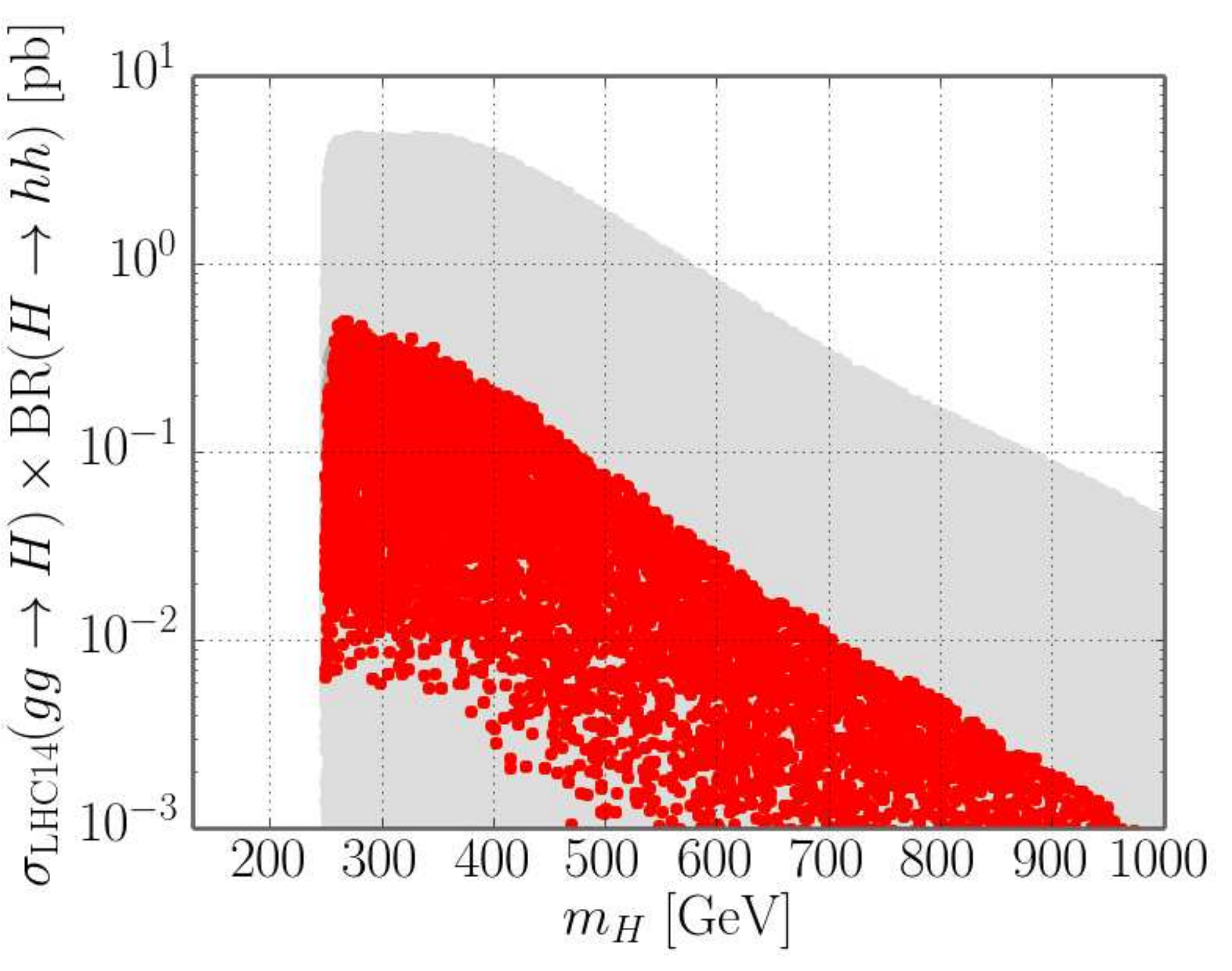}
}
\caption{\label{fig:lhcsig_abs} LHC signal rates of the heavy Higgs boson $H$ decaying into SM particles (\emph{a,c}) or into two light Higgs bosons, $H\to hh$, (\emph{b,d}), in dependence of the heavy Higgs mass, $m_H$, for CM energies of $8~\TeV$ (\emph{a,b}) and $14~\TeV$ (\emph{c,d}).}
\end{figure}

{The LHC production cross section of the heavier Higgs boson $H$ is given by the SM Higgs production cross section multiplied by $\sin^2\al$. For convenience, we introduce the rate scale factors~\cite{Pruna:2013bma}}
\begin{align}
\kappa&\equiv \frac{\sigma}{\sigma_\text{SM}}\times\text{BR} (H\to \mathrm{SM})=\sin^4\al \,\frac{\Gamma_\text{SM,tot}}{\Gamma_\text{tot}}, \label{Eq:kappa}\\
\kappa'&\equiv\frac{\sigma}{\sigma_\text{SM}} \times\text{BR}(H\to hh)=\sin^2\al\,\frac{\Gamma_{H\rightarrow hh}}{\Gamma_\text{tot}},\label{Eq:kappaprime}
\end{align}
{for the heavy Higgs collider processes leading to SM particles or two light Higgs bosons in the final state, respectively. Here, $\text{BR} (H\to \mathrm{SM})$ comprises all possible Higgs decay modes to SM particles. Note, that $\kappa+\kappa'\,=\,\sin^2\al$ corresponds to the inclusive heavy Higgs production rate, normalized to the inclusive SM Higgs production rate~\cite{Dittmaier:2011ti,Dittmaier:2012vm,Heinemeyer:2013tqa}.}

The predicted signal rates normalized to the SM production cross section, Eqs.~\eqref{Eq:kappa} and~\eqref{Eq:kappaprime}, are shown as a function of the Higgs mass $m_H$ for the high mass region in Fig.~\ref{fig:lhcsig}. We furthermore display the current $95\%~\mathrm{C.L.}$ exclusion limits from the latest CMS combination of SM Higgs searches~\cite{CMS:aya}, as well as from direct searches for the $H\to hh$ process with $\gamma\gamma b\bar{b}$~\cite{CMS:2014ipa} and $b\bar{b}b\bar{b}$~\cite{CMS:2014eda} final states.
We see that at the current stage, the experimental searches with SM-like final states yield important constraints for $m_H \lesssim {300}~\GeV$. As discussed above, at larger masses the upper limit on the mixing angle, and thus on the maximal production cross section, stems either from $m_W$ or from perturbativity. Note, that the {displayed} CMS limit from the SM Higgs search combination~\cite{CMS:aya} is only based on $\le 5.1~\mathrm{fb}^{-1}$ of $7~\TeV$ and $\le 12.2~\mathrm{fb}^{-1}$ of $8~\TeV$ data, thus not exploiting the full available data from LHC run 1. Obviously, future LHC searches for a SM-like Higgs boson with reduced couplings in the full accessible mass range will play an important role in probing the singlet extended SM.
The direct searches for the $H\rightarrow hh$ process carried out by CMS in the final states $\gamma\gamma b\bar{b}$~\cite{CMS:2014ipa} and, in particular, $b\bar{b}b\bar{b}$~\cite{CMS:2014eda} draw near to the allowed region at masses $\sim 450~\GeV$. While they do not yield any relevant constraints at the current stage, these searches will become important in this model in the upcoming LHC runs, as they are complementary to the SM-like Higgs searches. For reference, we also provide the predicted LHC signal cross section for both the SM Higgs signatures and $H\to hh$ signature for CM energies of $8$ and $14~\TeV$ in Fig.~\ref{fig:lhcsig_abs}. {Note that, as discussed earlier (see footnote~\ref{Footnote:interference}), we do not include effects from the interference with the Higgs boson at $\sim 125~\GeV$ in these predictions.}

\begin{figure}
\centering

\subfigure[~Ratio of total width, $\Gamma_\text{tot}$, and the Higgs mass, $m_H$.]{
\includegraphics[width=0.45\textwidth]{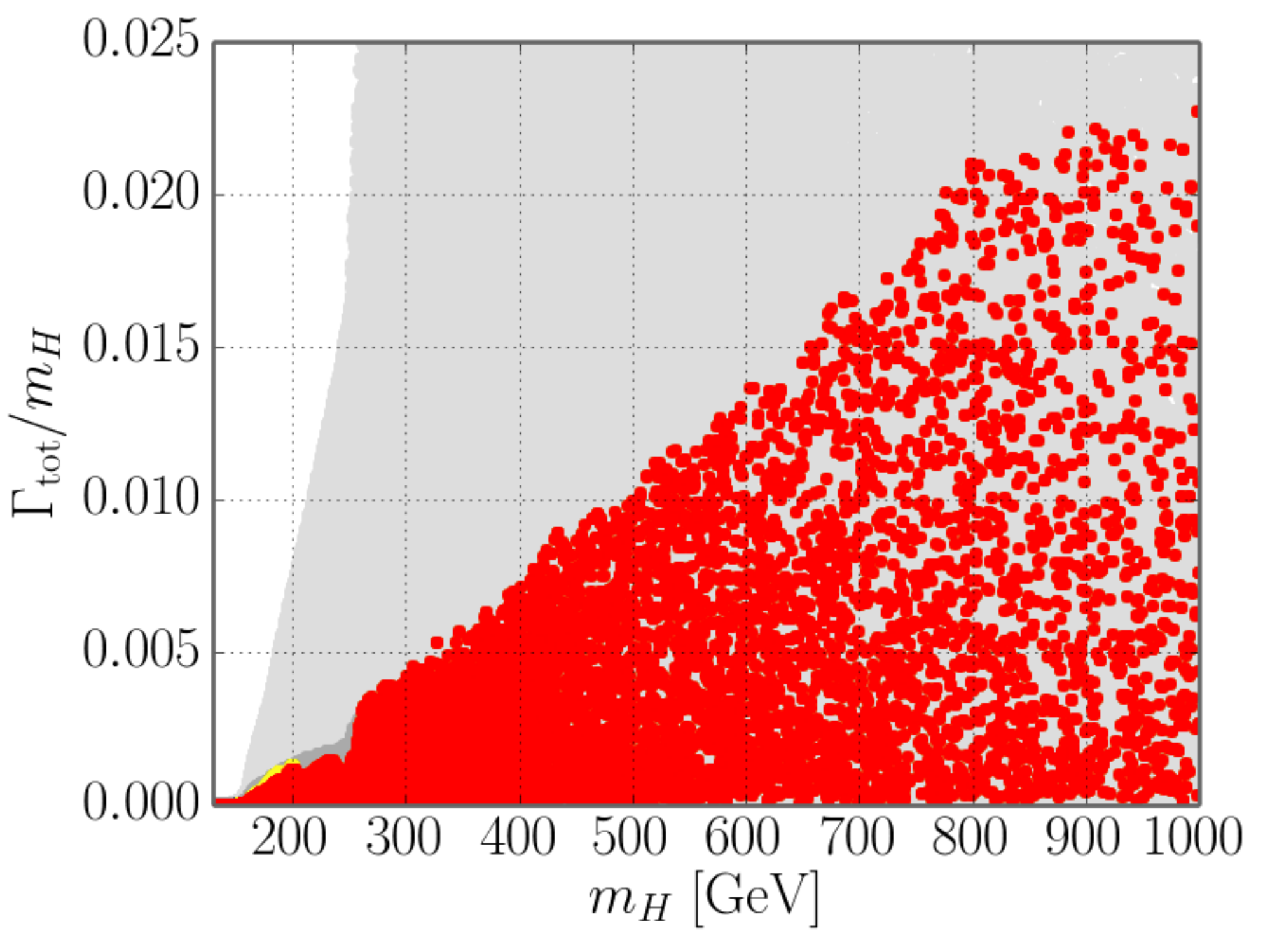}
}
\hfill
\subfigure[~Suppression of the total width with respect the total width of a SM Higgs boson at mass $m_H$.]{
\includegraphics[width=0.45\textwidth]{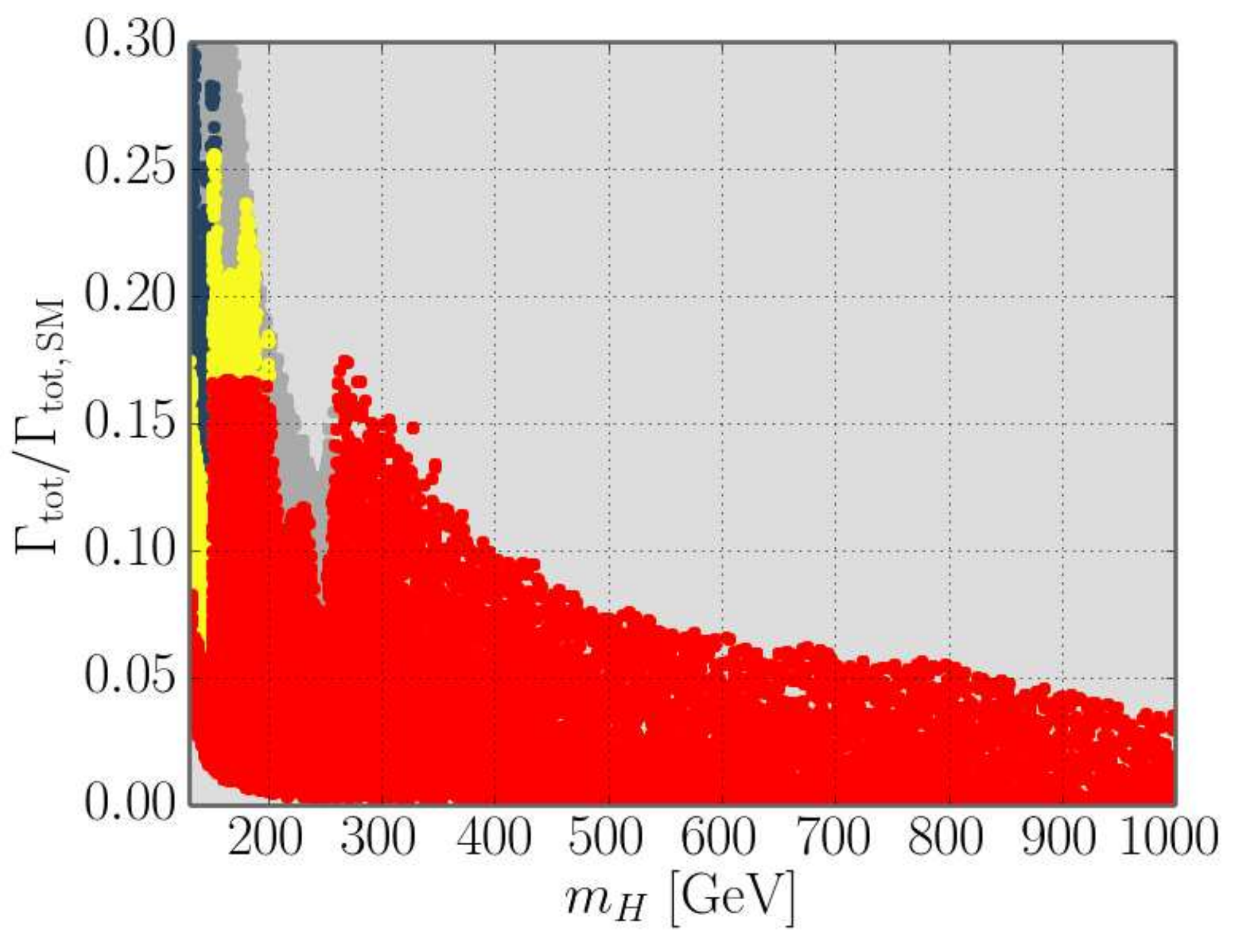}
}
\caption{\label{fig:widthshm} Total width, $\Gamma_\text{tot}$, as a function of the Higgs mass $m_H$. We display the ratio, $\Gamma_\text{tot}/m_H$ (a), as well as the suppression factor with respect to the SM Higgs width, $\Gamma_\text{tot}/\Gamma_\text{tot,~SM}$ (b). We obtain $\Gamma_\text{tot}/m_H \lesssim 0.02$, as well as a suppression of $25\,\%$ or lower of the total width compared to the SM prediction, in agreement with Ref.~\cite{Pruna:2013bma}.}
\end{figure}

\begin{figure}[!tb]
\centering
\subfigure[~($\Gamma_\text{tot}, \kappa$) plane for fixed Higgs mass values $m_H$.]{\includegraphics[width=0.49\textwidth, trim = -1.2cm 0 1.2cm 0 ]{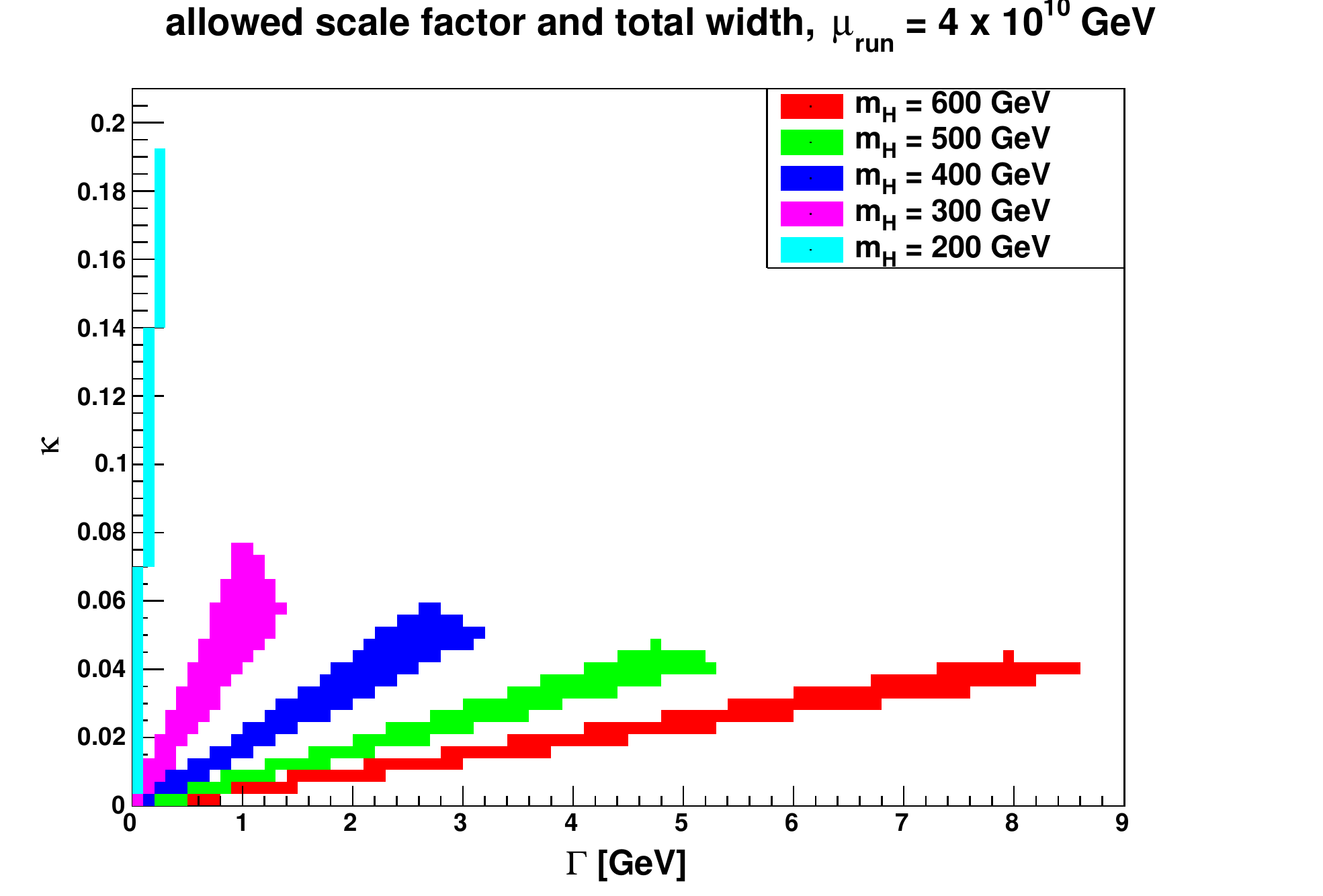}}
\hfill
\subfigure[~($\Gamma_\text{tot}, \kappa'$) plane for fixed Higgs mass values $m_H$.]{\includegraphics[width=0.49\textwidth, trim = -2.1cm 0 2.1cm 0]{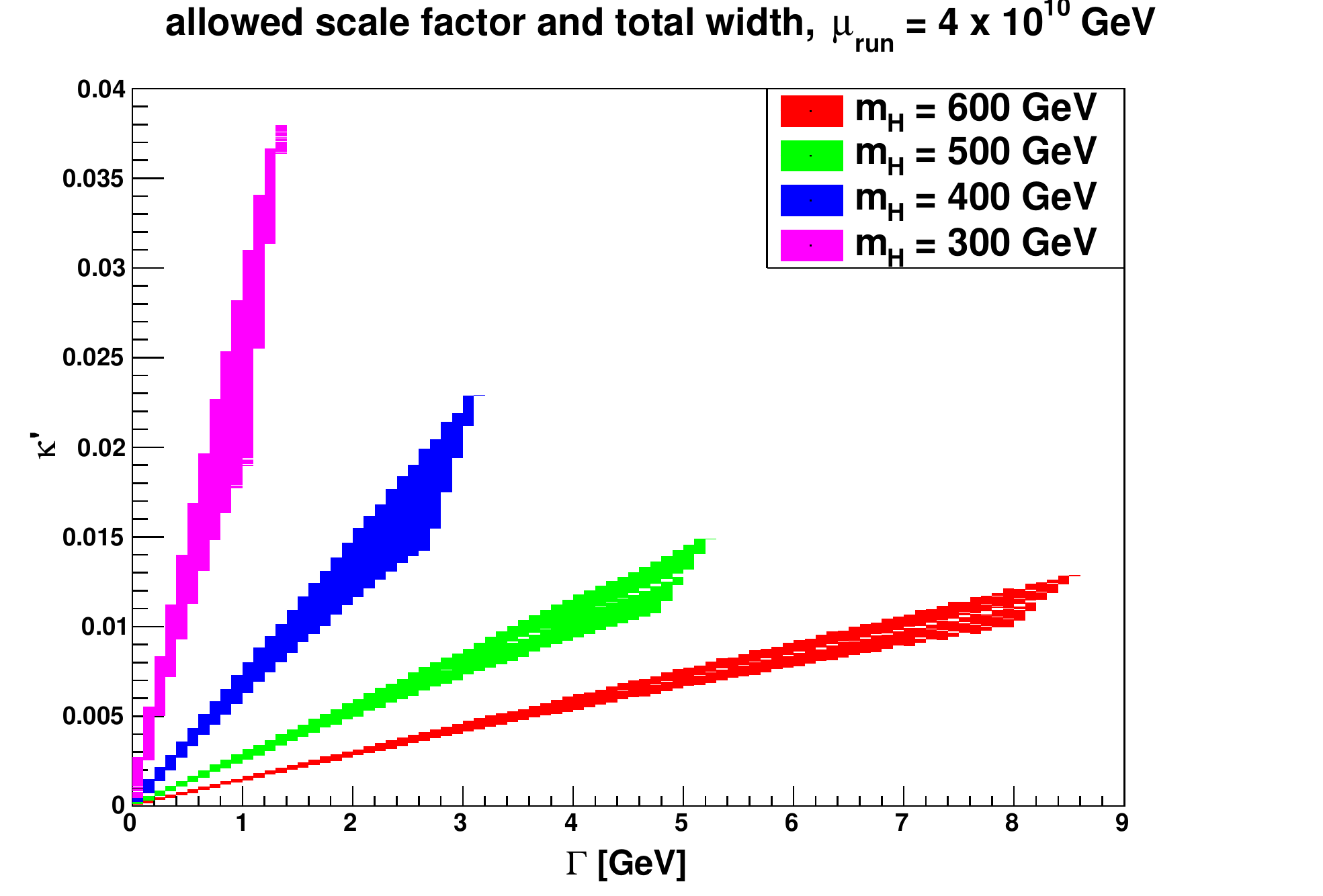}}
\subfigure[~($\Gamma_\text{tot}, \kappa$) plane from the full scan.]{\includegraphics[width=0.46\textwidth]{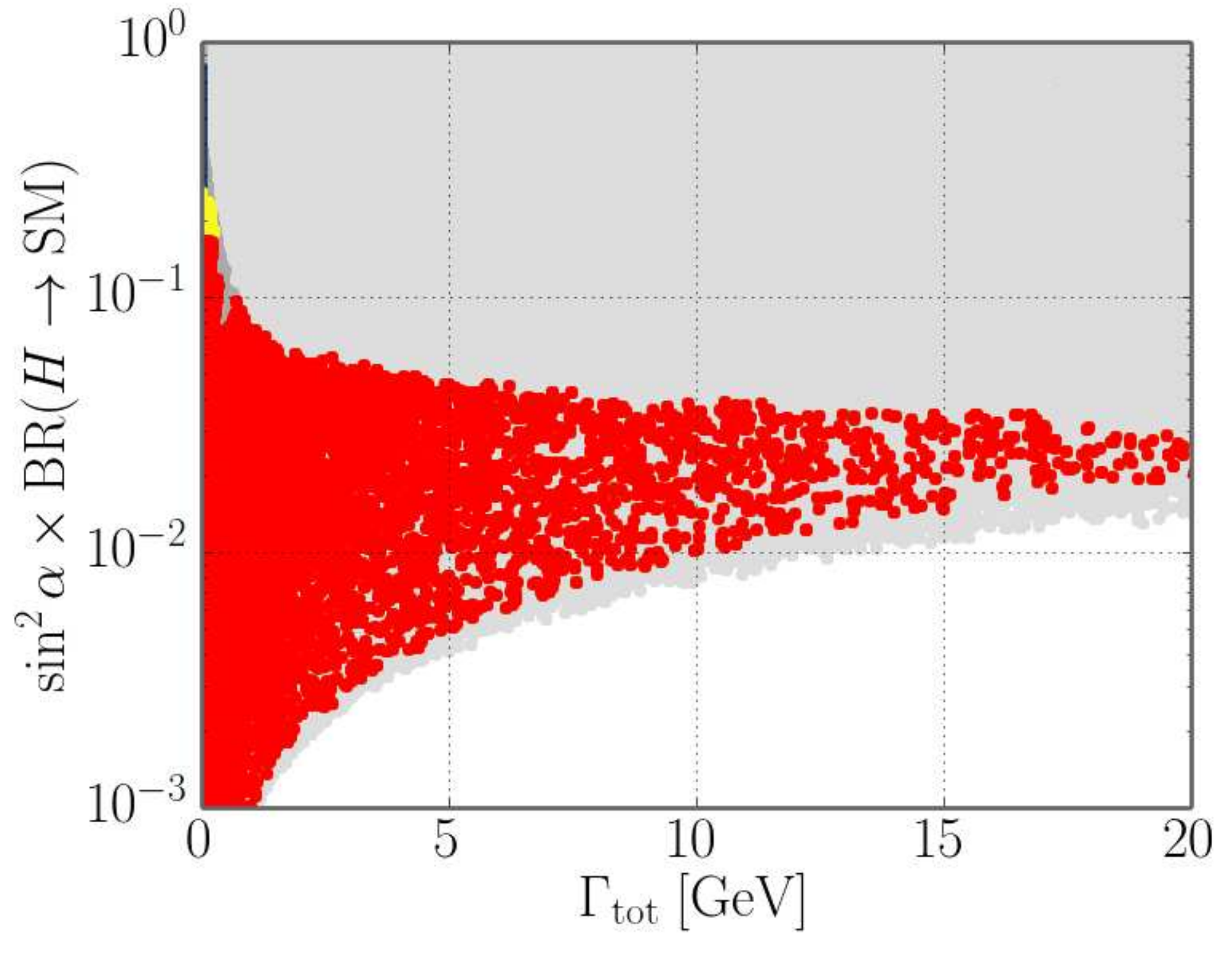}}
\hfill
\subfigure[~($\Gamma_\text{tot}, \kappa'$) plane from the full scan.]{\includegraphics[width=0.46\textwidth]{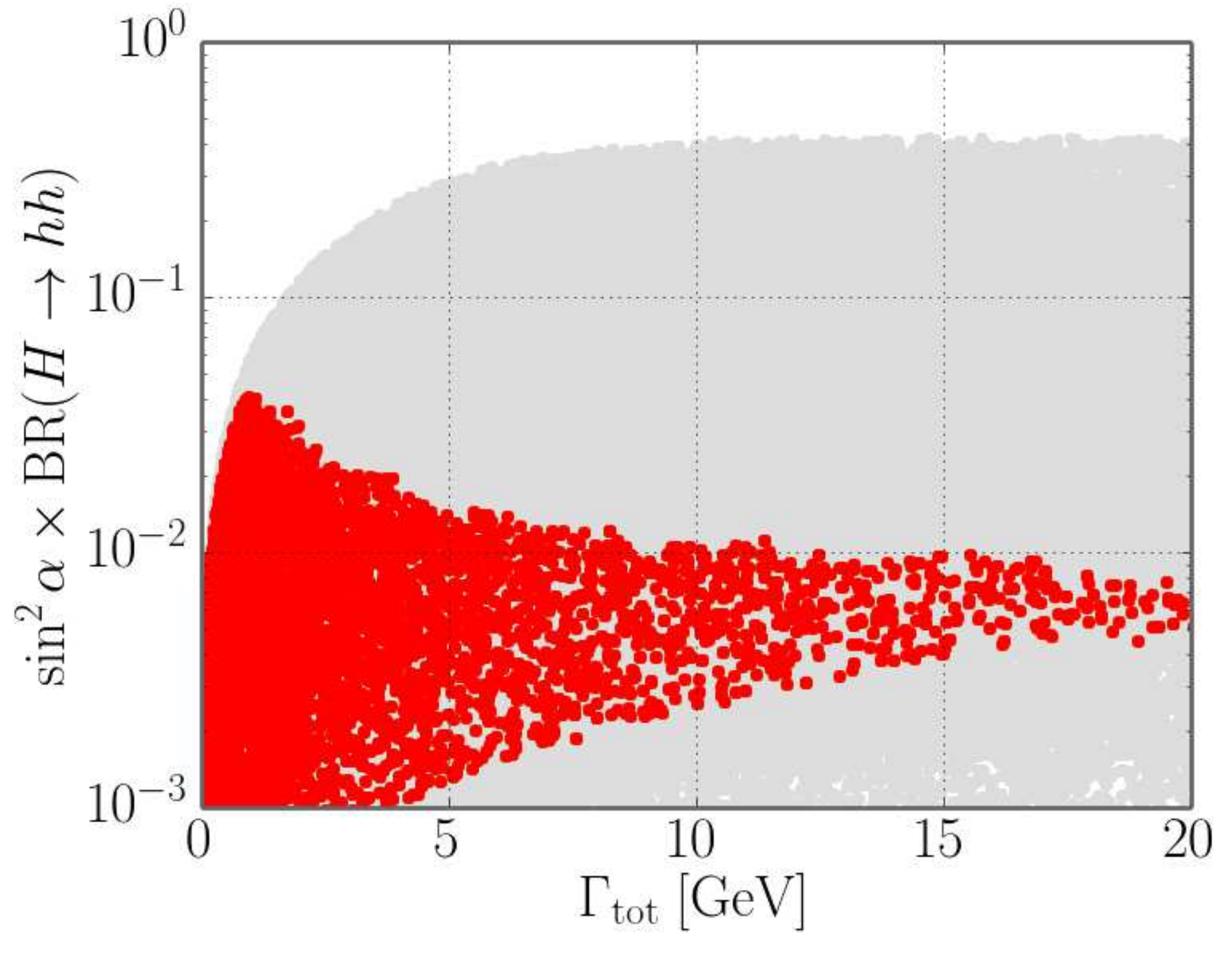}}
\caption{\label{fig:gk_scales} Allowed regions in the $(\Gamma_\text{tot}, \kappa)$ plane (\emph{a,c}) and $(\Gamma_\text{tot}, \kappa')$ plane  (\emph{b,d}). In (\emph{a,b}) the results are shown for various fixed values of $m_H$. $\kappa$ and $\kappa'$ denote the scaling factors for SM-like decays and the new physics channel $H\rightarrow hh$, respectively, cf.~Eq.~\eqref{Eq:kappa} and~\eqref{Eq:kappaprime}.} 
\end{figure}

In general, the {\sl total} width of the heavy Higgs boson is of high interest for collider searches. In the SM, the width of {the SM} Higgs {boson} rapidly rises with its mass. In Ref.~\cite{Pruna:2013bma} it was shown that in the singlet {extended SM} the total width of the heavy resonance, $\Gamma_\text{tot}\lb m_H \rb$, is highly suppressed {due to the small mixing angle required. The same behavior is observed here.} We show the ratio $\Gamma_\text{tot} / m_H$, as well as the suppression of the width, $\Gamma_\text{tot}/\Gamma_\text{SM}$, in Fig.~\ref{fig:widthshm}.
We see that the total width of the heavy Higgs only amounts to up to $\sim 20-25\%$ at lower masses $m_H\lesssim 200~\GeV$, while it is even further suppressed {to} below $5-15\%$ of the SM Higgs width for masses $m_H > 300~\GeV$. At $m_H = 1000~\GeV$, the total width is still below $25~\GeV$. In comparison to SM Higgs boson of the same mass, the total width of these resonances is therefore highly suppressed, which promises to enhance the validity of a narrow width approximation in this mass range.\footnote{See e.g.~Ref.~\cite{Goria:2011wa} for the discussion of finite width effects for SM-like Higgs bosons in the mass range $m_{h}\gtrsim 200~\GeV$.}

For completeness, we show the allowed parameter space in the ($\Gamma_\text{tot}, \kappa$) and ($\Gamma_\text{tot}, \kappa'$) planes in Fig.~\ref{fig:gk_scales}. {If these predictions are taken as independent input parameters in future Higgs boson collider searches}, a direct comparison with {the experimental results} renders additional constraints {and --- in case of a discovery --- could possibly lead to an exclusion of the entire model.}

\subsection{Low mass region}
\label{Sect:lowmass}

We now consider the low mass region, i.e.~we set the heavy Higgs mass to $m_H=125.14~\GeV$ {and investigate the parameter space with} $m_h\in [1,~120]~\GeV$. In contrast to the high mass region, results from LEP searches play an important role in this part of parameter space. 
As discussed in Section~\ref{sec:rge}, we here do not apply limits
from RGE running {of the couplings}. Before constraints from the signal rates are taken into account, this a priori leads to much larger allowed values for $\tan\be$, where the upper limit on $\tan\be$ stems from
perturbative unitarity, cf.~Fig.~\ref{fig:tbmax}. However, whenever the additional decay
$H\rightarrow hh$ is kinematically allowed, {$\tan\be$ values $\gtrsim 1$ generally} 
result in large branching ratios for this channel, cf.~Fig.~\ref{sf:lowmass_fixedmassBRHtohh}.
This immediately imposes a quite strong suppression of the {SM decays of the heavy} Higgs {state}, leading to strong bounds on the
minimal $|\sin\al|$ value from {the signal rates}, cf.~Section~\ref{Sect:HS}. {However, we should keep in mind that in parameter regions where $\tan\be \approx -\cos\al / \sin\al$, the branching ratio for $H\to hh$ decreases significantly, thus restoring the signal strength of the heavy Higgs boson to $\sin^2\alpha$ times the SM Higgs signal strength.}
 In the mass range where the additional decay is not allowed {and up to values of $m_h \lesssim 100~\GeV$}, the strongest limits on the mixing angle {stem} from LEP Higgs searches {in the channel $e^+e^- \to Zh \to Z(b\bar{b})$~\cite{Schael:2006cr}. For larger Higgs masses, the Higgs signal rates yield stricter limits on $|\sin\al|$ than the exclusion limits from LEP and LHC, cf.~also Fig.~\ref{fig:massoverlap_chi2}.}  We have summarized our finding in the low mass scenario in Tab.~\ref{tab:lowscale}.

\begin{table}
\begin{tabular}{| c || c | c | c | c|}
\toprule
$m_h~[\GeV]$& $|\sin\al|_\text{min, {HB}}$ & $|\sin\al|_\text{min, {HS}}$ &$(\tan\be)_\text{max}$&$(\tan\be)_{\text{no}~H\to hh} $\\
\colrule
{$120$} & {0.410}  & {0.918}    & {8.4}&-- \\
$110$&${0.819}$&${0.932}$&${9.3}$&--\\
$100$&${0.852}$&${0.891}$&$10.1$&--\\
$90$&${0.901}$& -- &$11.2$&-- \\
$80$&${0.974}$&--&$12.6$&--\\
$70$&${0.985} $&--&$14.4$&--\\
$60$&${0.978}$&${0.996}$&$16.8$&{0.21}\\
$50$&${0.981}$&${0.998}$&$20.2$&{0.20}\\
$40$&${0.984}$&${0.998}$&$25.2$&{0.18}\\
{$30$} &{0.988}&{0.998}& {33.6}&{0.16} \\
{$20$} &{0.993}&{0.998}&{50.4}&{0.12} \\
{$10$} &{0.997}&{0.998}&{100.8}&{0.08} \\
\botrule
\end{tabular}
\caption{\label{tab:lowscale} Limits on $\sin\al$ and $\tan\be$ in the low mass scenario {for various light Higgs masses $m_h$}. The limits on $\sin\al$ have been determined at $\tan\be=1$. {The lower limit on $\sin\al$ stemming from exclusion limits from LEP or LHC Higgs searches evaluated with \HB\ is given in the second column. If the lower limit on $\sin\al$ obtained from the test against the Higgs signal rates using \HS\ results in stricter limits, we display them in the third column.} The upper limit on $\tan\be$ in the fourth column stems from perturbative unitarity for the complete decoupling case ($|\sin\al|\,=\,1$), cf.~Fig.~\ref{fig:tbmax}. In the fifth column we give {the $\tan\be$ value for which $\Gamma_{H\rightarrow hh}=0$ is obtained, given the maximal mixing angle {allowed by the Higgs exclusion limits (second column)}. {At this $\tan\be$ value, the $|\sin\alpha|$ limit obtained from the Higgs signal rates (third column) is abrogated.}}}
\end{table}

{We now turn to the discussion of the full scan.} In order to {highlight} the importance of LEP constraints in the {low mass region}, we employ the following color coding for the plots:
\begin{itemize}
\setlength{\itemsep}{0pt}
\item {\sl Light gray:} points which fail theoretical constraints.
\item {\sl Dark gray:} points which are excluded by LHC Higgs searches.
\item {\sl Blue:} points allowed by LHC Higgs searches, but excluded by $>95\%~\mathrm{C.L.}$ by LEP searches.
\item {\sl Dark green:} points consistent with LEP constraints within $2\sigma$.
\item {\sl Light green:} points consistent with LEP constraints within $1\sigma$.
\item {\sl Yellow:} points favored within $2\sigma$ in the global fit (\HS\ $\chi^2$ + LEP $\chi^2$).
\item {\sl red:} points favored within $1\sigma$ in the global fit (\HS\ $\chi^2$ + LEP $\chi^2$).
\end{itemize}

\begin{figure}[t]
\centering
\subfigure[\label{fig:lowmass_sinatanb} ($\sin\al$, $\tan\be$) plane.]{\includegraphics[width=0.46\textwidth]{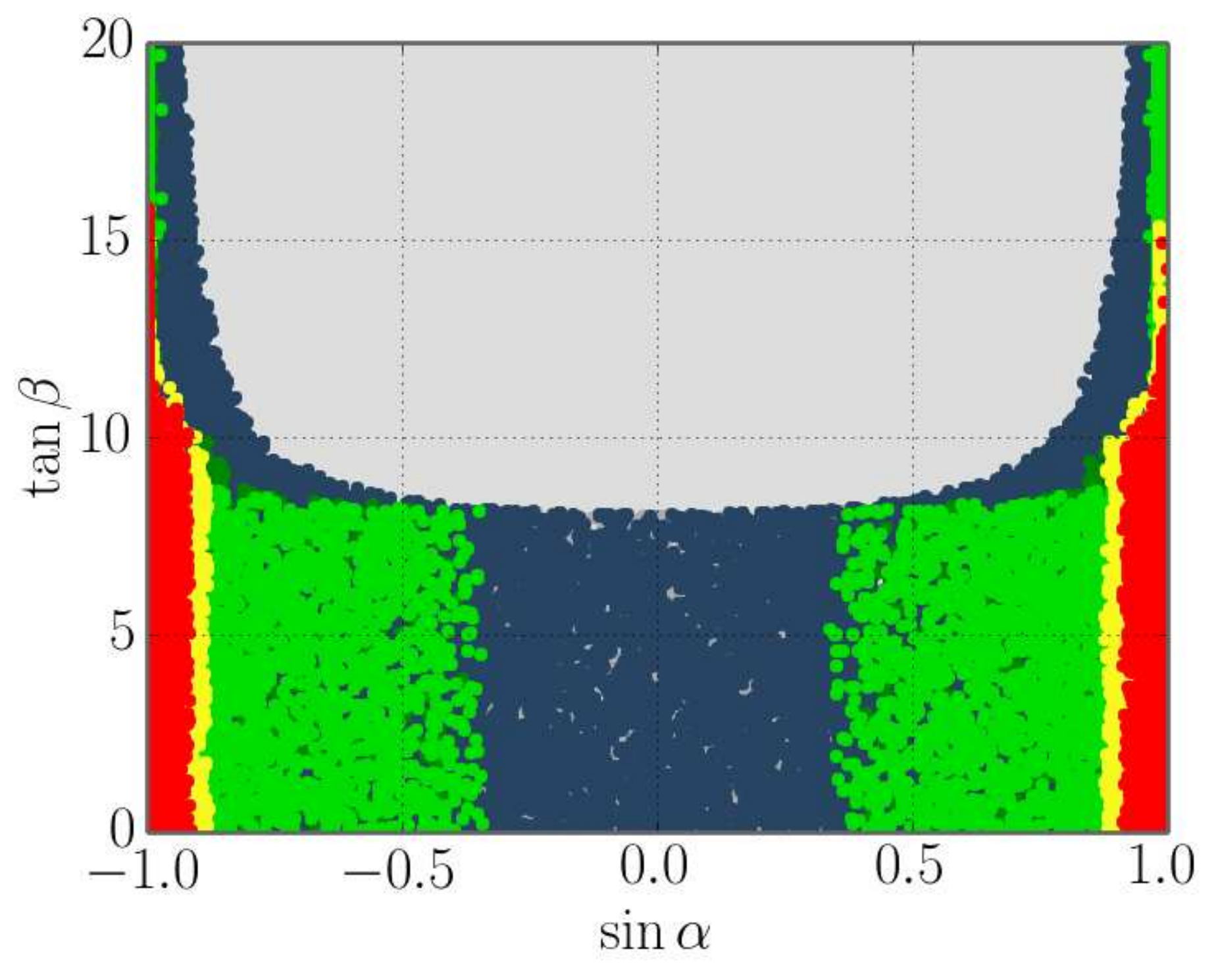}}
\hfill
\subfigure[\label{fig:lowmass_mh1tanb} ($m_h$, $\tan\be$) plane.]{ \includegraphics[width=0.46\textwidth]{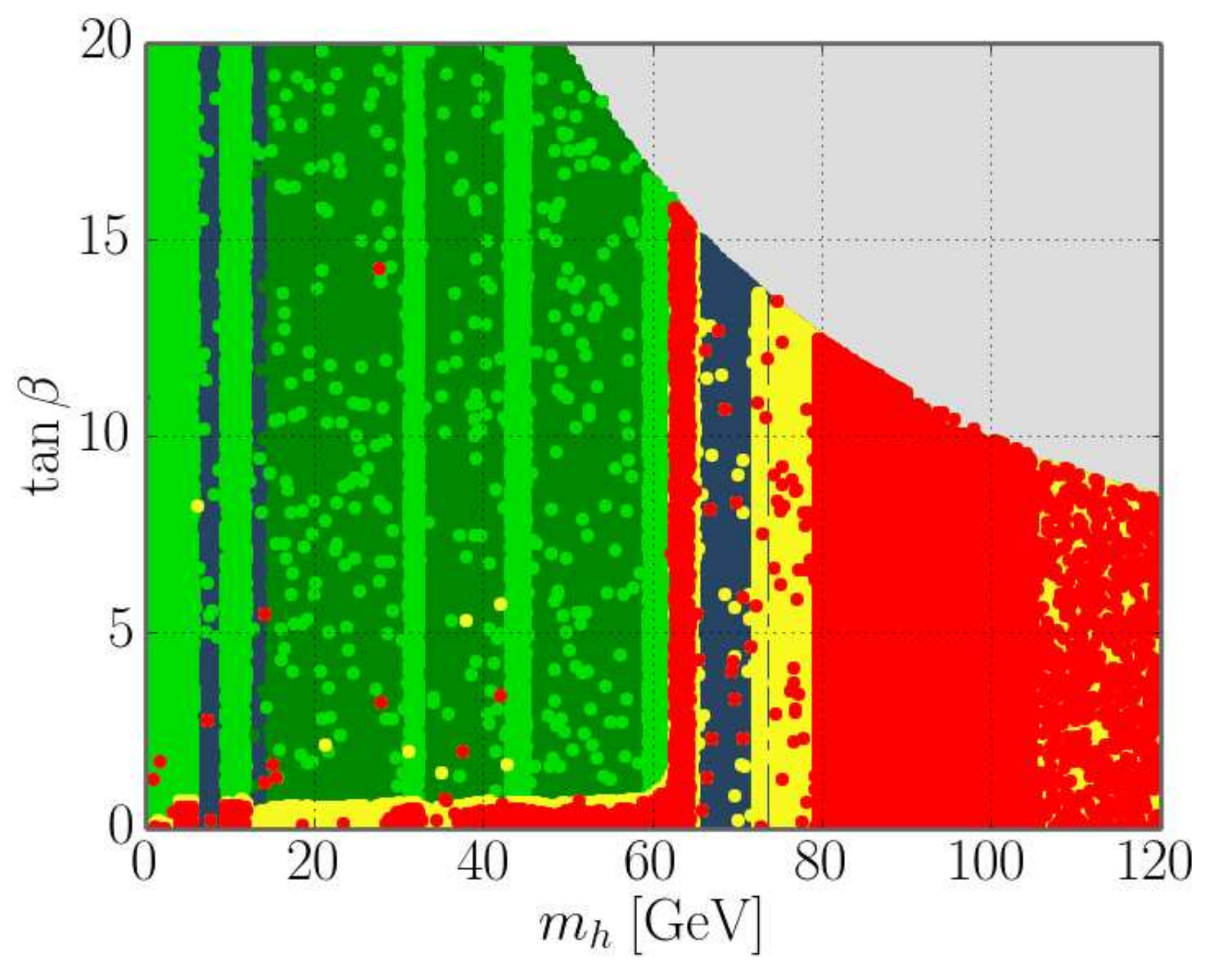}}
\subfigure[\label{fig:lowmass_mh1sina} ($m_h$, $\sin\al$) plane.]{ \includegraphics[width=0.46\textwidth]{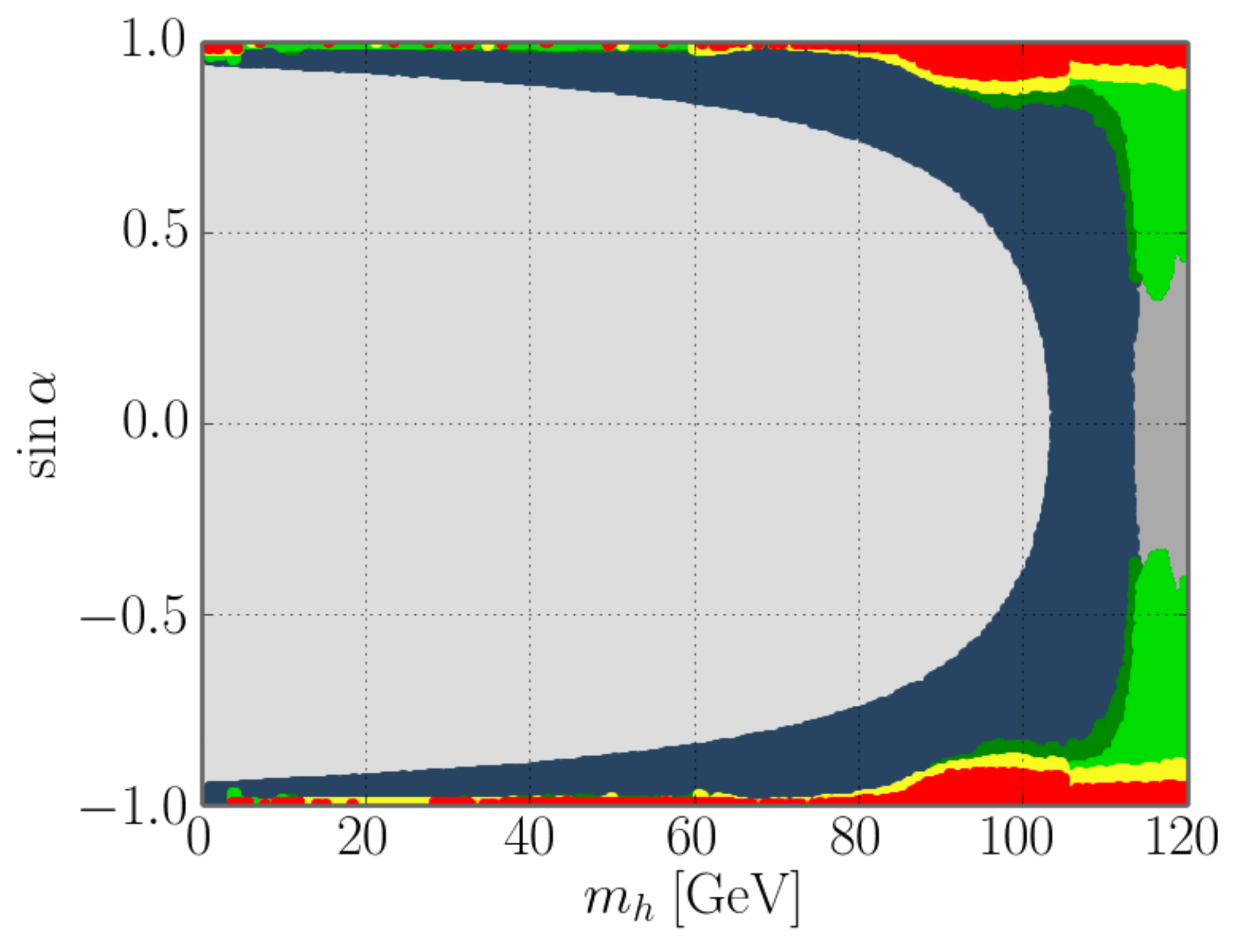}}
\caption{\label{fig:lowmh} Two-dimensional parameter correlations between $m_H$, $\tb$ and $\sa$ in the low mass region. See text for a description of the color coding.}
\end{figure}

{The results are shown in Fig.~\ref{fig:lowmh} in terms of two-dimensional scatter plots in the scan parameters.}  In Fig.~\ref{fig:lowmass_mh1sina}, we see that {most parameter points allowed by the global fit at the $2\sigma$ level are found for} $m_h\gtrsim 80\,\GeV$ {and $|\sin\al| \gtrsim 0.85$}. For {lower Higgs masses the mixing angle is constrained to values very close to the decoupling scenario ($|\sin\al |\approx1$). The LEP limits are particularly strong in the mass region between $65~\GeV$ and $72~\GeV$, cf.~Fig.~\ref{fig:sinaexp_a}, such that only a few valid points are found here, as can be seen best in Fig.~\ref{fig:lowmass_mh1tanb}.}
The semi-oval exclusion region in Figs.~\ref{fig:lowmass_sinatanb} and \ref{fig:lowmass_mh1sina} for large $\tan\be$ values and low {$|\sin\al|$} values, respectively, corresponds to a $2\sigma$ deviation {in} the electroweak {oblique} parameters $S$, $T$ and $U$.

In Fig.~\ref{fig:lowmass_mh1tanb}, we observe a drastic change in the distribution of allowed parameter points when going to Higgs masses $m_h < m_H/2 \approx 62~\GeV$, where the decay mode $H\to hh$ becomes kinematically accessible. As discussed earlier in Section~\ref{Sect:HS}, cf.~Fig.~\ref{fig:lowmass_fixedmassBRHtohh}, the decay $H\to hh$ easily becomes the dominant decay mode if $\tan\be \gtrsim 1$, unless the mixing angle is very close to $|\sin\al| = 1$. Hence, for $m_h < m_H/2$, most of the allowed points are found for small values of $\tan\be$, since the Higgs signal rates favor small values of $\mathrm{BR}(H\to hh)$. At larger Higgs masses, $m_h > m_H /2$, the favored points are equally distributed over the entire $\tan\be$ range allowed by perturbative unitarity.
\begin{figure}
\centering
\subfigure[\label{Fig:lowmass_kappaprime} Signal rate for the $H\to hh$ signature, normalized to the SM Higgs production cross section.]{\includegraphics[width=0.46\textwidth]{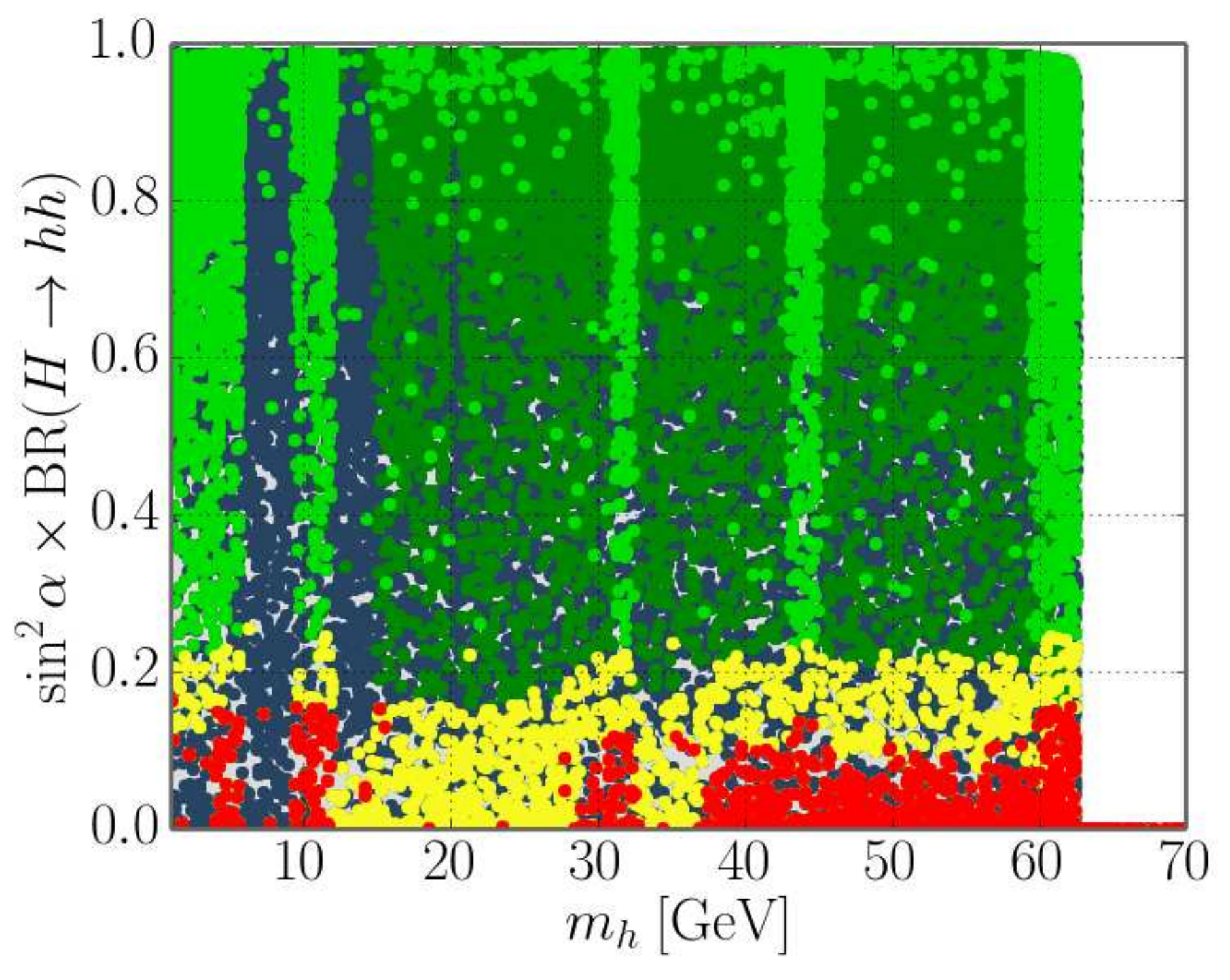}
}
\hfill
\subfigure[\label{Fig:lowmass_4mu} Signal rate for the $H \to hh \to \mu^+\mu^-\mu^+\mu^-$ signature at very low masses, normalized to the SM Higgs production cross section. The magenta line indicates the observed limit from a CMS $8~\TeV$ analysis~\cite{CMS:2013lea}.]{\includegraphics[width=0.477\textwidth]{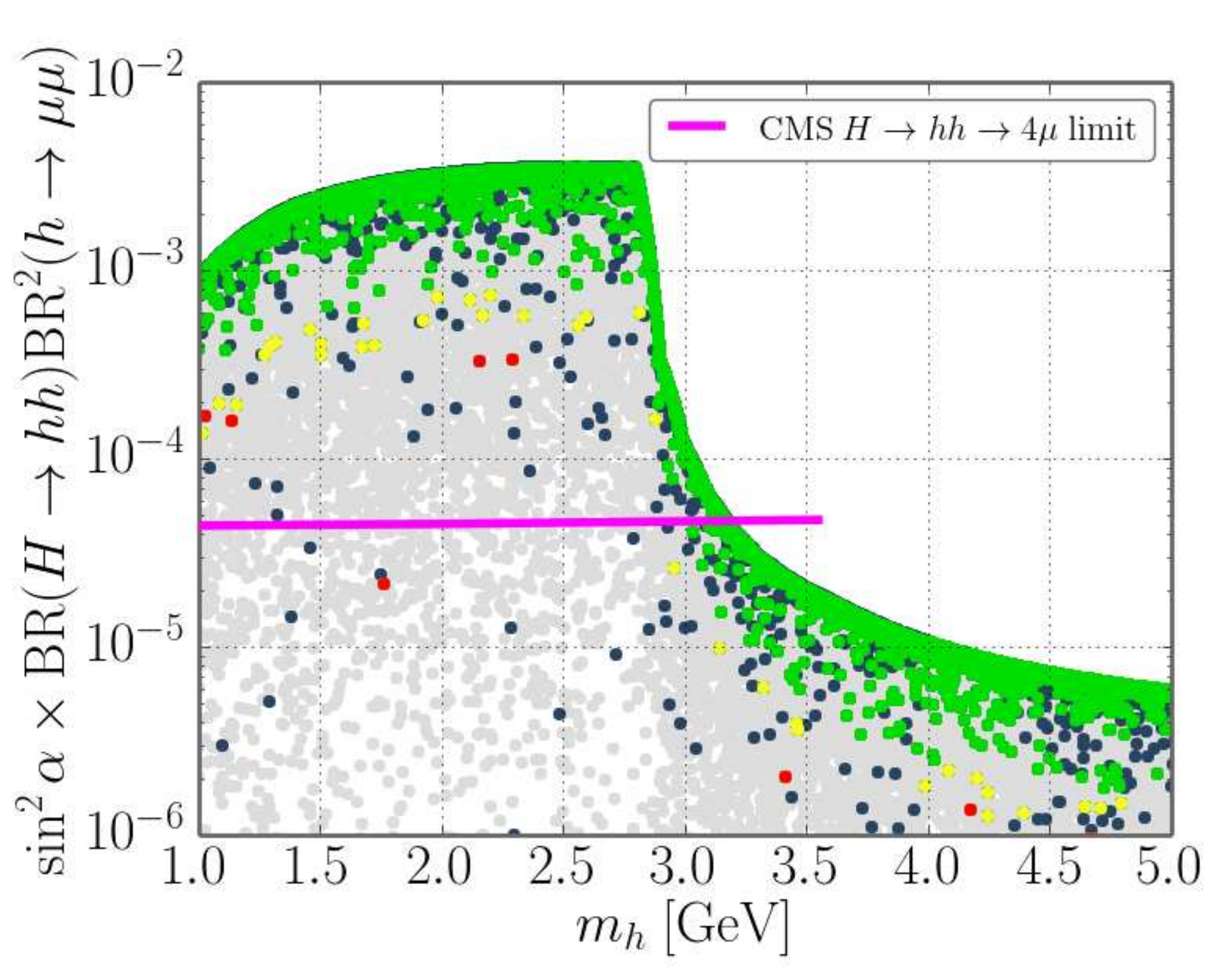}
}
\caption{\label{fig:lowlowmh} Signal rates for the $H\to hh$ signature in dependence of the light Higgs mass: (a) inclusive signal rate for Higgs production and successive decay $H\to hh$, normalized to the SM Higgs production rate; (b)~exclusive signal rate for the $\mu^+\,\mu^-\,\mu^+\,\mu^-$ final state for very low Higgs masses $m_h$.}
\end{figure}

It is interesting to investigate the allowed range of the $H\to hh$ signal rate in dependence of the light Higgs mass. This is shown in Fig.~\ref{Fig:lowmass_kappaprime}, where the signal rate is normalized to the SM Higgs boson production. Note, that due to the LEP constraints, the favored points feature a mixing angle $|\sin\al| \approx 1$ and thus the displayed signal rate closely resembles $\mathrm{BR}(H\to hh)$. We see that the maximally allowed $H\to hh$ signal rate is about $22\%$ and is roughly independent on the light Higgs mass\footnote{The reason why the density of allowed points still depends strongly on $m_h$ is that regions which are strongly constrained by LEP searches require a large fine-tuning of $\sin\al$ to render allowed points.
}. {This upper limit solely stems from the observed signal rates of the SM--like Higgs boson at $\sim 125~\GeV$. These constraints therefore also limit the total width of the heavy Higgs at $125.14\,\GeV$ to values $\sim\,3-5\,\MeV$, being in the vicinity of the SM total width of $\sim 4.1~\MeV$.}

We now discuss the case of very low Higgs masses, $m_h\lesssim 4~\GeV$. Here, the LEP constraints stem from the decay-mode independent analysis of $e^+e^- \to Zh$ by OPAL~\cite{Abbiendi:2002qp}, yielding a slightly weaker limit on the mixing angle, $|\sin\alpha| \gtrsim 0.965$, than at larger masses, cf.~Fig.~\ref{fig:sinaexp_a}. In the mass region $m_h \in [1, 3]~\GeV$, the branching fraction for the {light Higgs} decay $h\to \mu\mu$ amounts between $3-6\%$, thus allowing to search for the signature $(pp)\to H\to hh \to \mu^+\mu^- \mu^+\mu^-$ at the LHC. We show the predicted signal rate for this signature\footnote{We consider here only the Higgs production via gluon gluon fusion.} for the LHC at a center-of-mass energy of $8~\TeV$ in Fig.~\ref{Fig:lowmass_4mu}. A search for this signature has been performed by CMS~\cite{CMS:2013lea}, yielding the observed upper limit\footnote{This exclusion limit is not provided with \HBv{4.2.0}, because the expected limit from the CMS analysis is not publicly available.} displayed as magenta line in the figure. As can be seen, the CMS limit provides competitive constraints in this parameter region, excluding a sizable amount of the parameter region favored by the global fit. Future  LHC searches for the $4\mu$ signature therefore have a good discovery potential in this mass region. Other final states, composed of $\tau$ leptons, strange or charm quarks, could be exploited at a future linear $e^+e^-$ collider like the ILC.

A very light Higgs boson $h$ with mass values up to the $b\bar{b}$ threshold can also be probed at $B$-factories in the radiative decay $\Upsilon \to h \gamma$~\cite{Wilczek:1977zn}, with successive decay of the light Higgs boson to $\tau$-lepton, muon or hadron pairs. {Here, we provide a rough estimate of the present constraints. 

The decay rate for the $1^{--}$ bound state $\Upsilon(1s)$ to the Higgs-photon final state (normalized to the decay rate of $\Upsilon(1s) \to \mu^+\mu^-$) is given by~\cite{Wilczek:1977zn}
\begin{align}
\frac{\mathrm{BR}(\Upsilon(1s) \to h \gamma)}{\mathrm{BR}(\Upsilon(1s) \to \mu^+\mu^-)} = \frac{g_b^2 G_F m_b^2}{\sqrt{2}\pi\alpha}\left(1-\frac{m_h^2}{M^2_{\Upsilon(1s)}}\right) \times \mathcal{F},
\end{align}
where $G_F$ is the Fermi constant, $m_b$ the bottom quark mass, $\alpha$ the fine-structure constant and $\mathrm{BR}(\Upsilon(1s) \to \mu^+\mu^-)\approx 2.48\pm0.05\%$~\cite{Agashe:2014kda}. The factor $\mathcal{F}$ represents higher-order corrections. The one-loop QCD corrections have been calculated in Ref.~\cite{Vysotsky:1980cz,Nason:1986tr} and are known to reduce the leading-order estimate by up to $84\%$, see Ref.~\cite{Gunion:1989we} for an extended discussion. In our model, the rescaling factor of the bottom Yukawa coupling of the light Higgs is simply given by $g_b=\cos\alpha$.

Recent experimental searches have been carried out by BaBar~\cite{Aubert:2009cka,Lees:2012te,Aubert:2009cp,Lees:2012iw,Lees:2013vuj} and CLEO~\cite{Love:2008aa}, focussing on the search for a light $\mathcal{CP}$-odd Higgs boson motivated by certain next-to-minimal supersymmetric standard model (NMSSM) scenarios~\cite{Dermisek:2005ar,Dermisek:2006py,Domingo:2008rr}. The $90\%~\mathrm{C.L.}$ upper limits on the branching fraction of these search signatures are typically of $\sim\mathcal{O}(10^{-4}-10^{-6})$ and are listed for representative values of the light Higgs mass in Tab.~\ref{Tab:Bfactories} (cf.~also Refs.~\cite{Echenard:2012hq,Domingo:2010am,Bevan:2014iga} for more details). Generally, these limits underlie large statistical fluctuations, thus we prefer to use a roughly estimated mean value and indicate this by a '$\sim$` in front of the quoted number. Using the SM Higgs boson branching ratios for $h\to \mu^+\mu^-$, $h\to \tau^+\tau^-$, $h\to gg$ and $h \to s\bar{s}$ in this mass region, we can infer a $90\%~\mathrm{C.L}$ upper limit on the rescaling factor of the bottom Yukawa coupling, $g_b$, {which is} listed in Tab.~\ref{Tab:Bfactories}. If this upper limit is below $1$, we furthermore quote the resulting lower limit on the mixing angle $|\sin\alpha|$ in the table. The resulting limits cannot compete with those obtained from direct LEP searches, however, future $B$-physics facilities such as the Belle II experiment at the Super KEKB accelerator~\cite{Abe:2010gxa} will be able to probe the yet unexcluded region.}
\begin{table}
\centering
\begin{tabular}{| c || cccc | c | c |}
\toprule
$m_h~[\GeV]$  &  \multicolumn{4}{c|}{$90\%~\mathrm{C.L.}$ upper limit on $\mathrm{BR}(\Upsilon(1s)\to h\gamma, h\to \dots)$,} & $g_b^2$  & $|\sin\alpha|$ \\
			& $h \to \mu^+\mu^-$~\cite{Lees:2012iw} & $h\to \tau^+\tau^-$~\cite{Lees:2012te} & $h\to gg$~\cite{Lees:2013vuj} & $ h\to s\bar{s}$~\cite{Lees:2013vuj} & (upper limit) & (lower limit) \\
\colrule
$1.0$			&	$\sim4\cdot 10^{-6}$	& --                            & $\sim\mathbf{5 \cdot 10^{-6}}$	&  --	 & $\sim 0.25$ & $\sim0.87$ \\
$2.0$			&	$\sim\mathbf{5\cdot 10^{-6}}$ 	&  --                           & $\sim1 \cdot 10^{-4}$	& $\sim5 \cdot 10^{-5}$ & $\sim 1.16$& --	 \\
$3.0$			&	$\sim\mathbf{6\cdot 10^{-6}}$	&  --                           & $\sim2 \cdot 10^{-4}$	& $\sim8 \cdot 10^{-5}$  & $\sim 7.82$ & --	 \\
$4.0$			&	$\sim8\cdot 10^{-6}$	& $\mathbf{1.2 \cdot 10^{-5}}$ & $\sim4 \cdot 10^{-4}$ 	& $\sim3 \cdot 10^{-4}$ & $2.06$ & --	 \\
$5.0$			&	$\sim8\cdot 10^{-6}$	& $\mathbf{9.1 \cdot 10^{-6}}$ &  $\sim3 \cdot 10^{-4}$	& $\sim3 \cdot 10^{-4}$ & $0.68$ & $0.57$	 \\
$6.0$			&	$\sim1\cdot 10^{-5}$	& $\mathbf{2.3 \cdot 10^{-5}}$ &  $\sim5 \cdot 10^{-5}$	& $\sim8 \cdot 10^{-5}$ & $1.59$  & --	 \\
$7.0$			&	$\sim1\cdot 10^{-5}$	& $\mathbf{1.6 \cdot 10^{-5}}$ &  $\sim3 \cdot 10^{-4}$	& $\sim1 \cdot 10^{-4}$ & $1.33$ & -- 	 \\
$8.0$			&	$\sim2\cdot 10^{-5}$	& $\mathbf{3.2 \cdot 10^{-5}}$ &  $\sim1 \cdot 10^{-2}$	& $\sim4 \cdot 10^{-5}$ & $4.45$  & -- 	 \\
\botrule			
\end{tabular}
\caption{Constraints from $B$-factories on a light Higgs boson with mass $m_h$. The second to fifth column list the current experimental $90\%~\mathrm{C.L.}$ upper bounds on the decay rate of $\Upsilon(1s) \to h \gamma$ and successive Higgs decay (specified in the second title row). The inferred upper limit on the rescaling factor of the bottom Yukawa coupling in given in the sixth column, and --- if possible --- the lower limit on the singlet-doublet mixing angle $|\sin\alpha|$ is given in the last column. We indicate the most relevant constraint for the model (yielding the listed limits on the model parameters) by bold numbers.}
\label{Tab:Bfactories}
\end{table}

Finally, we want to comment that {despite of} the quite strong constraints in the low mass region, a substantial number of low mass Higgs {bosons} could already have been directly produced {at the LHC}. Table \ref{tab:lowlhc} exemplarily lists the maximal{ly allowed} {LHC} cross sections for direct production in gluon gluon fusion for a selected range of light Higgs masses {at CM energies of $8$ and $14~\TeV$.}\footnote{We thank M. Grazzini for providing us with the production cross sections for $m_h <80\,\GeV$.} We {encourage} the LHC experiments to {explore the feasibility of experimental searches within the low mass region and to potentially} extend the searches for directly produced scalars into this mass range.

\begin{table}
\begin{tabular}{| c || c | c |}
\toprule
$m_h~[\GeV]$& $\sigma_{gg}^{8\,\TeV}[\pb]$& $\sigma_{gg}^{14\,\TeV}[\pb]$ \\
\colrule
{$120$} &3.28 &8.40\\
$110$&3.24&8.12\\
$100$&6.12&14.96\\
$90$&6.82&16.26\\
$80$&2.33&5.41\\
$70$&2.97&6.73\\
$60$&0.63&1.38\\
$50$&0.45&0.96\\
$40$&0.74&1.50\\
\botrule
\end{tabular}
\caption{\label{tab:lowlhc} Maximal{ly allowed} cross sections, $\sigma_{gg}=\lb \cos^2\al \rb_\text{max}\times\sigma_{gg,\text{SM}}$, for direct light Higgs production {at the LHC at CM energies of $8$ and $14~\TeV$} after all current constraints have been taken into account. {The SM Higgs} production cross sections have been taken from Ref.~\cite{Heinemeyer:2013tqa,grazzini}.}
\end{table}


\subsection{Intermediate mass region}\label{sec:inter}

{For the intermediate mass region, which contains the special case of
  mass-degenerate Higgs states, we treat both Higgs masses as free parameters in
  the fit, $m_h, m_H \in [120,130]\,\GeV$. {Note, that the following discussion is based on a few simplifying assumptions about overlapping Higgs signals in the experimental analyses. It should be clear that a precise investigation of the near mass-degenerate Higgs scenario can only be performed by analyzing the LHC data directly and is thus restricted to be done by the experimental collaborations (see e.g.~Ref.~\cite{Khachatryan:2014ira} for such an analysis). Nevertheless, we want to point out this interesting possibility here and encourage the LHC experiments for further investigations.}

If the Higgs states have very similar masses, their signals cannot be clearly distinguished in the experimental analyses and (to first approximation) the sum of the signal rates has to be considered for the comparison with the measured rates. Moreover, the observed peak in the invariant mass distribution in the $H\to\gamma\gamma$ and $H\to ZZ^*\to 4\ell$ channels, which is fitted to determine the Higgs mass, would actually comprise two (partially) overlapping Higgs resonances, where the height of each resonance is governed by the corresponding  signal strength. Therefore, for each Higgs analysis where a mass measurement has been performed, cf.~Tab.~\ref{Tab:HSobservables}, we calculate a signal strength weighted mean value of the Higgs masses\footnote{Testing overlapping signals of multiple Higgs bosons against mass measurements by employing a mass average calculation is the default procedure in \HS\ since version \texttt{1.3.0}.},}
\begin{align}
\overline{m} = \frac{ \mu_{h} \cdot m_h +\mu_H \cdot m_H}{\mu_h + \mu_H},
\end{align}
to be tested against the measurement, where the SM normalized signal strengths are given by
\begin{align}
\mu_{h/H} = \frac{\sum_a \epsilon^a \, \sigma_a (m_{h/H}) \times \mathrm{BR}_a (m_{h/H})}{\sum_a \epsilon^a \, \sigma_\text{SM}^a (\hat{m}) \times \mathrm{BR}^a_\text{SM} (\hat{m})}.
\label{Eq:signalstrength}
\end{align}
Here, $\hat{m}$ denotes the mass value hypothesized by the experiment during to signal rate measurement.
The index $a$ runs over all signal channels, i.e.~Higgs production times decay mode, considered in the experimental analysis, and $\epsilon^a$ denotes the corresponding efficiencies. The predicted cross sections $\sigma$ and partial widths are obtained from rescaling the respective SM quantities~\cite{Dittmaier:2011ti,Dittmaier:2012vm,Heinemeyer:2013tqa} by $\csqa$ and $\ssqa$ for $h$ and $H$, respectively. As mentioned earlier in Section~\ref{Sect:HS}, the SM normalized signal strengths $\mu_{h/H}$ contain a slight mass dependence\footnote{This mass dependence is neglected per default in \HS\ since additional complications arise if theoretical mass uncertainties are present. This is however not the case here, since we use the Higgs masses directly as input parameters. The evaluation of the signal strength according to Eq.~\eqref{Eq:signalstrength} can be activated in \HS\ by setting \texttt{normalize\_rates\_to\_reference\_position=.True.}~in the file \texttt{usefulbits\_HS.f90}.} since the SM cross sections and branching ratios are not constant over the relevant mass range.

\begin{table}
\begin{tabular}{| c || c | c | c |}
\toprule
$m~[\GeV]$& $|\sin\al|_\text{HB}$ & $|\sin\al|_\text{HS}$ &$(\tan\be)_\text{max}$\\
\colrule
$130$& {$< 0.806$}  & {$< 0.370$}    & {$7.76$}  \\
$129$& {$< 0.881$}  & {$< 0.373$}    & {$7.81$}  \\
$128$& {$< 0.988$}  & {$< 0.377$}    & {$7.88$} \\
$127$& {--}  & $< 0.381$   &  {$7.94$}\\
$126$&  {--}  & $< 0.552$   &  {$8.00$}\\
$125$&  {--}  & --  &   {$8.07$}\\
$124$&  {--}  &$> 0.793$   &  {$8.13$}\\
$123$ & {--}  & $> 0.864$    &   {$8.20$}\\
{$122$} & {--}  & $> 0.904 $    &  {$8.26$}\\
{$121$} & {--}  & $> 0.913 $    &   {$8.34$}\\
{$120$} & {$> 0.410$}  & {$> 0.918$}    &   {$8.41$}\\
\botrule
\end{tabular}
\caption{\label{tab:intermediatescale} Limits on $\sin\al$ and $\tan\be$ in the intermediate mass scenario. We fix one Higgs mass at $125.14~\GeV$ and vary the mass of the other Higgs state, $m$. {The limit on $\sin\al$ that stems from LHC Higgs searches evaluated with \HB\ is given in the second column (if available). The limit on $\sin\al$ obtained from the test against the Higgs signal rates with \HS\  is given in the third column. Note, that depending on the mass hierarchy, we have either an upper or lower limit on $\sin\al$, indicated by the ``$<$'' and ``$>$'', respectively.} The upper limit on $\tan\be$ is given in the fourth column and always stems from perturbative unitarity, {see also~Fig.~\ref{fig:tbmax}. Note, that we do \emph{not} impose constraints from perturbativity and vacuum stability at a high energy scale via RGE evolution of the couplings here.}}
\end{table}

We present limits on $\sin\al$ and $\tan\be$ for various choices of the Higgs mass $m$ in the intermediate mass region in Tab.~\ref{tab:intermediatescale}. We fixed the other Higgs mass to $125.14~\GeV$. Depending on whether or not the Higgs mass $m$ is larger than $125.14~\GeV$, we obtain either an upper or lower limit on $\sin\al$ from the LHC Higgs search exclusion limits or signal rate measurements, which are listed separately. In the case of nearly degenerate Higgs masses, $m = 125~\GeV$, no limit on $\sin\al$ can be obtained, since the Higgs signals completely overlap. We find that no limits from $95\%~\mathrm{C.L.}$ exclusions from Higgs searches can be obtained for Higgs masses within $121~\GeV$ and $127~\GeV$. Moreover, the limits inferred from the signal rates become weaker the closer $m$ is to $125.14~\GeV$ due to the signal overlap. In the full intermediate mass region, the limits inferred from the Higgs signal rates supersede the limits obtained from null results in LHC Higgs searches.

The upper limits on $\tan\be$ listed in Tab.~\ref{tab:intermediatescale} correspond to the perturbative unitarity bound (cf.~Fig.~\ref{fig:tbmax}). Similarly as in the low mass region, we do not impose constraints from perturbativity and vacuum stability at a high energy scale here. If these were additionally required, $\tan\be$ would be limited to values $\lesssim 1.86$ for $m \ge 125.14~\GeV$. For lower Higgs masses $m$ no valid points would be found. It should be noted, however, that the collider phenomenology does not depend on $\tan\be$ in the intermediate mass region, since Higgs-to-Higgs decays are kinematically not accessible.

\begin{figure}
\centering
\subfigure[\;($m_h$, $m_H$) plane.\label{fig:mhmH}]{\includegraphics[width=0.46\textwidth]{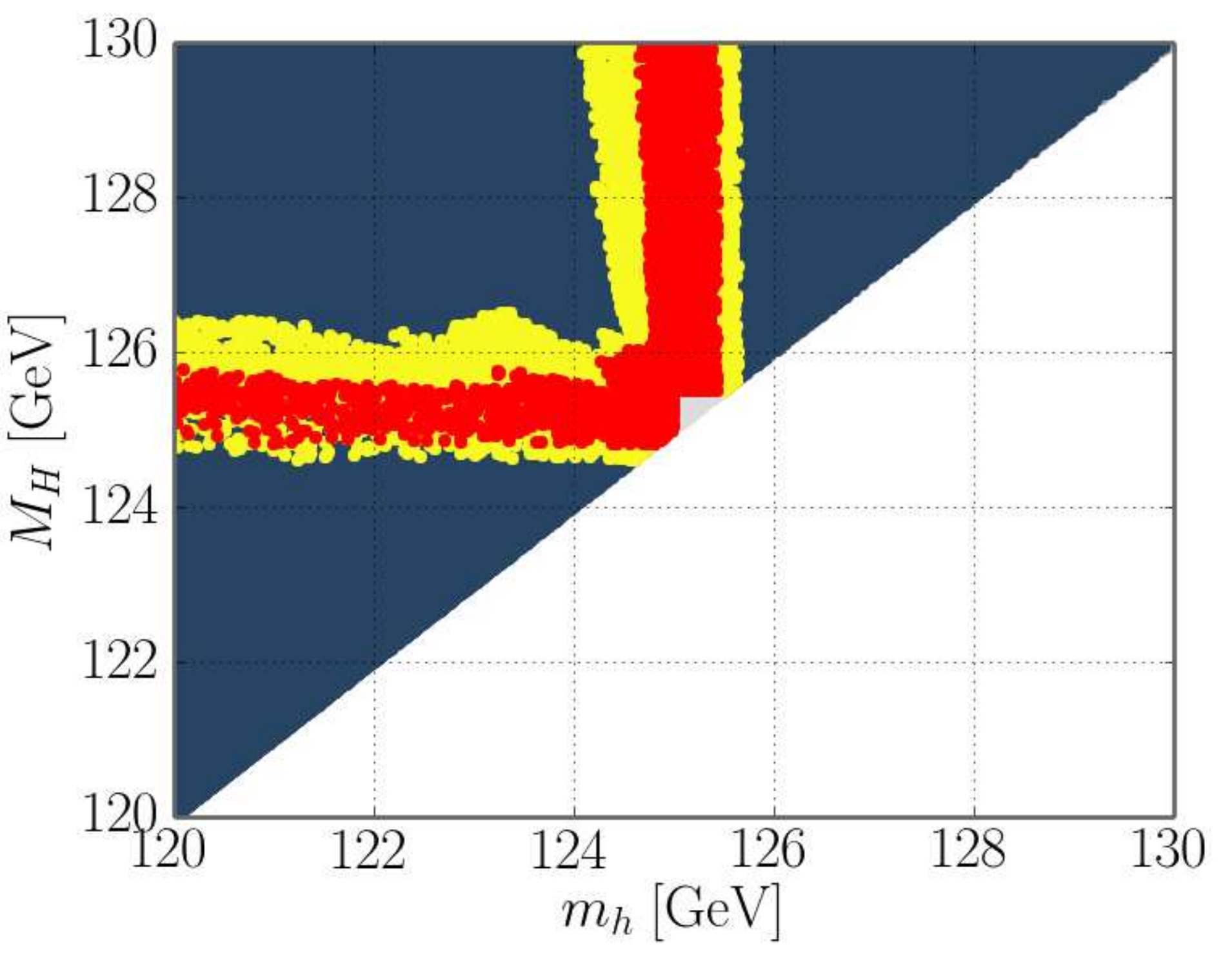}}
\hfill
\subfigure[\;($\sin\al$, $\tan\beta$) plane.\label{fig:sinatb}]{\includegraphics[width=0.46\textwidth]{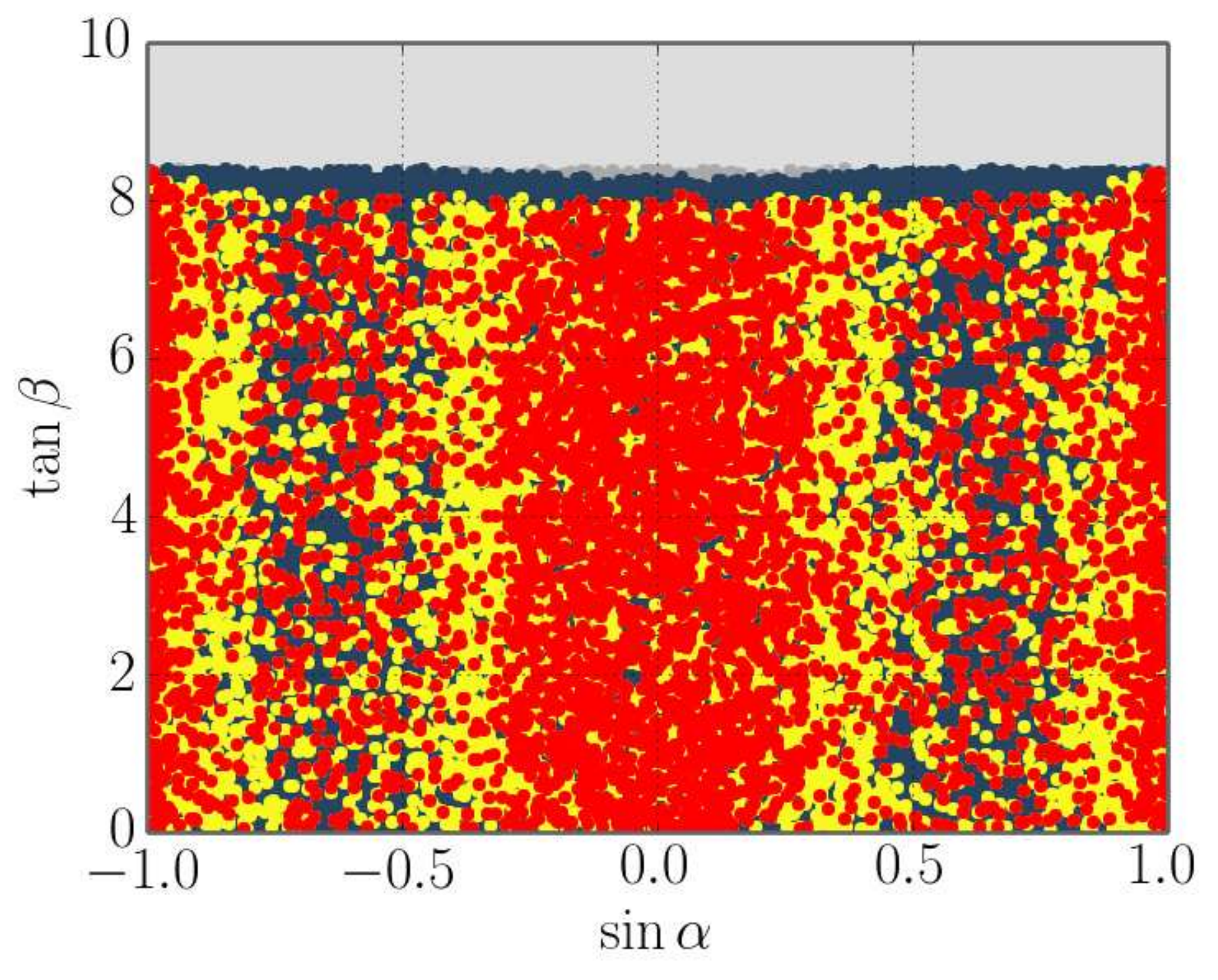}}
\subfigure[\;($m_h$, $\sin\al$) plane.\label{fig:mhsina}]{\includegraphics[width=0.46\textwidth]{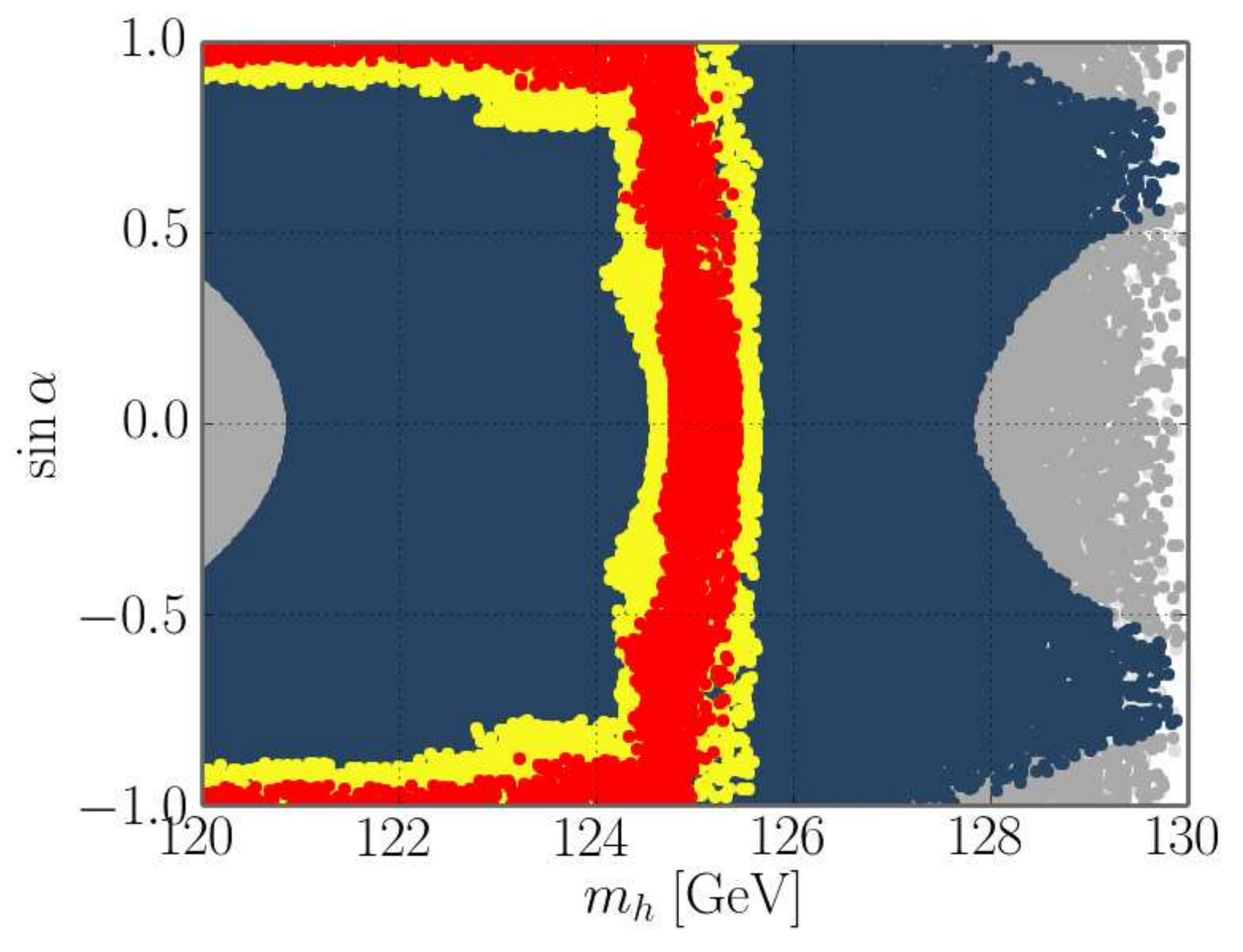}}
\hfill
\subfigure[\;($m_H$, $\sin\al$) plane.\label{fig:mHsina}]{\includegraphics[width=0.46\textwidth]{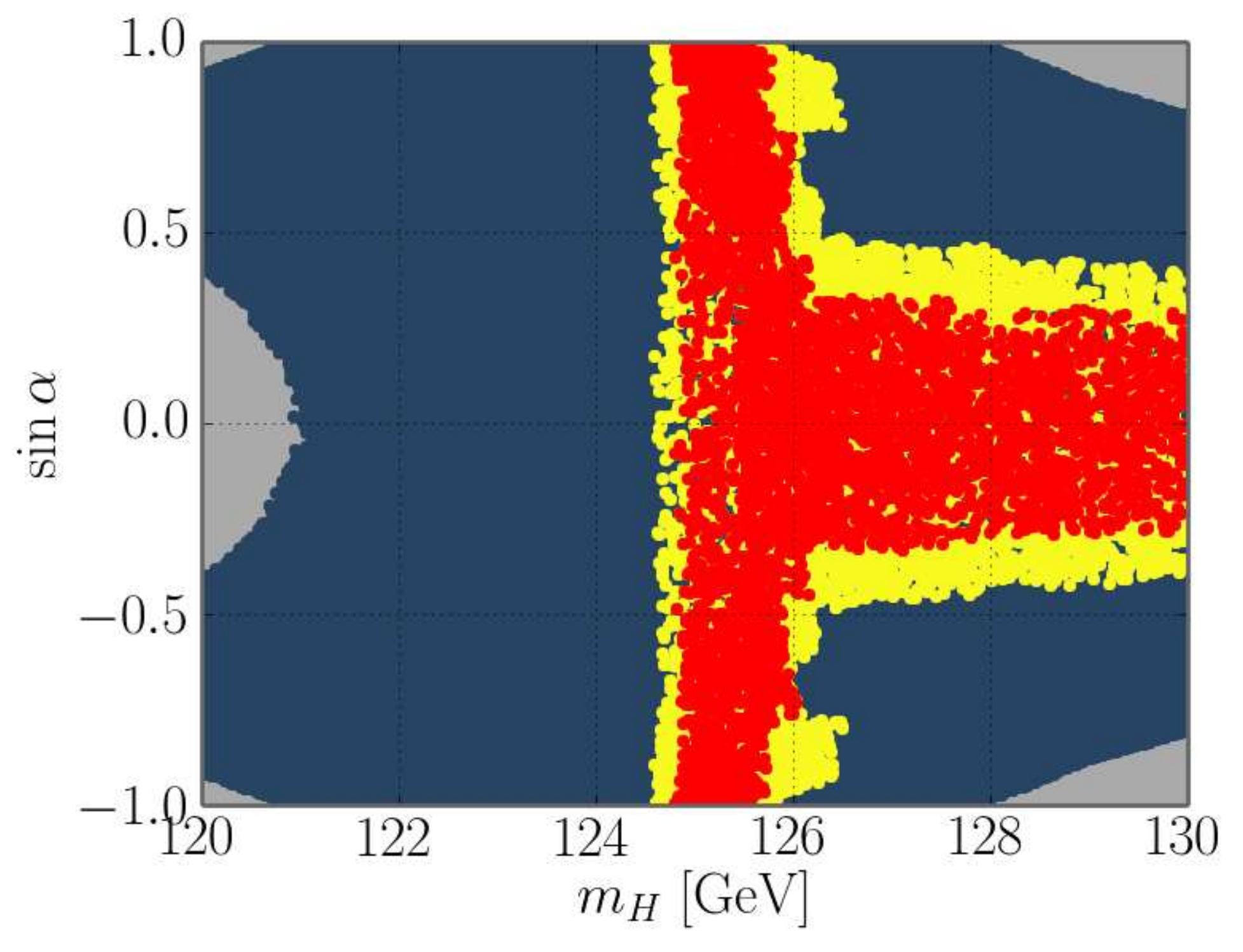}}
\subfigure[\;($m_h$, $\tan\beta$) plane.\label{fig:mhtb}]{\includegraphics[width=0.46\textwidth]{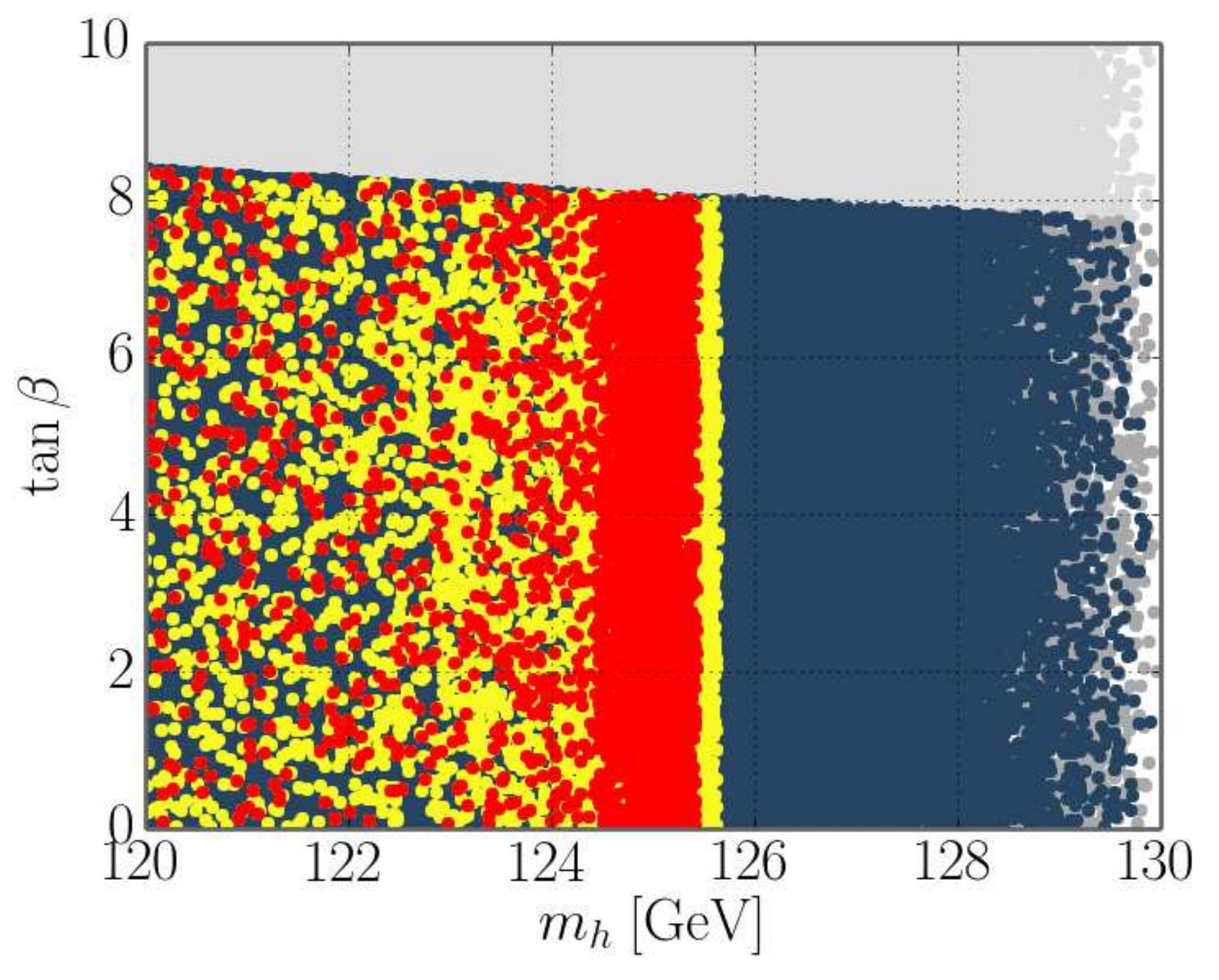}}
\hfill
\subfigure[\;($m_H$, $\tan\beta$) plane.\label{fig:mHtb}]{\includegraphics[width=0.46\textwidth]{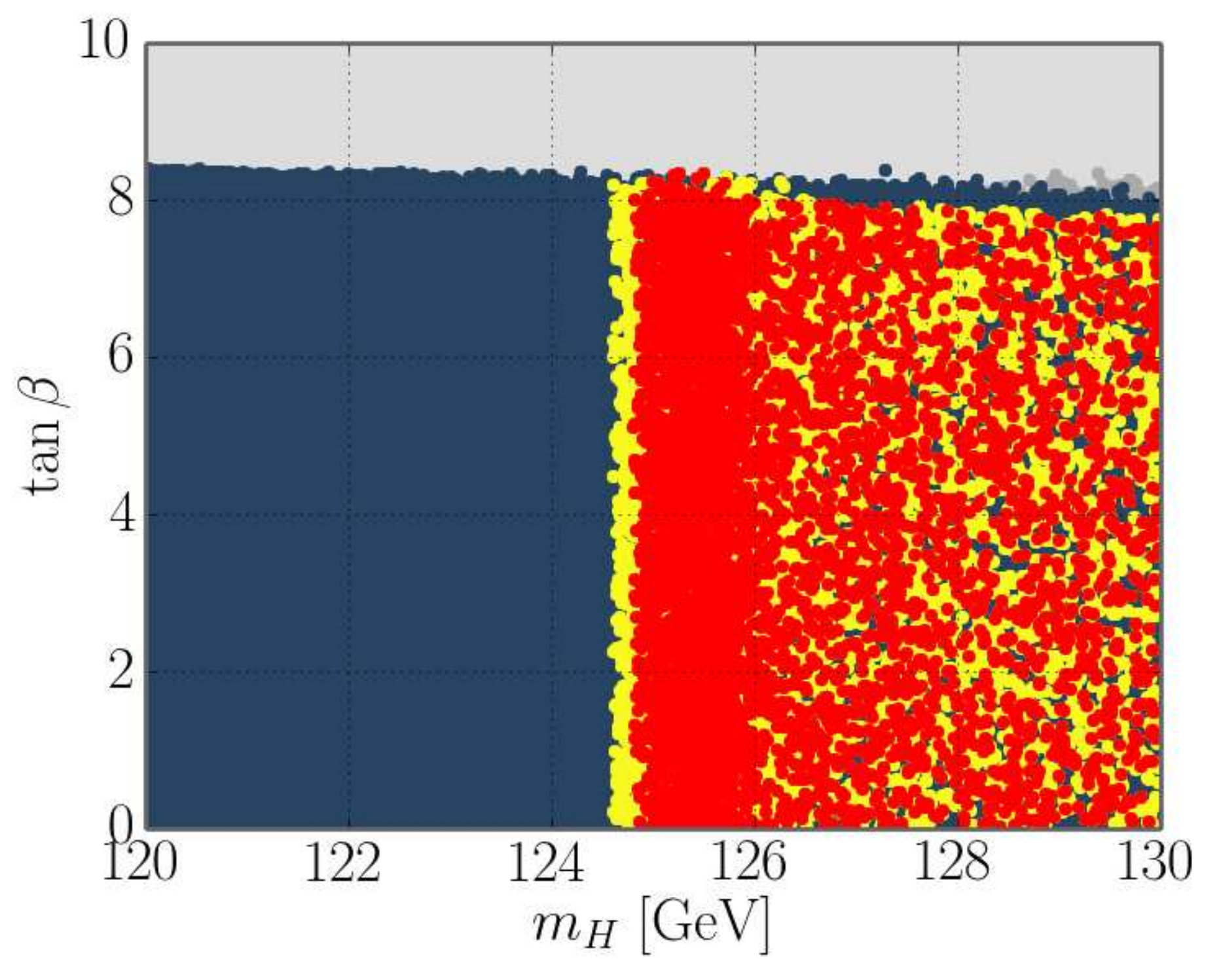}}
\caption{Two-dimensional correlations between $m_h$, $m_H$, $\sin\al$ and $\tan\beta$ in the intermediate mass region.
}
\label{Fig:IntermediateMassRegion_2D}
\end{figure}

The results from the full four-dimensional scan are presented in Fig.~\ref{Fig:IntermediateMassRegion_2D} in terms of two-dimensional scatter plots, using the same color coding as in {the high mass region (see~e.g.~Fig.~\ref{fig:mHres})}. The correlation between the two Higgs masses, Fig.~\ref{fig:mhmH}, shows that allowed parameter points with Higgs bosons in the full intermediate mass region are found, however, at least one of the Higgs masses is always required to be roughly between $124~\GeV$ and $126.5~\GeV$. We can furthermore learn from Figs.~\ref{fig:mhsina} and~\ref{fig:mHsina}, that allowed points with one Higgs mass being 
below $124~\GeV$ (above $126.5~\GeV$) feature $|\sin\alpha|$ values close to $1~(0)$, such that the other Higgs state at around $125~\GeV$ has SM Higgs--like signal strengths. In the near-degenerate case, $m_h \approx m_H \approx  125~\GeV$, all mixing angles $\sin\al$ are allowed and the model appears indistinguishable from the SM at current collider experiments.

Fig.~\ref{Fig:IntermediateMassRegion_2D} also shows the correlations of $\tan\beta$ with the mixing angle $\sin\al$, Fig.~\ref{fig:sinatb}, and the Higgs masses, Figs.~\ref{fig:mhtb} and~\ref{fig:mHtb}. As stated earlier, $\tan\beta$ does not influence the collider phenomenology in the intermediate mass range, thus we find allowed parameter points in the full $\tan\beta$ range up to the maximal value given by perturbative unitarity.

A direct search for the second Higgs boson in the intermediate mass region at the LHC seems challenging. Even if the mass splitting between the two Higgs states is large enough to be resolved by the experimental analyses, we expect the second resonance to be much smaller than the established signal. Nevertheless we would like to encourage the LHC experiments to perform dedicated resonance searches, in particular in the mass region slightly above the current signal, $m_H \sim (125.5-126.5)~\GeV$, since in this case {larger} values of the mixing angle are still allowed while an improvement of the vacuum stability at the high--scale may be obtained.
More promising prospects to resolve the near mass--degenerate Higgs scenario have future experimental facilities like the ILC~\cite{Asner:2013psa,Dawson:2013bba} or a muon collider~\cite{Alexahin:2013ojp,Dawson:2013bba}, where the latter provides excellent opportunities to measure the mass and the total width of the discovered Higgs boson via a line-shape scan.

\section{Conclusions}
\label{Sec:Conclusions}
\noindent

In this work, we have investigated the theoretical and experimental limits on the parameter space of a real singlet extension of the SM Higgs sector, {considering mass values of the second Higgs boson ranging from $1~\GeV$ to $1~\TeV$, i.e.~within the accessible mass range of past, current and future collider experiments}. This study complements a previous work~\cite{Pruna:2013bma} that was restricted to $m_H\in[600\,\GeV ,1\,\TeV]$ and {moreover} did not include {constraints from} direct {Higgs collider} searches. In the present work, either the heavy or the light {Higgs state} can take the role of the {discovered} SM-like Higgs boson at $125~\GeV$. We found that {up to Higgs masses $m \lesssim 300~\GeV$, exclusion limits from direct Higgs collider searches at LEP and the LHC, as well as the requirement of consistency with the measured SM-like Higgs signal rates} pose quite strong constraints. At higher Higgs masses, strong limits stem from electroweak precision observables, {in particular from} the $W$ boson mass calculated at NLO, as well as {from requiring} perturbativity of the couplings and vacuum stability. {The latter two are tested both at the electroweak scale and at a high scale $\mu\sim 4 \times 10^{10}~\GeV$ using} the $\beta$-functions of the theory (see e.g.~Ref.~\cite{Pruna:2013bma} and references therein).

 {We performed a exhaustive scan in the three model parameters --- specified by the Higgs mixing angle, the second Higgs mass and the ratio of the Higgs VEVs ---  and provided a detailed discussion of the viable parameter space and the relative importance of the various constraints.} We translated these {results} into predictions for collider observables {for the second yet undiscovered Higgs boson,} which are currently {investigated} by the LHC experiments. {In particular, we focussed on the} global rescaling factor $\kappa$ for the SM Higgs decay modes, {the signal rate for the Higgs-to-Higgs decay signature $H\to hh$} as well as the total width $\Gamma$ of the new scalar. A typical feature of the model is that the total width of the new scalar is quite suppressed with respect to the SM Higgs boson at such masses. {At very light Higgs boson masses below $10~\GeV$ we found that new results from LHC searches for the signature $H\to h h \to 4\mu$ are complementary to LEP Higgs searches and thus probe an unexplored parameter region. Also future $B$-factories should be able to probe these parameter regions through the decay $\Upsilon \to h \gamma$.}

{We furthermore investigated the intermediate mass region, where both Higgs masses are between $120~\GeV$ and $130~\GeV$, and discussed some of the experimental challenges in probing this scenario. Dedicated LHC searches for an additional resonance in the invariant mass spectra of the $H\to \gamma\gamma$ (see Ref.~\cite{Khachatryan:2014ira} for a CMS analysis) and $H\to ZZ^* \to 4\ell$ channel in the vicinity of the discovered Higgs boson as well as future precision experiments at the ILC or a muon collider may shed more light onto this case.}

{The discovery of additional Higgs states is one of the main goals of the upcoming runs of the LHC. In this model, two distinct and complementary signatures of the second Higgs state arise. Firstly, the $H\to hh$ decay signature, where the best sensitivity for the LHC is obtained for heavy Higgs masses between $250~\GeV$ and roughly $500~\GeV$. These signatures have been recently explored by ATLAS and CMS~\cite{CMS:2014ipa,CMS:2014eda,atlHtohh} but the analyses are not yet sensitive to constrain the parameter space. Secondly, Higgs searches designed for a SM Higgs boson are sensitive probes of the parameter space. We strongly encourage the experimental collaborations to continue these searches in the full accessible mass range. However, some of the features of the second Higgs state discussed in this work, such as the strong reduction of the total width, should be taken into account in upcoming analyses.}
Finally, we hope that the predictions of LHC {signal} cross sections at {a} CM energy of $14~\TeV$ {will be found useful for designing} some interesting benchmark points for the experimental analyses of {this} model.


\section*{Acknowledgements} 
\noindent
We thank Klaus~Desch and Sven~Heinemeyer for inspiring remarks and for motivating us to perform this study. We furthermore acknowledge helpful discussions with Philip~Bechtle, Howie~Haber, Antonio~Morais, Marco~Sampaio, Rui~Santos and Martin~Wiebusch. The code for testing perturbative unitarity has been adapted from Ref.~\cite{Pruna:2013bma}. TS is supported in part by U.S.~Department of Energy grant number DE-FG02-04ER41286, and in part by a Feodor-Lynen research fellowship sponsored by the Alexander von Humboldt Foundation.


\appendix
\section{Minimization and vacuum stability conditions}\label{app:hpot}

In this appendix we briefly guide the reader through the steps from Eq.~\eqref{potential} to Eq.~\eqref{bound_pot}, using the definition of the scalar fields given in Eq.~\eqref{unit_higgs}. We basically follow the discussion as presented in Ref.~\cite{Pruna:2013bma}.

With the definition of the VEVs according to Eq.~\eqref{unit_higgs}, the extrema of $V$ are determined using the following set of equations:
\begin{align}\label{minimisation}
\frac{\partial V}{\partial v}(v,x) &= v \cdot \left( -m^2\,+\, \lambda_1 v^2 +
\frac{\lambda_3}{2}x^2 \right)=0 \\
 \frac{\partial V}{\partial x}(v,x) &= x \cdot \left( -\mu^2\,+\, \lambda_2 x^2 +
\frac{\lambda_3}{2}v^2 \right)=0
\end{align}
The physically interesting solutions have $v, x>0$:
\begin{eqnarray}\label{min_sol1}
v^2 &=& \frac{\lambda_2 m^2 - \frac{\lambda_3}{2} \mu ^2}{\lambda_1
  \lambda_2 - \frac{\lambda_3^{\phantom{o}2}}{4}}, \\
\nonumber  \\ 
\label{min_sol2}
x^2 &=& \frac{\lambda_1 \mu^2 - \frac{\lambda_3}{2} m ^2}{\lambda_1
  \lambda_2 - \frac{\lambda_3^{\phantom{o}2}}{4}}.
\end{eqnarray}
Alternatively, we use Eq.~\eqref{minimisation} to eliminate $m^2$ and $\mu^2$, leading to
\begin{\eqn}
m^2\,=\,\lam_1\,v^2+\frac{\lam_3}{2}x^2,\qquad\,\mu^2\,=\,\lam_2\,x^2+\frac{\lam_3}{2}v^2.
\end{\eqn}
Since the denominator in Eqs.~\eqref{min_sol1}--\eqref{min_sol2} is always positive (assuming that the potential is well-defined), the numerators need to be positive as well in order to guarantee a positive-definite non-vanishing solution for $v$ and $x$.

For the determination of the extrema we evaluate the Hessian
matrix:
\begin{equation}\label{hessian}
\mathcal{H}(v,x)\equiv \left(
\begin{aligned}
\frac{\partial^2 V}{\partial v^2} &\ & \frac{\partial^2 V}{\partial v
  \partial x} \\
\frac{\partial^2 V}{\partial v \partial x} &\ & \frac{\partial^2
  V}{\partial x^2}
\end{aligned}
\right) = \left(
\begin{aligned}
2 \lambda_1 v^2 &\ & \lambda_3 v x \\
\lambda_3 v x &\ & 2 \lambda_2 x^2
\end{aligned}
\right) .
\end{equation}
From this equation, it is straightforward to verify that the solutions
are minima if and only if Eqs.~\eqref{bound_pot} and~\eqref{pos_pot} are
satisfied.
\section{RGEs for SM gauge couplings and the top quark Yukawa coupling}\label{sec:rge_sm}

This section basically follows the discussion in Ref.~\cite{Pruna:2013bma}.
In the SM, all one-loop RGEs for gauge couplings are of the form
\begin{\eqn*}
\frac{dx}{dt}\,=\,a\,x^2.
\end{\eqn*}
The exact analytic solution for this equation is given by
\begin{\eqn}\label{eq:rge_sol}
x\lb t \rb\,=\,\frac{x\lb t\,=\,t_0 \rb}{1-a\,x(t=t_0)\,\lb t-t_0 \rb},
\end{\eqn}
where for $t\,=\,\log\lb \frac{\lambda^2}{\lambda_\text{ref}^2} \rb$ we have
\begin{\eqn*}
t-t_0\,=\,2\,\log\lb \frac{\lambda}{\lambda_0} \rb.
\end{\eqn*}
For positive values of $a$, the coupling reaches the Landau pole when the denominator in Eq.~\eqref{eq:rge_sol} goes to 0; for negative values, $x\,\rightarrow\,0$ for $t\,\rightarrow\,\infty$.\\

The Yukawa coupling terms are in turn given by
\begin{\eqn*}
\frac{dx}{dt}\,=\,a\,x+b\,x^3
\end{\eqn*}
with the solution
\begin{\eqn*}
x\lb t \rb\,=\,\frac{\sqrt{a\,C'(t_0)}\,e^{a\,(t-t_0)}}{\sqrt{1-b\,e^{2\,a(t-t_0)}\,C'(t_0)}},
\end{\eqn*}
with $C'(t_0)\,=\,\frac{x^2_0}{a+b\,x_0^2}$, where $x(t=t_0)\,\equiv\,x_0$ defines the initial value. For the top quark Yukawa coupling we have
\begin{eqnarray*}
16\,\pi^2\,a &=&-4\,g_s^2-\frac{9}{8}g^2-\frac{17}{24}g'^2,\\ 
16\,\pi^2\,b\,&=&\,\frac{9}{4}.
\end{eqnarray*}
However, taking the explicit scale-dependence of the SM gauge couplings into account, the above solution needs to be modified such that $a\,(t-t_0)$ is replaced by $\int ^t_{t_0} a(t')\,dt'$. {In this work we} chose to solve the RGE of the top quark Yukawa coupling numerically.

\newpage
\bibliography{paper}

\begin{thebibliography}{100}

\bibitem{atlres}
ATLAS Collaboration, G.~Aad {\em et~al.},
\newblock Phys.Lett. {\bf B716}, 1 (2012), arXiv:1207.7214.

\bibitem{cmsres}
CMS Collaboration, S.~Chatrchyan {\em et~al.},
\newblock Phys.Lett. {\bf B716}, 30 (2012), arXiv:1207.7235.

\bibitem{Aad:2013xqa}
ATLAS Collaboration, G.~Aad {\em et~al.},
\newblock Phys.Lett. {\bf B726}, 120 (2013), arXiv:1307.1432.

\bibitem{Aad:2014eha}
ATLAS Collaboration, G.~Aad {\em et~al.},
\newblock (2014), arXiv:1408.7084.

\bibitem{Aad:2014eva}
ATLAS Collaboration, G.~Aad {\em et~al.},
\newblock (2014), arXiv:1408.5191.

\bibitem{Aad:2014xzb}
ATLAS Collaboration, G.~Aad {\em et~al.},
\newblock (2014), arXiv:1409.6212.

\bibitem{Khachatryan:2014ira}
CMS Collaboration, V.~Khachatryan {\em et~al.},
\newblock Eur.Phys.J. {\bf C74}, 3076 (2014), arXiv:1407.0558.

\bibitem{Chatrchyan:2014vua}
CMS Collaboration, S.~Chatrchyan {\em et~al.},
\newblock Nature Phys. {\bf 10} (2014), arXiv:1401.6527.

\bibitem{Chatrchyan:2013mxa}
CMS Collaboration, S.~Chatrchyan {\em et~al.},
\newblock Phys.Rev. {\bf D89}, 092007 (2014), arXiv:1312.5353.

\bibitem{Chatrchyan:2013iaa}
CMS Collaboration, S.~Chatrchyan {\em et~al.},
\newblock JHEP {\bf 1401}, 096 (2014), arXiv:1312.1129.

\bibitem{Higgs:1964ia}
P.~W. Higgs,
\newblock Phys.Lett. {\bf 12}, 132 (1964).

\bibitem{Higgs:1964pj}
P.~W. Higgs,
\newblock Phys.Rev.Lett. {\bf 13}, 508 (1964).

\bibitem{Englert:1964et}
F.~Englert and R.~Brout,
\newblock Phys.Rev.Lett. {\bf 13}, 321 (1964).

\bibitem{Guralnik:1964eu}
G.~Guralnik, C.~Hagen, and T.~Kibble,
\newblock Phys.Rev.Lett. {\bf 13}, 585 (1964).

\bibitem{Kibble:1967sv}
T.~Kibble,
\newblock Phys.Rev. {\bf 155}, 1554 (1967).

\bibitem{Aad:2014aba}
ATLAS Collaboration, G.~Aad {\em et~al.},
\newblock Phys.Rev. {\bf D90}, 052004 (2014), arXiv:1406.3827.

\bibitem{CMS:2014ega}
{CMS Collaboration},
\newblock (2014),
\newblock CMS-PAS-HIG-14-009.

\bibitem{Asner:2013psa}
D.~Asner {\em et~al.},
\newblock (2013), arXiv:1310.0763.

\bibitem{Schabinger:2005ei}
R.~Schabinger and J.~D. Wells,
\newblock Phys.Rev. {\bf D72}, 093007 (2005), arXiv:hep-ph/0509209.

\bibitem{Patt:2006fw}
B.~Patt and F.~Wilczek,
\newblock (2006), arXiv:hep-ph/0605188.

\bibitem{Barger:2007im}
V.~Barger, P.~Langacker, M.~McCaskey, M.~J. Ramsey-Musolf, and G.~Shaughnessy,
\newblock Phys.Rev. {\bf D77}, 035005 (2008), arXiv:0706.4311.

\bibitem{Bhattacharyya:2007pb}
G.~Bhattacharyya, G.~C. Branco, and S.~Nandi,
\newblock Phys.Rev. {\bf D77}, 117701 (2008), arXiv:0712.2693.

\bibitem{Dawson:2009yx}
S.~Dawson and W.~Yan,
\newblock Phys.Rev. {\bf D79}, 095002 (2009), arXiv:0904.2005.

\bibitem{Bock:2010nz}
S.~Bock {\em et~al.},
\newblock Phys.Lett. {\bf B694}, 44 (2010), arXiv:1007.2645.

\bibitem{Fox:2011qc}
P.~J. Fox, D.~Tucker-Smith, and N.~Weiner,
\newblock JHEP {\bf 1106}, 127 (2011), arXiv:1104.5450.

\bibitem{Englert:2011yb}
C.~Englert, T.~Plehn, D.~Zerwas, and P.~M. Zerwas,
\newblock Phys.Lett. {\bf B703}, 298 (2011), arXiv:1106.3097.

\bibitem{Englert:2011us}
C.~Englert, J.~Jaeckel, E.~Re, and M.~Spannowsky,
\newblock Phys.Rev. {\bf D85}, 035008 (2012), arXiv:1111.1719.

\bibitem{Batell:2011pz}
B.~Batell, S.~Gori, and L.-T. Wang,
\newblock JHEP {\bf 1206}, 172 (2012), arXiv:1112.5180.

\bibitem{Englert:2011aa}
C.~Englert, T.~Plehn, M.~Rauch, D.~Zerwas, and P.~M. Zerwas,
\newblock Phys.Lett. {\bf B707}, 512 (2012), arXiv:1112.3007.

\bibitem{Gupta:2011gd}
R.~S. Gupta and J.~D. Wells,
\newblock Phys.Lett. {\bf B710}, 154 (2012), arXiv:1110.0824.

\bibitem{Dolan:2012ac}
M.~J. Dolan, C.~Englert, and M.~Spannowsky,
\newblock Phys.Rev. {\bf D87}, 055002 (2013), arXiv:1210.8166.

\bibitem{Bertolini:2012gu}
D.~Bertolini and M.~McCullough,
\newblock JHEP {\bf 1212}, 118 (2012), arXiv:1207.4209.

\bibitem{Batell:2012mj}
B.~Batell, D.~McKeen, and M.~Pospelov,
\newblock JHEP {\bf 1210}, 104 (2012), arXiv:1207.6252.

\bibitem{Lopez-Val:2013yba}
D.~Lopez-Val, T.~Plehn, and M.~Rauch,
\newblock JHEP {\bf 1310}, 134 (2013), arXiv:1308.1979.

\bibitem{Heinemeyer:2013tqa}
The LHC Higgs Cross Section Working Group, S.~Heinemeyer {\em et~al.},
\newblock (2013), arXiv:1307.1347.

\bibitem{Chivukula:2013xka}
R.~S. Chivukula, A.~Farzinnia, J.~Ren, and E.~H. Simmons,
\newblock Phys.Rev. {\bf D88}, 075020 (2013), arXiv:1307.1064.

\bibitem{Englert:2013tya}
C.~Englert and M.~McCullough,
\newblock JHEP {\bf 1307}, 168 (2013), arXiv:1303.1526.

\bibitem{Cooper:2013kia}
B.~Cooper, N.~Konstantinidis, L.~Lambourne, and D.~Wardrope,
\newblock Phys.Rev. {\bf D88}, 114005 (2013), arXiv:1307.0407.

\bibitem{Caillol:2013gqa}
C.~Caillol, B.~Clerbaux, J.-M. Frere, and S.~Mollet,
\newblock Eur.Phys.J.Plus {\bf 129}, 93 (2014), arXiv:1304.0386.

\bibitem{Coimbra:2013qq}
R.~Coimbra, M.~O. Sampaio, and R.~Santos,
\newblock Eur.Phys.J. {\bf C73}, 2428 (2013), arXiv:1301.2599.

\bibitem{Pruna:2013bma}
G.~M. Pruna and T.~Robens,
\newblock Phys.Rev. {\bf D88}, 115012 (2013), arXiv:1303.1150.

\bibitem{Dawson:2013bba}
S.~Dawson {\em et~al.},
\newblock (2013), arXiv:1310.8361.

\bibitem{Lopez-Val:2014jva}
D.~Lopez-Val and T.~Robens,
\newblock Phys.Rev. {\bf D90}, 114018 (2014), arXiv:1406.1043.

\bibitem{Englert:2014aca}
C.~Englert and M.~Spannowsky,
\newblock Phys.Rev. {\bf D90}, 053003 (2014), arXiv:1405.0285.

\bibitem{Englert:2014ffa}
C.~Englert, Y.~Soreq, and M.~Spannowsky,
\newblock (2014), arXiv:1410.5440.

\bibitem{Chen:2014ask}
C.-Y. Chen, S.~Dawson, and I.~Lewis,
\newblock (2014), arXiv:1410.5488.

\bibitem{Karabacak:2014nca}
D.~Karabacak, S.~Nandi, and S.~K. Rai,
\newblock Phys.Lett. {\bf B737}, 341 (2014), arXiv:1405.0476.

\bibitem{Profumo:2014opa}
S.~Profumo, M.~J. Ramsey-Musolf, C.~L. Wainwright, and P.~Winslow,
\newblock (2014), arXiv:1407.5342.

\bibitem{Basso:2010jm}
L.~Basso, S.~Moretti, and G.~M. Pruna,
\newblock Phys.Rev. {\bf D82}, 055018 (2010), arXiv:1004.3039.

\bibitem{Strassler:2006im}
M.~J. Strassler and K.~M. Zurek,
\newblock Phys.Lett. {\bf B651}, 374 (2007), arXiv:hep-ph/0604261.

\bibitem{Strassler:2006ri}
M.~J. Strassler and K.~M. Zurek,
\newblock Phys.Lett. {\bf B661}, 263 (2008), arXiv:hep-ph/0605193.

\bibitem{Bechtle:2008jh}
P.~Bechtle, O.~Brein, S.~Heinemeyer, G.~Weiglein, and K.~E. Williams,
\newblock Comput.~Phys.~Commun. {\bf 181}, 138 (2010), arXiv:0811.4169.

\bibitem{Bechtle:2011sb}
P.~Bechtle, O.~Brein, S.~Heinemeyer, G.~Weiglein, and K.~E. Williams,
\newblock Comput.~Phys.~Commun. {\bf 182}, 2605 (2011), arXiv:1102.1898.

\bibitem{Bechtle:2013gu}
P.~Bechtle {\em et~al.},
\newblock PoS {\bf CHARGED2012}, 024 (2012), arXiv:1301.2345.

\bibitem{Bechtle:2013wla}
P.~Bechtle {\em et~al.},
\newblock Eur.~Phys.~J.~C {\bf 74}, 2693 (2013), arXiv:1311.0055.

\bibitem{Bechtle:2013xfa}
P.~Bechtle, S.~Heinemeyer, O.~St{\aa}l, T.~Stefaniak, and G.~Weiglein,
\newblock Eur.Phys.J. {\bf C74}, 2711 (2014), arXiv:1305.1933.

\bibitem{Bechtle:2014ewa}
P.~Bechtle, S.~Heinemeyer, O.~St{\aa}l, T.~Stefaniak, and G.~Weiglein,
\newblock JHEP {\bf 1411}, 039 (2014), arXiv:1403.1582.

\bibitem{Bowen:2007ia}
M.~Bowen, Y.~Cui, and J.~D. Wells,
\newblock JHEP {\bf 0703}, 036 (2007), arXiv:hep-ph/0701035.

\bibitem{Altarelli:1990zd}
G.~Altarelli and R.~Barbieri,
\newblock Phys.~Lett.~B {\bf 253}, 161 (1991).

\bibitem{Peskin:1990zt}
M.~E. Peskin and T.~Takeuchi,
\newblock Phys.Rev.Lett. {\bf 65}, 964 (1990).

\bibitem{Peskin:1991sw}
M.~E. Peskin and T.~Takeuchi,
\newblock Phys.Rev. {\bf D46}, 381 (1992).

\bibitem{Maksymyk:1993zm}
I.~Maksymyk, C.~Burgess, and D.~London,
\newblock Phys.Rev. {\bf D50}, 529 (1994), arXiv:hep-ph/9306267.

\bibitem{Lee:1977eg}
B.~W. Lee, C.~Quigg, and H.~Thacker,
\newblock Phys.Rev. {\bf D16}, 1519 (1977).

\bibitem{Luscher:1988gc}
M.~Luscher and P.~Weisz,
\newblock Phys.Lett. {\bf B212}, 472 (1988).

\bibitem{Basso:2010jt}
L.~Basso, A.~Belyaev, S.~Moretti, and G.~Pruna,
\newblock Phys.Rev. {\bf D81}, 095018 (2010), arXiv:1002.1939.

\bibitem{Lerner:2009xg}
R.~N. Lerner and J.~McDonald,
\newblock Phys.Rev. {\bf D80}, 123507 (2009), arXiv:0909.0520.

\bibitem{Degrassi:2012ry}
G.~Degrassi {\em et~al.},
\newblock JHEP {\bf 1208}, 098 (2012), arXiv:1205.6497.

\bibitem{Costa:2014qga}
R.~Costa, A.~P. Morais, M.~O.~P. Sampaio, and R.~Santos,
\newblock (2014), arXiv:1411.4048.

\bibitem{Alcaraz:2006mx}
ALEPH Collaboration, DELPHI Collaboration, L3 Collaboration, OPAL
  Collaboration, LEP Electroweak Working Group, J.~Alcaraz {\em et~al.},
\newblock (2006), arXiv:hep-ex/0612034.

\bibitem{Aaltonen:2012bp}
CDF Collaboration, T.~Aaltonen {\em et~al.},
\newblock Phys.Rev.Lett. {\bf 108}, 151803 (2012), arXiv:1203.0275.

\bibitem{D0:2013jba}
D0 Collaboration, V.~M. Abazov {\em et~al.},
\newblock Phys.Rev. {\bf D89}, 012005 (2014), arXiv:1310.8628.

\bibitem{Awramik:2003rn}
M.~Awramik, M.~Czakon, A.~Freitas, and G.~Weiglein,
\newblock Phys.Rev. {\bf D69}, 053006 (2004), arXiv:hep-ph/0311148.

\bibitem{Baak:2014ora}
Gfitter Group, M.~Baak {\em et~al.},
\newblock Eur.Phys.J. {\bf C74}, 3046 (2014), arXiv:1407.3792.

\bibitem{Hagiwara:1994pw}
K.~Hagiwara, S.~Matsumoto, D.~Haidt, and C.~Kim,
\newblock Z.Phys. {\bf C64}, 559 (1994), arXiv:hep-ph/9409380.

\bibitem{Goria:2011wa}
S.~Goria, G.~Passarino, and D.~Rosco,
\newblock Nucl.Phys. {\bf B864}, 530 (2012), arXiv:1112.5517.

\bibitem{Hagiwara:2005wg}
K.~Hagiwara {\em et~al.},
\newblock Phys.Rev. {\bf D73}, 055005 (2006), arXiv:hep-ph/0512260.

\bibitem{Uhlemann:2008pm}
C.~Uhlemann and N.~Kauer,
\newblock Nucl.Phys. {\bf B814}, 195 (2009), arXiv:0807.4112.

\bibitem{Wiesler:2012rkl}
D. Wiesler, DESY-THESIS-2012-048.

\bibitem{Maina:2015ela}
E.~Maina,
\newblock (2015), arXiv:1501.02139.

\bibitem{Logan:2014ppa}
H.~E. Logan,
\newblock (2014), arXiv:1412.7577.

\bibitem{Kalinowski:2008fk}
J.~Kalinowski, W.~Kilian, J.~Reuter, T.~Robens, and K.~Rolbiecki,
\newblock JHEP {\bf 0810}, 090 (2008), arXiv:0809.3997.

\bibitem{Kauer:2012hd}
N.~Kauer and G.~Passarino,
\newblock JHEP {\bf 1208}, 116 (2012), arXiv:1206.4803.

\bibitem{Chatrchyan:2012tx}
CMS Collaboration, S.~Chatrchyan {\em et~al.},
\newblock Phys.Lett. {\bf B710}, 26 (2012), arXiv:1202.1488.

\bibitem{Aad:2012an}
ATLAS Collaboration, G.~Aad {\em et~al.},
\newblock Phys.Rev. {\bf D86}, 032003 (2012), arXiv:1207.0319.

\bibitem{CMS:aya}
{CMS Collaboration},
\newblock (2012),
\newblock CMS-PAS-HIG-12-045.

\bibitem{ATLAS:2013nma}
{ATLAS Collaboration},
\newblock (2013),
\newblock ATLAS-CONF-2013-013, ATLAS-COM-CONF-2013-018.

\bibitem{Chatrchyan:2012dg}
CMS Collaboration, S.~Chatrchyan {\em et~al.},
\newblock Phys.Rev.Lett. {\bf 108}, 111804 (2012), arXiv:1202.1997.

\bibitem{CMS:xwa}
{CMS Collaboration},
\newblock (2013),
\newblock CMS-PAS-HIG-13-002.

\bibitem{ATLAS:2013wla}
{ATLAS Collaboration},
\newblock (2013),
\newblock ATLAS-CONF-2013-030, ATLAS-COM-CONF-2013-028.

\bibitem{CMS:bxa}
{CMS Collaboration},
\newblock (2013),
\newblock CMS-PAS-HIG-13-003.

\bibitem{Aad:2014ioa}
ATLAS Collaboration, G.~Aad {\em et~al.},
\newblock Phys.Rev.Lett. {\bf 113}, 171801 (2014), arXiv:1407.6583.

\bibitem{Schael:2006cr}
ALEPH Collaboration, DELPHI Collaboration, L3 Collaboration, OPAL
  Collaboration, LEP Working Group for Higgs Boson Searches, S.~Schael {\em
  et~al.},
\newblock Eur.Phys.J. {\bf C47}, 547 (2006), arXiv:hep-ex/0602042.

\bibitem{Abbiendi:2002qp}
OPAL Collaboration, G.~Abbiendi {\em et~al.},
\newblock Eur.Phys.J. {\bf C27}, 311 (2003), arXiv:hep-ex/0206022.

\bibitem{CMS:2013lea}
CMS Collaboration, {CMS Collaboration},
\newblock (2013),
\newblock CMS-PAS-HIG-13-010.

\bibitem{CMS:2013eua}
CMS Collaboration, {CMS Collaboration},
\newblock (2013),
\newblock CMS-PAS-HIG-13-025.

\bibitem{ATLAS-WW-Note}
{ATLAS collaboration},
\newblock (2014),
\newblock ATLAS-CONF-2014-060, ATLAS-COM-CONF-2014-078.

\bibitem{ATLAS-tautau-Note}
{ATLAS collaboration},
\newblock (2014),
\newblock ATLAS-CONF-2014-061, ATLAS-COM-CONF-2014-080.

\bibitem{Alexahin:2013ojp}
Y.~Alexahin {\em et~al.},
\newblock (2013), arXiv:1308.2143.

\bibitem{CMS:2014ipa}
CMS Collaboration, {CMS Collaboration},
\newblock (2014),
\newblock CMS-PAS-HIG-13-032.

\bibitem{CMS:2014eda}
CMS Collaboration, {CMS Collaboration},
\newblock (2014),
\newblock CMS-PAS-HIG-14-013.

\bibitem{Dittmaier:2011ti}
LHC Higgs Cross Section Working Group, S.~Dittmaier {\em et~al.},
\newblock (2011), arXiv:1101.0593.

\bibitem{Dittmaier:2012vm}
S.~Dittmaier {\em et~al.},
\newblock (2012), arXiv:1201.3084.

\bibitem{Wilczek:1977zn}
F.~Wilczek,
\newblock Phys.Rev.Lett. {\bf 39}, 1304 (1977).

\bibitem{Agashe:2014kda}
Particle Data Group, K.~Olive {\em et~al.},
\newblock Chin.Phys. {\bf C38}, 090001 (2014).

\bibitem{Vysotsky:1980cz}
M.~Vysotsky,
\newblock Phys.Lett. {\bf B97}, 159 (1980).

\bibitem{Nason:1986tr}
P.~Nason,
\newblock Phys.Lett. {\bf B175}, 223 (1986).

\bibitem{Gunion:1989we}
J.~F. Gunion, H.~E. Haber, G.~L. Kane, and S.~Dawson,
\newblock Front.Phys. {\bf 80}, 1 (2000).

\bibitem{Aubert:2009cka}
BaBar Collaboration, B.~Aubert {\em et~al.},
\newblock Phys.Rev.Lett. {\bf 103}, 181801 (2009), arXiv:0906.2219.

\bibitem{Lees:2012te}
BaBar Collaboration, J.~Lees {\em et~al.},
\newblock Phys.Rev. {\bf D88}, 071102 (2013), arXiv:1210.5669.

\bibitem{Aubert:2009cp}
BaBar Collaboration, B.~Aubert {\em et~al.},
\newblock Phys.Rev.Lett. {\bf 103}, 081803 (2009), arXiv:0905.4539.

\bibitem{Lees:2012iw}
BaBar collaboration, J.~Lees {\em et~al.},
\newblock Phys.Rev. {\bf D87}, 031102 (2013), arXiv:1210.0287.

\bibitem{Lees:2013vuj}
BaBar Collaboration, J.~Lees {\em et~al.},
\newblock Phys.Rev. {\bf D88}, 031701 (2013), arXiv:1307.5306.

\bibitem{Love:2008aa}
CLEO Collaboration, W.~Love {\em et~al.},
\newblock Phys.Rev.Lett. {\bf 101}, 151802 (2008), arXiv:0807.1427.

\bibitem{Dermisek:2005ar}
R.~Dermisek and J.~F. Gunion,
\newblock Phys.Rev.Lett. {\bf 95}, 041801 (2005), arXiv:hep-ph/0502105.

\bibitem{Dermisek:2006py}
R.~Dermisek, J.~F. Gunion, and B.~McElrath,
\newblock Phys.Rev. {\bf D76}, 051105 (2007), arXiv:hep-ph/0612031.

\bibitem{Domingo:2008rr}
F.~Domingo, U.~Ellwanger, E.~Fullana, C.~Hugonie, and M.-A. Sanchis-Lozano,
\newblock JHEP {\bf 0901}, 061 (2009), arXiv:0810.4736.

\bibitem{Echenard:2012hq}
B.~Echenard,
\newblock Adv.High Energy Phys. {\bf 2012}, 514014 (2012), arXiv:1209.1143.

\bibitem{Domingo:2010am}
F.~Domingo,
\newblock JHEP {\bf 1104}, 016 (2011), arXiv:1010.4701.

\bibitem{Bevan:2014iga}
BaBar Collaboration, Belle Collaboration, A.~Bevan {\em et~al.},
\newblock Eur.Phys.J. {\bf C74}, 3026 (2014), arXiv:1406.6311.

\bibitem{Abe:2010gxa}
Belle-II Collaboration, T.~Abe {\em et~al.},
\newblock (2010), arXiv:1011.0352.

\bibitem{grazzini}
M. Grazzini. Private communication.

\bibitem{atlHtohh}
{ATLAS collaboration},
\newblock (2014),
\newblock ATLAS-CONF-2014-005, ATLAS-COM-CONF-2014-007.

\end{thebibliography}

\end{document}